\documentclass[12pt]{article}
\usepackage{amssymb, amsmath,amsfonts}
\usepackage{graphicx}
\usepackage[pdftex, bookmarks=true,colorlinks,linkcolor=red,urlcolor=blue,citecolor=blue]{hyperref}
\input epsf

\makeatletter\@addtoreset{equation}{section}\makeatother

\def\be{\begin{equation}}
\def\ee{\end{equation}}
\def\bea{\begin{eqnarray}}
\def\eea{\end{eqnarray}}

\makeatletter\@addtoreset{equation}{section}\makeatother

\newcommand{\preprint}[1]{\begin{table}[t]  
             \begin{flushright}               
             {#1}                             
             \end{flushright}                 
             \end{table}}                     
\renewcommand{\title}[1]{\vbox{\center\LARGE{#1}}\vspace{5mm}}
\renewcommand{\author}[1]{\vbox{\center#1}\vspace{5mm}}
\newcommand{\address}[1]{\vbox{\center\em#1}}

\begin{document}

\unitlength = .8mm

\begin{titlepage}
\vspace{.5cm}
\preprint{IFT-UAM/CSIC-15-038}
 
\begin{center}
\hfill \\
\hfill \\
\vskip 1cm

\title{Anomalous magnetoconductivity and relaxation times in holography}

\vskip 0.5cm
{Amadeo Jimenez-Alba}\footnote{Email: {\tt amadeo.j@gmail.com}},
 {Karl Landsteiner}\footnote{Email: {\tt karl.landsteiner@uam.es}}, 
 {Yan Liu}\footnote{Email: {\tt yan.liu@csic.es} } and 
 {Ya-Wen Sun}\footnote{Email: {\tt yawen.sun@csic.es}}

\address{Instituto de F\'\i sica Te\'orica UAM/CSIC, C/ Nicol\'as Cabrera 13-15,\\
Universidad Aut\'onoma de Madrid, Cantoblanco, 28049 Madrid, Spain}

\end{center}

\vskip 1.5cm

\abstract{We study the magnetoconductivity induced by the axial anomaly via the chiral magnetic effect in strongly coupled holographic models.
An important ingredient in our models is that the axial charge is non-conserved beyond the axial anomaly. We achieve this either
by explicit symmetry breaking via a non-vanishing non-normalisable mode of
an axially charged scalar or using a St\"uckelberg field to make the AdS-bulk gauge field massive.
The DC magnetoconductivites can be calculated analytically. They take a universal form
in terms of gauge field mass at the horizon and quadratic dependence on the magnetic field.
The axial charge relaxation time grows
linearly with magnetic field in the large $B$ regime. Most strikingly positive magnetoconductivity is still present even when the relaxation times
are short $\tau_5 \approx 1/(\pi T)$ and the axial charge can not be thought of as an approximate symmetry. In the $U(1)_A$ explicit  breaking model, we also observe
that the chiral separation conductivity and the axial magnetic conductivity for the consistent axial current vanish in the limit of strong symmetry breaking.}

\vfill

\end{titlepage}


\section{Introduction}

Anomaly induced transport phenomena have been in the focus of much theoretical and experimental research lately.
One particular example is the so-called chiral magnetic effect (CME) \cite{Fukushima:2008xe}.\footnote{Early version of the CME  have appeared in \cite{Vilenkin:1980fu}. It has been derived in a variety of approaches ranging from hydrodynamics to the gauge gravity duality, e.g. 
\cite{Son:2009tf,Newman:2005hd}.
}  It states that in the presence of an imbalance
in the number of left-handed and right-handed charged fermions an electric current is generated in a magnetic field whose origin
can be traced back to the presence of the axial anomaly
\begin{equation}\label{eq:axialanomaly}
 \partial_\mu J_5^\mu= \frac{c}{8} \epsilon^{\mu\nu\rho\sigma} F_{\mu\nu} F_{\rho\sigma}\,
\end{equation}
with the anomaly constant $c=1/(2\pi^2)$ for a single Weyl fermion.  
The chiral magnetic effect is most conveniently expressed in the form
\begin{equation}\label{eq:cmecov}
 {\bf{J}} = c\mu_5\bf{B}\,,
\end{equation}
where the axial chemical potential $\mu_5$ parametrises the imbalance in the number of left- and right-handed fermions.
The subtleties in the interpretation of this formula have been discussed extensively in the literature \cite{Landsteiner:2012kd, Yamamoto:2015fxa,Kharzeev:2013ffa}.
Here it shall suffice to emphasise that $\mu_5$ is a parameter that characterises a state in which the chiral imbalance has
been induced in a dynamical way, e.g. via the very axial anomaly by switching on parallel electric and magnetic fields. Indeed
such a field configuration will induce an axial charge whose Fourier transform is
\begin{equation}
 \rho_5 = \frac{i}{\omega} c {\bf E}\cdot {\bf B}\,.
\end{equation}
We also relate axial charge and axial chemical potential via an axial charge susceptibility $\mu_5 = \chi_5 \rho_5$
and using (\ref{eq:cmecov}) we obtain the electric field response in the current
\begin{equation}\label{eq:negmagresinf}
 J_i =  \sigma_E E_i + \frac i \omega \frac{ c^2B_i B_j}{ \chi_5} E_j\,.
\end{equation}
Here $\sigma_E$ denotes the quantum critical conductivity  (also named as charge-conjugation symmetric conductivity, or Ohmic conductivity). In \cite{Landsteiner:2014vua} this effect has been studied in both hydrodynamics and holography and the result indeed confirmed the above formula. It was found that the formula applies even in the large
magnetic field regime in which the anomalous hydrodynamics of \cite{Son:2009tf} is not applicable anymore. More precisely
it turned out that 
$\sigma_E$ became anisotropic and its longitudinal component is exponentially suppressed for large $B$, see (\ref{sigmaE-prvious}).
At the same time the axial susceptibility becomes linear in the magnetic field, consistent with the expectation of Landau level
physics at weak coupling.  The delta function
peak indicated by the imaginary pole in (\ref{eq:negmagresinf}) therefore showed a cross over to a scaling behaviour that
was linear in $B$. This indeed is consistent with weak coupling results \cite{Nielsen:1983rb, sonspivak, Gorbar:2013dha}.

Thus if axial charge was conserved up to the axial anomaly this would result in a
perfect superconductor. The pole in the imaginary part implies indeed a delta function singularity located at zero frequency in the real part
via the Kramers-Kronig relation.

Alas nature seems to abhor axial symmetries even beyond the electromagnetic axial anomaly. Fermions are typically
massive. Another way the axial symmetry is broken is via the QCD contribution to the anomaly.\footnote{We assume the electromagnetic fields to be external not dynamical contributions in the axial anomaly. Instabilities arising from dynamical gauge field have been discussed in \cite{Adam:2001ma}.} In both cases it is more realistic to allow for a non-conservation
of the axial charge and introduce an axial charge relaxation time $\tau_5$. In this case the frequency is shifted 
$\omega \rightarrow \omega + i/\tau_5$
in (\ref{eq:negmagresinf}) resulting in a finite DC magnetoconductivity
\begin{equation}\label{eq:negmagresfin}
 J_i =  \sigma_E E_i + \tau_5 \frac{ c^2 B_i B_j}{\chi_5} E_j\,.
\end{equation}

In the studies up to now the relaxation time $\tau_5$ has been introduced via a naive relaxation time approximation. Basically it was added ``by hand'' and did not represent dynamics inherent to the holographic system.

Recently a number of experiments on Dirac or Weyl-(semi-)metals have reported strong positive magnetoconductivity along the magnetic field \cite{{weyl-exp}, Li:2014bha, Huang, Zhang, Xiong}. They furthermore always show a quadratic dependence on the magnetic field strength.\footnote{There is a regime for very small 
magnetic field in which the conductivity first decreases with magnetic field. This seems to be a disorder effect called ``weak anti-localization''
and most likely represents physics unrelated to the anomaly.} These materials are candidates for realising the axial anomaly in a solid state setup.
They do however also have an intrinsic axial charge relaxation time since the axial charge is realised as an accidental low energy symmetry.

This motivates us to study the anomaly induced magnetoconductivity in models in which the relaxation
time is an intrinsic property of the model. In particular we will concentrate on two models with different symmetry breaking mechanisms.

The first model uses a massless bulk gauge field for the axial current. We break the axial symmetry by introducing a tree level coupling to
a non-normalisable mode of an axially charged scalar field. We chose the bulk mass of the scalar so that the dimension of the breaking parameter
is one and can be interpreted as a mass $M$ for the dual fermions. The weak coupling interpretation is now that the divergence of the axial current is of the form
\begin{equation}
 \partial_\mu J_5^\mu = M \bar \psi\gamma_5 \psi +  c_\mathrm{em} \epsilon^{\mu\nu\rho\beta} F_{\mu\nu}F_{\rho\beta}\,.
\end{equation}
Although the system has the same ingredients as a holographic
superconductor its dynamics is quite different due to the fact that it is the non-normalisable mode that is non-vanishing. In particular
the DC conductivity is finite and the quasinormal mode spectrum does not contain sound modes.
The axial charge relaxation time is determined by the the mass parameter $M$. When $\tau_5 T\gg 1$ we can think
of the axial symmetry as approximately conserved. In this regime we indeed find linear scaling of the susceptibility and at the same time
linear scaling of the relaxation time $\tau_5$ with large magnetic field. Consequently the magnetoconductivity scales quadratic with
magnetic field which is at odds with the naive relaxation time approximation that assumes independence of $\tau_5$ on $B$.
When $\tau_5 T \simeq 1$ we do not really have justification anymore to think of the axial charge as approximately conserved.
Generically the quasinormal
frequencies for non-conserved operators have imaginary values of the order of $T$ and therefore there is no hierarchy between the
relaxation of the axial charge and a generic perturbation. In this regime the derivation of the magnetoconductivity (\ref{eq:negmagresfin})
outlined above becomes invalid.
Somewhat surprisingly we still find a magnetoconductivity that depends quadratically
in the magnetic field strength in this regime! In fact we can find an exact analytic expression for the magnetoconductivity that
confirms the exact quadratic dependence on the magnetic field for all values of the charge relaxation time.

The second model
uses a massive gauge field. As is well-known global symmetries correspond to massless gauge field in holography. The mass of the gauge field
gives the dual current operator an anomalous dimension $\Delta = -1+\sqrt{1+m^2}$. Since conformal symmetry implies that the dimension of a conserved current is precisely
$d$ in $d+1$ dimensions this means that a massive gauge field corresponds to a non-conserved vector operator in a conformal theory.
In QCD the axial anomaly for $U(1)_A$ receives not only electromagnetic but also a gluonic contribution. Holography effectively replaces the gluon
dynamics by gravity in anti-de Sitter space. It has been argued some time ago in \cite{Klebanov:2002gr} that the gravity dual of this gluonic contribution
to the anomaly is a St\"uckelberg mass term for a vector field. Therefore the divergence of the axial current is dual to the weak coupling form
\begin{equation}
 \partial_\mu J_{5}^\mu = \epsilon^{\mu\nu\rho\beta} \Big( c_\mathrm{em} F_{\mu\nu}F_{\rho\beta} + c_\mathrm{strong}\mathrm{tr}\left( G_{\mu\nu}G_{\rho\beta}\right) \Big)
\end{equation}
where the anomaly coefficients $c_\text{em}$ and $c_\text{strong}$ depend on the fermion spectrum.\footnote{In particular to make sure that the gluonic contribution is not
subleading in the supergravity limit one needs to assume typically Veneziano limit with the number of flavor to be of the same order as the number of colours \cite{Gursoy:2014ela}.}
Another way of viewing such models is by noting that the anomaly is a dimension four operator and therefore should couple to a marginal scalar field, this is
the St\"uckelberg field. One might also include another real scalar field that can serve as the dilaton, the source for the kinetic term of the non-abelian
gauge fields, which also would allow to break the underlying conformal symmetry via a dilaton flow.
Such models have already been introduced and studied in \cite{Gursoy:2014ela, Jimenez-Alba:2014iia}. We will refine that analysis of the magnetoconductivity
of \cite{Jimenez-Alba:2014iia}.
We find an analytic result for the magnetoconductivity and it turns out that it is of the same form as for the previous model, except that now the mass of the
gauge field is constant throughout the bulk spacetime. Again the magnetic field dependence is quadratic independently of the value of the gauge field mass.
We also calculate the magnetic field dependence of the axial relaxation time and the axial charge susceptibility. 
Whereas the relaxation time shows linear dependence in $B$ the susceptibility has a scaling that is determined by 
the dimension of the axial current $B^{1+\Delta}$. This is an indication that for this model the simple hydrodynamic  reasoning does not apply straightforwardly since it would predict a dependence of the form $\tau_5/ \chi_5$.
We also emphasise that when the current operator has dimension $3+\Delta$ the source for the time
component $J_5^0$ is not an axial chemical potential but rather a true coupling in the theory.

This paper is organised as follows: in section \ref{sec2} we study the model with an axially charged scalar field. We switch on a non-normalisable mode that serves as an axial symmetry breaking parameter (fermion mass). We compute the magnetoconductivity analytically and check the expression numerically. We numerically
study the large $B$ behaviour of the relaxation time and the axial susceptibility and find linear scaling for both in the regime $T \tau_5 \gg 1$. This is the regime in which the hydrodynamic reasoning applies and it indicates indeed quadratic dependence of the magnetoconductivity on $B$. In the regime $T \tau_5 \simeq 1$ hydrodynamics does not apply anymore since the axial charge is not even approximately conserved but the analytic
formula is of course still valid in this regime. 
 In section \ref{sec3} we study the model with a St\"uckelberg axion. We find the analytic result for the magnetoconductivity and check it numerically.
In this model we can also find an analytic result for the axial charge susceptibility. It shows scaling with $B^{1+\Delta}$ for large $B$. Numerically we also find linear scaling in $B$ of the axial charge relaxation time.
In section \ref{sec4} we present a comparison of the two models and an alternative derivation of the DC conductivity based on near far matching method. 
We finish in section \ref{sec5} with a discussion of our results. Technical details of the calculations are summarised in the appendix.

\section{Holographic $U(1)_\text{A}$  explicit breaking model}
\label{sec2}

It has long been known that for a chiral anomalous fluid in presence of magnetic field, a large longitudinal magnetoconductivity due to the chiral anomaly is induced.
\cite{Nielsen:1983rb, sonspivak, Gorbar:2013dha}. In a previous study in the framework of hydrodynamics \cite{Landsteiner:2014vua}, in order to get a finite longitudinal DC magnetoconductivity,
it was necessary to include energy, momentum and also (axial) charge dissipation. However, in the limit of zero densities (i.e. $\mu/(\epsilon+p)=\mu_5/(\epsilon+p)=0$), axial charge dissipation
alone suffices to give a finite value.


In this section, we consider the holographic axial charge dissipation effect in the magnetoconductivity by breaking the axial charge $U(1)_A$ symmetry with a scalar source. We will also introduce two $U(1)$ symmetries and because the electric $U(1)_V$ symmetry is still conserved, the scalar field only couples to the axial gauge field in the bulk. In this way the anomaly caused buildup
of axial charge will be compensated by the dissipation sourced by this scalar operator resulting in a finite DC longitudinal magnetoconductivity. We will consider the following action\footnote{We have set the curvature scale $L=1$.}
\bea
S&=&\int d^5x \sqrt{-g}\bigg[\frac{1}{2\kappa^2}\Big(R+12\Big)-\frac{1}{4}\mathcal{F}^2-\frac{1}{4}F^2+\frac{\alpha}{3}\epsilon^{\mu\nu\rho\sigma\tau}A_\mu \Big(F_{\nu\rho} F_{\sigma\tau}+3 \mathcal{F}_{\nu\rho} \mathcal{F}_{\sigma\tau}\Big)\nonumber\\&&~~~-(D_\mu\Phi)^*(D^\mu\Phi)-m_s^2\Phi^*\Phi\bigg]\,
\eea
with
$$
\mathcal{F}_{\mu\nu}=\partial_\mu V_\nu-\partial_\nu V_\mu\,,
~~F_{\mu\nu}=\partial_\mu A_\nu-\partial_\nu A_\mu\,,~~D_\mu=\nabla_\mu-iqA_\mu\,
$$
where the gauge fields $V_\mu$ and $A_\mu$ correspond to the vector and axial $U(1)$ currents respectively and $\Phi$ is a complex scalar field with mass $m_s$. Similar models have been
studied before in e.g. \cite{Amado:2014mla,Jimenez-Alba:2014pea} to  describe the dual anomalous superconductor with $U(1)_V\times U(1)_A$ symmetry where a charged scalar field is introduced to spontaneously break $U(1)_V$ while not $U(1)_A$.  
In this paper we shall turn on a non-zero source associated to the dual scalar operator in order to break the $U(1)_A$ symmetry explicitly, which introduces an {\em axial charge dissipation} mechanism in our system.

Note that we do have a conserved current associated to $U(1)_V,$ which means that the electric charge is always conserved. With the mass term for the scalar field, the dual scalar operator has a scaling dimension $\Delta_{\Phi}=2\pm \sqrt{4+m_s^2}$ and to make sure that the scaling dimension of the axial current does not change, $m_s^2$ has to be negative and above the BF bound (i.e. $-4\leq m_s^2<0$). Without loss of generality, in the following we will choose $m_s^2=-3$, which is the most appropriate value as at this mass the scalar operator reproduces exactly the scaling dimension of the mass term for free fermions.\footnote{Of course the properties of the holographic system could be far away from the free massive fermion system. However it is very intriguing that we do find some similar properties of our holographic model compared to free Fermi gas picture.  This is similar to the fermionic holographic case with special mass of the probe fermion such that the dimension of the dual fermionic operator approaches that of a free fermion where the dual system has Fermi surface but not exactly Landau's Fermi liquid theory  \cite{{Cubrovic:2009ye},Faulkner:2009wj}.} Thus the conformal dimension of the dual scalar operator is 3 and the corresponding source is of dimension 1. This reminds us to the four dimensional free massive fermion systems.

We will study the system in the probe limit, which means only axial charge dissipation is required to get a finite longitudinal DC magnetoconductivity. In the probe limit, fields live in the Schwartzschild black hole background in the bulk
\be
ds^2=r^2 \big(-f(r)dt^2+dx^2\big)+\frac{dr^2}{r^2 f(r)}\,,~~~~f(r)=1-\frac{r_0^4}{r^4}\,,
\ee
with the dual thermodynamical quantities
$$\epsilon=3r_0^4\,,~~~ s=4\pi r_0^3\,,~~~ T=\frac{r_0}{\pi}\,.$$
The equations of motion for these matter fields in the Schwarzschild-$AdS_5$ background are
\bea\label{eq:eomtwou1-1}
\nabla_\nu \mathcal{F}^{\nu\mu}+2\alpha\epsilon^{\mu\tau\beta\rho\sigma} F_{\tau\beta}\mathcal{F}_{\rho\sigma}&=&0\,,\\
\label{eq:eomtwou1-2}\nabla_\nu F^{\nu\mu}+\alpha\epsilon^{\mu\tau\beta\rho\sigma} \big(F_{\tau\beta}F_{\rho\sigma}
+\mathcal{F}_{\tau\beta}\mathcal{F}_{\rho\sigma}\big)+i q\big(\Phi (D^\mu\Phi)^*-\Phi^*(D^\mu\Phi)\big)&=&0\,,\\
\label{eq:eomtwou1-3}D_\mu D^\mu\Phi-m_s^2\Phi&=&0\,.
\eea 

Let us briefly comment on 
the Ward identity for the current. 
The dual {\em consistent} currents are obtained as the variation of the total action with respect to the gauge fields,
\begin{align}\label{1pexpV}
J^\mu_{(\text{con})} &= \lim_{r\rightarrow\infty}\sqrt{-g}\Big(\mathcal{F}^{\mu r}+4\alpha\epsilon^{r \mu\beta\rho\sigma} A_{\beta}\mathcal{F}_{\rho\sigma}  \Big) + \text{c.t.}\,,\\
\label{1pexp}
 J^\mu_{5(\text{con})} &= \lim_{r\rightarrow\infty}\sqrt{-g}\Big(F^{\mu r}+\frac{4\alpha}{3}\epsilon^{r \mu\beta\rho\sigma} A_{\beta}F_{\rho\sigma}  \Big) + \text{c.t.}\,.   
\end{align}
From the on-shell condition, we have
\begin{align}
\partial_\mu J_{(\text{con})}^\mu &= 0\,,\\
\label{1pdexp}
\partial_\mu J_{5(\text{con})}^\mu &= \lim_{r\rightarrow\infty}\sqrt{-g}\left( -\frac{\alpha}{3}\epsilon^{r \mu\beta\rho\sigma} \left(F_{\mu\beta}F_{\rho\sigma}+3\mathcal{F}_{\mu\beta}\mathcal{F}_{\rho\sigma}\right) -iq\left[ \Phi (D^r \Phi)^*-\Phi^*(D^r \Phi)\right] \right) + \text{c.t.}\,.
\end{align}

The counterterm contribution is not explicitly shown because it does not add any valuable information. 
The last term in (\ref{1pdexp}) contributes to the 1-point function only if the non-normalisable mode of the scalar field is switched on. It is tempting 
to interpret it as the contribution of the fermion mass to the Ward identity via $M \bar{\psi}\gamma^5\psi$.  
The {\em covariant} version the current is obtained by removing the Chern-Simons term from (\ref{1pexpV} - \ref{1pexp}), i.e. 
\begin{align}
J^\mu&=\lim_{r\rightarrow\infty}\sqrt{-g} \mathcal{F}^{\mu r}+ \text{c.t.}\,,\\
J^\mu_{5}&=\lim_{r\rightarrow\infty}\sqrt{-g} F^{\mu r}+ \text{c.t.}\,.
\end{align}
Since the covariant current is the one which has been widely used in the framework of hydrodynamics,  we will use the covariant current in most parts of our paper and briefly comment on the behaviour of the consistent current in subsection \ref{ssec:anocoe}. 



\subsection{Background in the probe limit}

Since we are going to study the magneto response, we turn on a background magnetic field in the $U(1)_V$ sector.
We consider the following background with non zero components: 
\be\label{eq:bgansatz}
V_\mu=\big(V_t(r),By, 0,V_z(r),0\big)\,,~~~A_\mu=\big(A_t(r),0, 0,A_z(r),0\big)\,,~~~ \Phi(r)=\phi(r)\,.\ee
It is useful to note that this ansatz is invariant under transformation $(r,r_0, V_t, V_z, A_t, A_z)\to b^{-1} (r,r_0, V_t, V_z, A_t, A_z),$ $(t, x, y, z)\to b (t,x,y,z)$ and $B \to b^{-2} B.$ We can set $r_0$ to be 1 by choosing $b=r_0.$ Plugging this ansatz into the equations (\ref{eq:eomtwou1-1} - \ref{eq:eomtwou1-3}) we obtain five second order ODEs for five unknown real functions $V_t, V_z, A_t, A_z, \phi$.  The equations can be found in (\ref{eq:twou1forbg1} - \ref{eq:twou1forbg5}) of the appendix \ref{app:eqexplicit}. We can furthermore reduce them  
into four second order ODEs as (\ref{eq:twou1forbg1a}), (\ref{eq:twou1forbg2}), (\ref{eq:twou1forbg3}) and  (\ref{eq:twou1forbg5}) for four functions $V_t, \phi, A_t, A_z.$ Once we have the solutions for these four fields, $V_z$ is totally fixed by imposing normalisable boundary condition. 

For $m_s^2=-3,$ near the conformal boundary $r\to\infty$ we have
\bea
 V_t&\simeq& \mu -\frac{\rho}{2r^2}\,,\\
\phi&\simeq& \frac{M}{r} \Big(1+\frac{\lambda_1}{r^2}\ln r\Big)+\frac{\varphi}{r^3}\,,\\
 A_t&\simeq& \mu_5 \Big(1+\frac{\mu_1}{r^2}\ln r\Big)-\frac{\rho_5}{2r^2}\,,\\
A_z&\simeq& s_0\Big(1+\frac{s_1}{r^2}\ln r\Big)+\frac{s_2}{r^2} \,
\eea
with $\lambda_1=\frac{1}{2}(\mu_5^2-s_0^2) q^2$, $\mu_1=s_1=-q^2M^2.$
The interpretation for the boundary value is the following:  $\mu, \mu_5$ are the chemical potential and axial chemical potential respectively. $M$ is the source of the scalar operator which explicitly breaks the $U(1)_A$ symmetry and the corresponding response can be obtained to be $\mathcal{O}=\frac{\delta S_\text{ren}}{\delta M}=2\varphi$.\footnote{Note that $S_\text{ren}=S+S_\text{c.t}$ where the counterterm for this special mass is \be\label{eq:ct} S_\text{c.t}=\int_\text{bnd}d^4x\sqrt{-\gamma}\Big(-|\Phi|^2+\frac{1}{2}\big(\log r^2-1\big)\big[\frac{1}{4}F^2+\frac{1}{4}\mathcal{F}^2+|D_\mu\Phi|^2\big]\Big)\ee and $\gamma_{\mu\nu}$ is the induced metric near the boundary.}
The parameter $s_0=0$ is the analogue of a superfluid velocity. In our context it can also be interpreted as the holographic analogue of the separation of the Weyl nodes
in momentum space of a Weyl semi-metal.

For this system, the solutions of four second order ODEs  (\ref{eq:twou1forbg1a}), (\ref{eq:twou1forbg2}), (\ref{eq:twou1forbg3}) and  (\ref{eq:twou1forbg5}) are specified by eight integration constants, which correspond to the eight independent parameters at the conformal boundary $\mu, \rho, \mu_5, \rho_5, M, \varphi, s_0, s_2$. Near the horizon, there are four independent parameters $ A_t'(r_0), V_t'(r_0), A_z(r_0), \phi(r_0)$ after specifying regular horizon boundary conditions for $A_t$, $V_t$, $\phi$ and $A_z$. Thus for fixed values of $m_s^2, q$, we have a {\em four}-parameter family of black hole solutions, which correspond to $\mu,$ $\mu_5,$ $M$, $s_0$. 


For our purpose of breaking the axial charge conservation symmetry explicitly at the boundary, we focus on solutions with nonzero $M$, which sources this explicit breaking of axial charge conservation. Solutions with finite chemical potential $\mu$ and axial chemical potential $\mu_5$ can only be obtained numerically. In fact, as we already know from \cite{Landsteiner:2014vua}, the infinite axial charge transfer due to chiral anomaly exists even at both $\mu=0$ and $\mu_5=0$. Thus in this paper we will mainly focus on the $\mu=\mu_5=0$ case and sometimes also the $\mu=0,\mu_5 \neq 0$ and $\mu_5=0, \mu \neq 0$ cases for simplicity. We also set the parameter $s_0=0$. In the following we list the bulk background solutions for $\mu=\mu_5=0$, $\mu=0,\mu_5 \neq 0$ and $\mu_5=0, \mu \neq 0$ respectively.

\begin{itemize}
\item The solution for $\mu=\mu_5=0$ can be obtained easily by choosing $V_t=V_z=A_t=A_z=0$ and in this case we can still have non trivial solution for $\phi$ with source $M$. $\phi$ can be solved to be  
\be\label{specialphi} \phi=\frac{M}{T}\bigg(\frac{2}{\pi}\bigg)^{3/2}\frac{r_0\Gamma[3/4]}{r\Gamma[1/4]}\text{EllipticK}\Big[\frac{1-\frac{r_0^2}{r^2}}{2}\Big]\,,\ee
where the function EllipticK gives the complete elliptic integral of the first kind. As $\mu=\mu_5=0$, the response of the scalar operator is completely determined by the source $M$ and we have 
\be\label{eq:hbrelation}
 \frac{M}{\phi_0}=\frac{\Gamma[1/4]}{\sqrt{2\pi}\Gamma[3/4]}r_0\simeq 1.18 r_0\,,~~~~
\frac{\mathcal{O}}{M}=-\frac{\Gamma[3/4]^2}{4\Gamma[5/4]^2}r_0^2\simeq -0.456 r_0^2\,,\ee 
with $M,\mathcal{O}$ the source and response associated to the scalar operator and $\phi_0$ the horizon value of the scalar field. In this case note that for $M=0$ we only have the trivial solution $\phi=0$.

\item
The second simple while interesting solution is for $V_t=A_z=0$ while $A_t, V_z \neq 0$ (i.e. $\mu=0, \mu_5\neq 0$). In this case solutions are governed by the equations (\ref{eq:twou1forbg1a}) and  (\ref{eq:twou1forbg5}) and we can solve the system numerically. The plot for the scalar response as a function of $T$ or $\mu_5$ can be found in Fig. \ref{fig:phase} for several fixed values of source $M$. In this case, when $M=0$ there can be nontrivial solutions below a certain critical  temperature $T_c$, which correspond to spontaneous symmetry breaking. For $T>T_c$ and $M=0$, the $U(1)_A$ symmetry is not broken
and we have the same configuration as \cite{Landsteiner:2014vua}. When $\mu_5/T\to 0,$ the system becomes the previous case with $\mu=\mu_5=0.$ We have checked that the numerical solutions for the response in this limit satisfy precisely the relation (\ref{eq:hbrelation}).

\begin{figure}[ht]
\begin{center}
\includegraphics[width=0.49\textwidth]{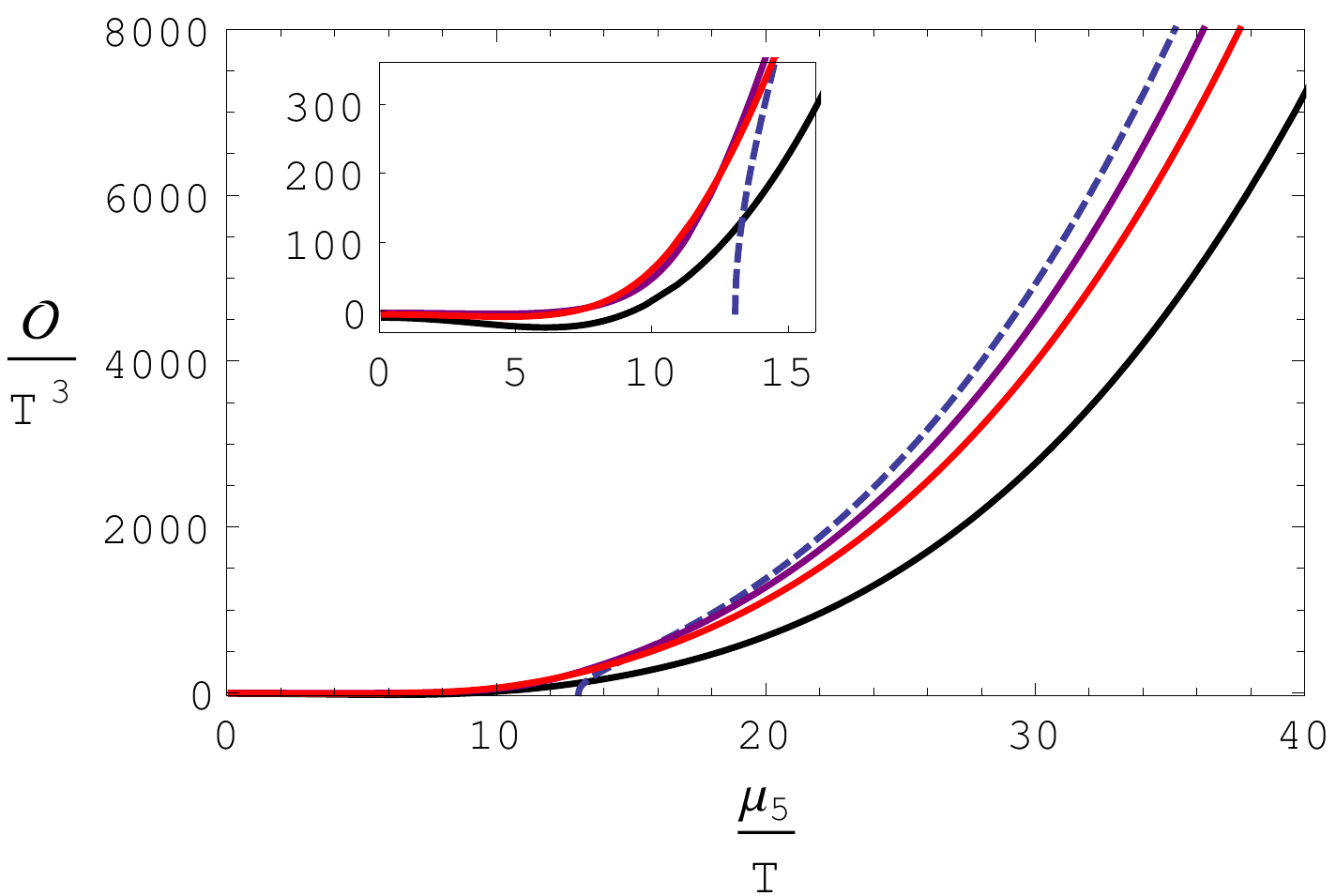}
\includegraphics[width=0.49\textwidth]{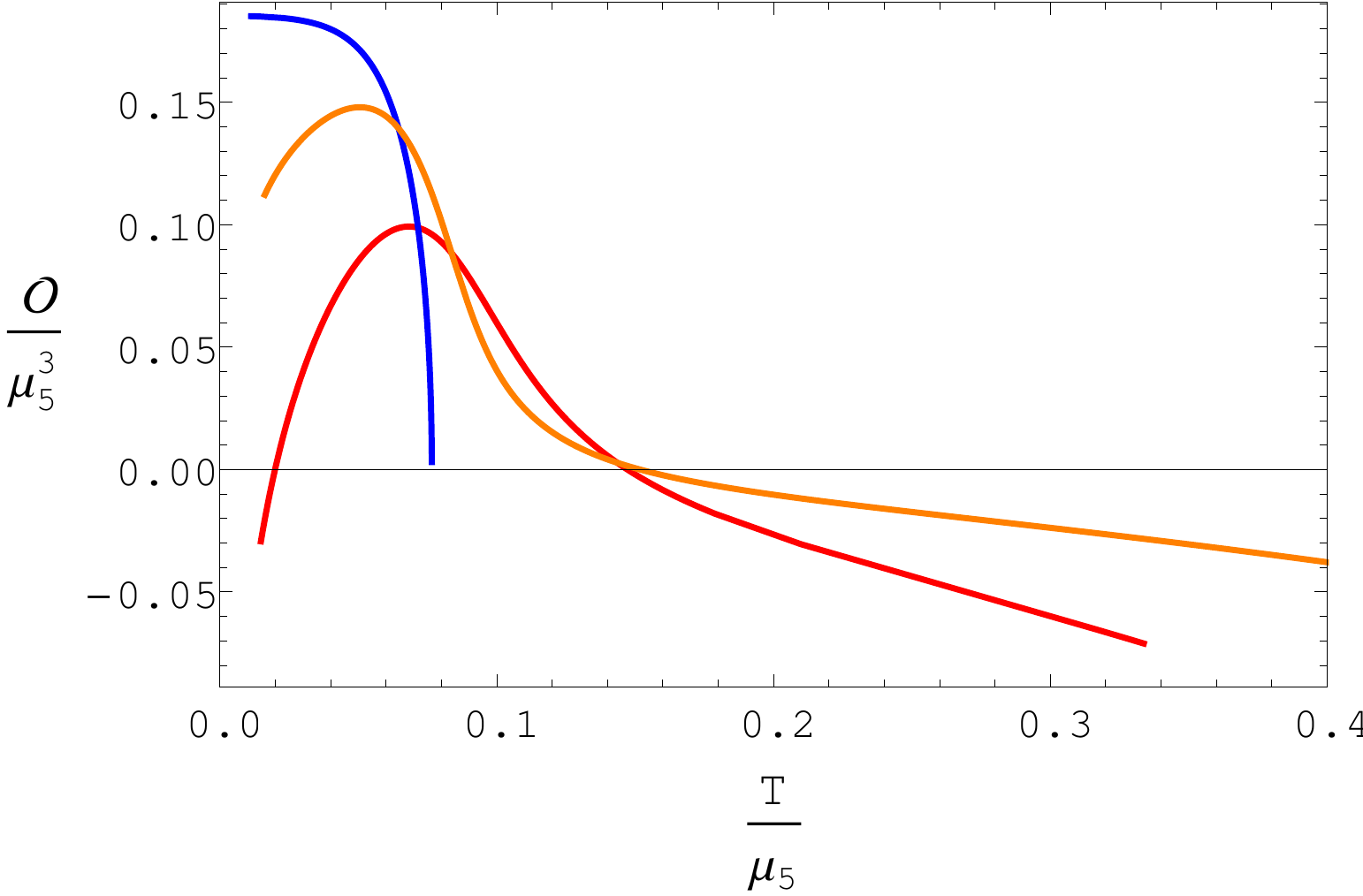}
\parbox{15cm}{\caption{\small The response $\mathcal{O}$ as a function of $T$ and $\mu_5$ for different sources $M$. Left: Fixing $T$ and $B$ to be $8B\alpha/(\pi^2 T^2)=0.1$ and varying $\mu_5$ for $qM/T=0$ (blue), $0.5$ (purple), $1$ (red) and $2$ (black). Right: Fixing $\mu_5$ and varying $T$ and $B$ at the same time with fixed ratio $8B\alpha/(\pi^2 T^2)=0.1$, $qM/\mu_5=0$ (blue), $0.04$ (orange), $0.1$ (red).  }\label{fig:phase}}
\end{center}
\end{figure}

\item
The third kind of simple while interesting solutions is for $A_t=V_z=0$ while in general $V_t, A_z$ non zero (i.e. $\mu_5=0, \mu\neq 0$).  In this case for $M=0$ the solution for $\phi$ will be trivial despite $\mu\neq 0$: no solution with spontaneous symmetry breaking will exist due to the zero axial chemical potential. The solutions of $\phi$ for $M\neq 0$ are different from the solutions in the $\mu=\mu_5=0$ case due to nonzero $A_z$ for $B\neq 0$ case and the solutions can be obtained numerically, which we will not give in detail here. When $B=0$ we have a simple solution $V_t=\mu(1-\frac{r_0^2}{r^2}), V_z=A_t=A_z=0$ and $\phi$ is the same as (\ref{specialphi}).
\end{itemize}
In the following, we will mainly consider the magnetoconductivity\footnote{Recently the 
magnetoconductivity has been widely studied in e.g.  \cite{Blake:2014yla,{Amoretti:2015gna},{Blake:2015ina},{Kim:2015wba},{Zhou:2015dha}} for AdS$_4/$CFT$_3$ with momentum dissipations. The {\em longitudinal} magnetoconductivity is generally monotonically non-increasing as a function of the magnetic field \cite{wannier}, i.e. positive magnetoresistivity. Though there can be exceptions, e.g. systems with paramagnetic impurities can have isotropic positive magnetoconductivity, which is caused by $B$ dependent scatterings. Here we wish to emphasise that the negative (anisotropic)  longitudinal magneoresistivity is driven by axial anomaly and can be the smoking gun for the existence of axial anomaly.} 
for the first case, i.e. $\mu=\mu_5=0$ and will comment on the other two cases as well for Hall conductivity and the anomalous transport. A detailed study of the most general cases will be reported in a follow up work. 

\subsection{Magnetoconductivity and relaxation time}

Without any axial charge dissipation, we know that the dual anomalous system will have an infinite DC longitudinal conductivity in presence of 
a background magnetic field \cite{Landsteiner:2014vua}. To show that our explicit breaking $U(1)_A$ model indeed encodes the axial charge dissipation, we need to compute the electric conductivity and the axial charge relaxation time. We consider the following fluctuations at zero momentum on top of the  background (\ref{eq:bgansatz})
$$\delta V_\mu= v_\mu (r) e^{-i \omega t}\,,~~~~\delta A_\mu= a_\mu (r) e^{-i \omega t}\,,~~~~ \delta\Phi= \Phi_1(r,t)+i \Phi_2 (r,t)=(\phi_1+i\phi_2) e^{-i \omega t}\,.$$  
For the purpose of calculating the electric conductivity, we will impose the sourceless boundary condition for $\delta \Phi$ and $\delta A_{\mu}$ while $\delta E_z=i w v_z$ at the boundary. We can always choose the gauge $v_r=a_r=0$ (or $v_r=\phi_2=0$) using the fact that the equations in the bulk are invariant under  the transformation \be\label{gaugetransf} \delta V_\mu\to \delta V_\mu+\partial_\mu\Lambda_1\,,~~\delta A_\mu\to\delta A_\mu+\partial_\mu\Lambda\,,~~ \Phi_1\to\Phi_1-q\Lambda \Phi_2\,,~~\Phi_2\to\Phi_2+q\Lambda ( \phi+\Phi_1)\,. \ee Note that in the gauge transformations above, the order of the gauge transformations should be at the level of perturbations, thus the terms $q\Lambda \Phi_2$ and $q\Lambda \Phi_1$ can be ignored as they are second order in perturbations. For these fluctuations at zero momentum,  the equations for the longitudinal fluctuations ${a_t, v_z, v_t, v_z, \phi_1,\phi_2}$ and transverse fluctuations ${a_x, a_y, v_x, v_y}$ decouple from each other. In the $\mu=\mu_5=0$ background, the fluctuation $\phi_1$ also decouples from other longitudinal modes, which means that only the fluctuation of the phase mode $\phi_2/(q\phi)$ plays a role here. These equations can be found in appendix \ref{secEqfluc1} and \ref{secEqfluc2}. We will focus on the longitudinal fluctuations to study the effect of axial charge dissipations on the longitudinal magnetoconductivity in this paper. The discussion on transverse magnetoresistivity and Hall conductivity can be found in appendix \ref{secEqfluc2}. 


For the case $\mu=\mu_5=0$ (i.e. $A_t=A_z=V_t=V_z=0$),  we have three independent equations of motion for $a_t, v_z, \phi_2$ as follows
\bea
\label{eq:flu1} a_t''+\frac{3}{r}a_t'-\frac{2 q^2\phi^2}{r^2 f}a_t+\frac{8B\alpha }{r^3}v_z'-i\omega\frac{2q\phi\phi_2}{r^2f}&=&0\,,
 \\
\label{eq:flu2}\omega a_t'+\frac{8B\alpha \omega}{r^3}v_z+2i q r^2 f\big(-\phi_2\phi'+\phi\phi_2'\big)&=&0\,,\\
\label{eq:flu3}v_z''+\bigg(\frac{3}{r}+\frac{f'}{f}\bigg)v_z'+\frac{\omega^2}{r^4 f^2}v_z+\frac{8B\alpha}{r^3f}a_t'
&=&0\,.\eea

Note that the second equation is the equation of motion for $a_r$ and is of first derivative in the fields. The second order equation for $\phi_2$ (i.e. the last equation in appendix \ref{secEqfluc1} with $A_t=A_z=0$) consistently follows from (\ref{eq:flu1}), (\ref{eq:flu2}) and the background equation of $\phi$. One immediate observation is that $\Sigma_5=0$ which is defined as the response of $J_{Az}$ to the electric field $E_z$ and this result is the same as the case without axial charge dissipation \cite{Landsteiner:2014vua}.

When $r\to \infty$, we have 
\be
v_z=v_z^{(0)}+\frac{v_z^{(1)}}{r^2}\ln r+\frac{v_z^{(2)}}{r^2}\,,~~~a_t=a_t^{(0)}+\frac{a_t^{(1)}}{r^2}\ln r+\frac{a_t^{(2)}}{r^2}\,.
\ee
From (\ref{eq:flu2}), we have 
\be\phi_2=q\phi\Big(c_0-\int_r^\infty \frac{i\omega(r^3 a_t'+8B\alpha v_z)}{2q^2 r^5 f\phi^2}\Big)\, ,\ee where $c_0$ is an integration constant which can be chosen to be zero considering the sourceless boundary condition for $\phi_2$. 
Thus when $r\to \infty,$ we have
\be
\phi_2=q\phi\bigg(c_0+\frac{i\omega a_t^{(1)}}{2q^2M^2}\frac{\ln r}{r^2}+\frac{i\omega(2 a_t^{(2)}-8B\alpha v_z^{(0)})}{4r^2q^2M^2}\bigg)\,.\ee 
From (\ref{eq:flu1}) and (\ref{eq:flu3}), we have 
$a_t^{(1)}=-q^2M^2a_t^{(0)}-i\omega q^2M^2 c_0$ and $v_t^{(1)}=\frac{\omega^2}{2}v_t^{(0)},$ which show that the boundary constants $a_t^{(1)}$, $v_t^{(1)}$ are fully determined by  $a_t^{(0)}$ and $v_t^{(0)}$. Note that we always use the combination $\phi_2/(q\phi)$ which is in fact the phase mode fluctuation $-\theta(r)$ if we write $\delta \phi=\phi(r)e^{i\theta(r)}$.

To get the magnetoconductivity by solving (\ref{eq:flu1} - \ref{eq:flu3}), we impose infalling boundary conditions 
for $v_z, a_t, \phi_2$ at the horizon and sourceless condition for $a_t, \phi_2$ at the conformal boundary, i.e. $c_0=a_t^{(0)}=0$. At the horizon, the infalling boundary conditions give 
\bea\label{infallingboundary}
v_z&=&(r-r_0)^{-\frac{i\omega}{4r_0}}\bigg(v_{(0)}-\frac{8B\alpha a_{(1)}(4r_0-i\omega)+v_{(0)}\omega(\omega-2ir_0)}{8r_0^2(2 r_0-i\omega)}(r-r_0)+\dots\bigg)\,,\nonumber \\
\label{eq:bndat}a_t&=&(r-r_0)^{-\frac{i\omega}{4r_0}}\bigg(a_{(1)} (r-r_0)+\dots\bigg)\,,\\
\phi_2/(q \phi)&=&  (r-r_0)^{-\frac{i\omega}{4r_0}}\bigg(-\frac{\pi\Gamma[1/4]^2\big(a_{(1)}r_0^2(4r_0-i\omega)+32B\alpha v_{(0)}\big)}{16q^2 r_0^3 \Gamma[3/4]^2(M/T)^2}+\dots\bigg)\,,\nonumber
\eea where the dots denote terms which are higher order corrections.

In numerics, notice that we have only two linearly independent solutions for  (\ref{eq:flu1} - \ref{eq:flu3}) which are fully determined by the near horizon values $a_{(1)}, v_{(0)}$.  Using the fact that the 
system is invariant under the residual symmetry $a_t\to a_t+i\omega\Lambda, \phi_2\to\phi_2 -q \Lambda \phi$, where $\Lambda$ is a constant independent of $r$, we will be able to generate solutions with $c_0=0$ for each independent numerical solution. Then we can use their linear combination to set $a_t^{(0)}=0$ in numerical calculations \cite{Amado:2009ts}.

\begin{figure}[h!]
\begin{center}
\includegraphics[width=0.49\textwidth]{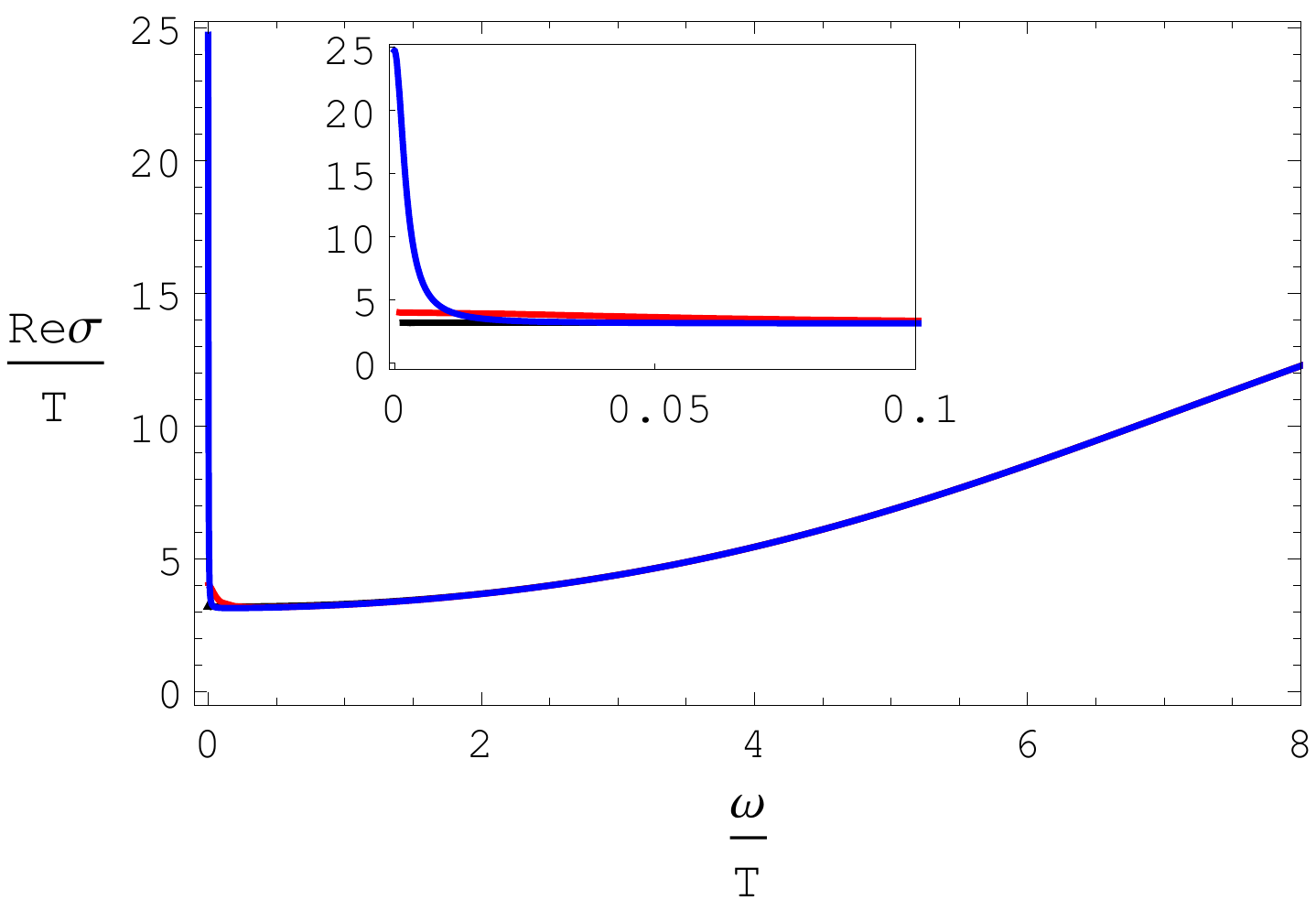}
\includegraphics[width=0.49\textwidth]{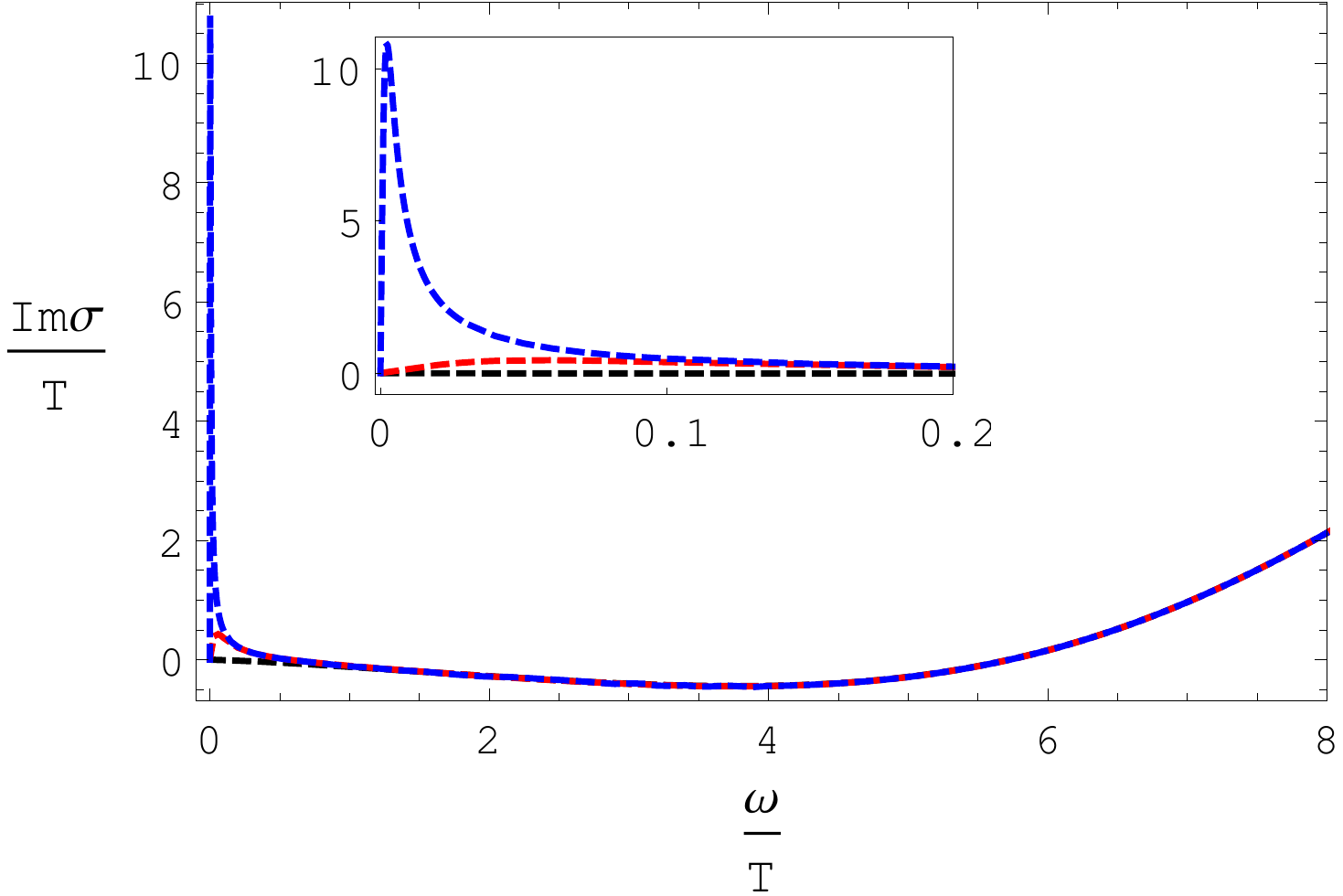}
\includegraphics[width=0.49\textwidth]{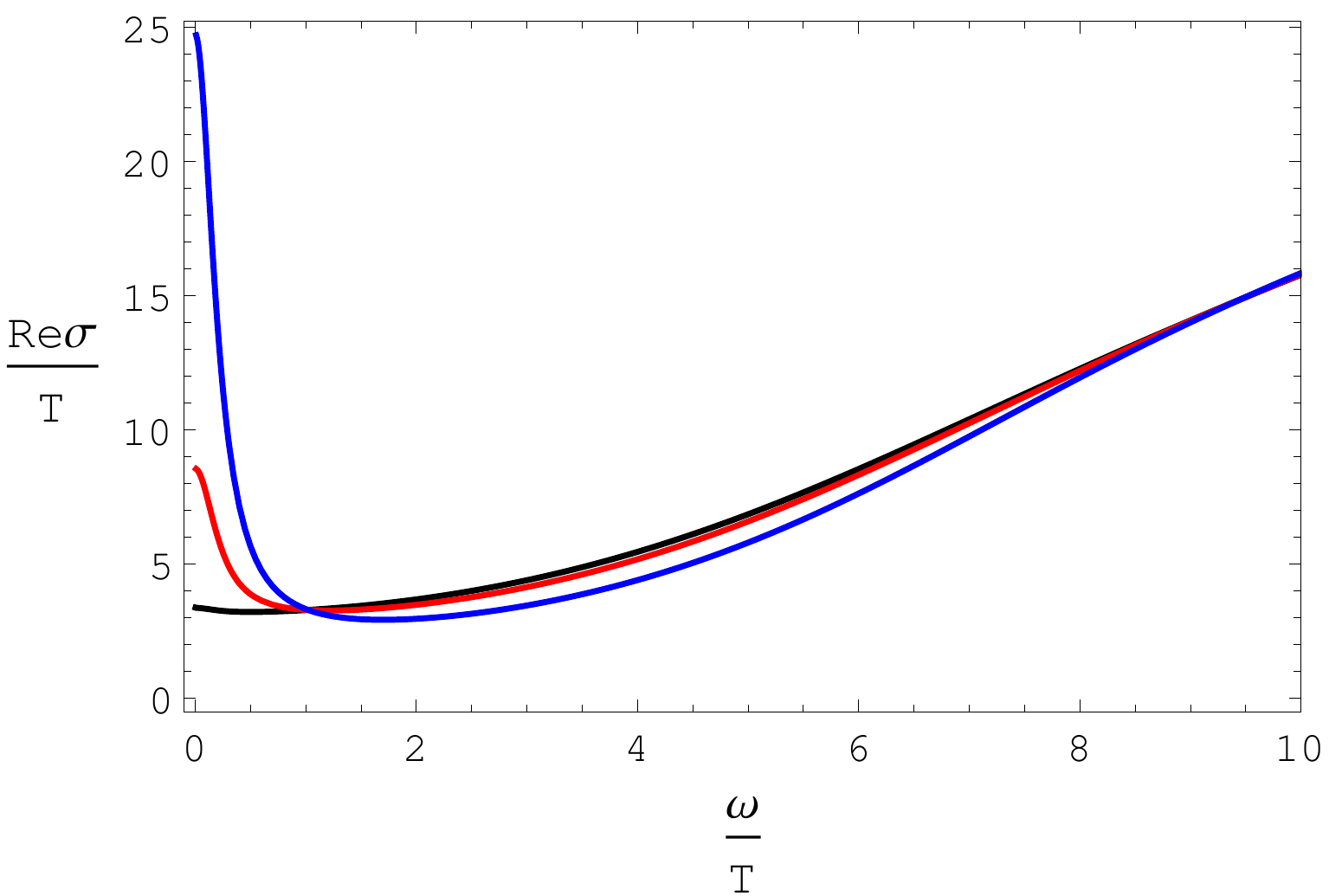}
\includegraphics[width=0.49\textwidth]{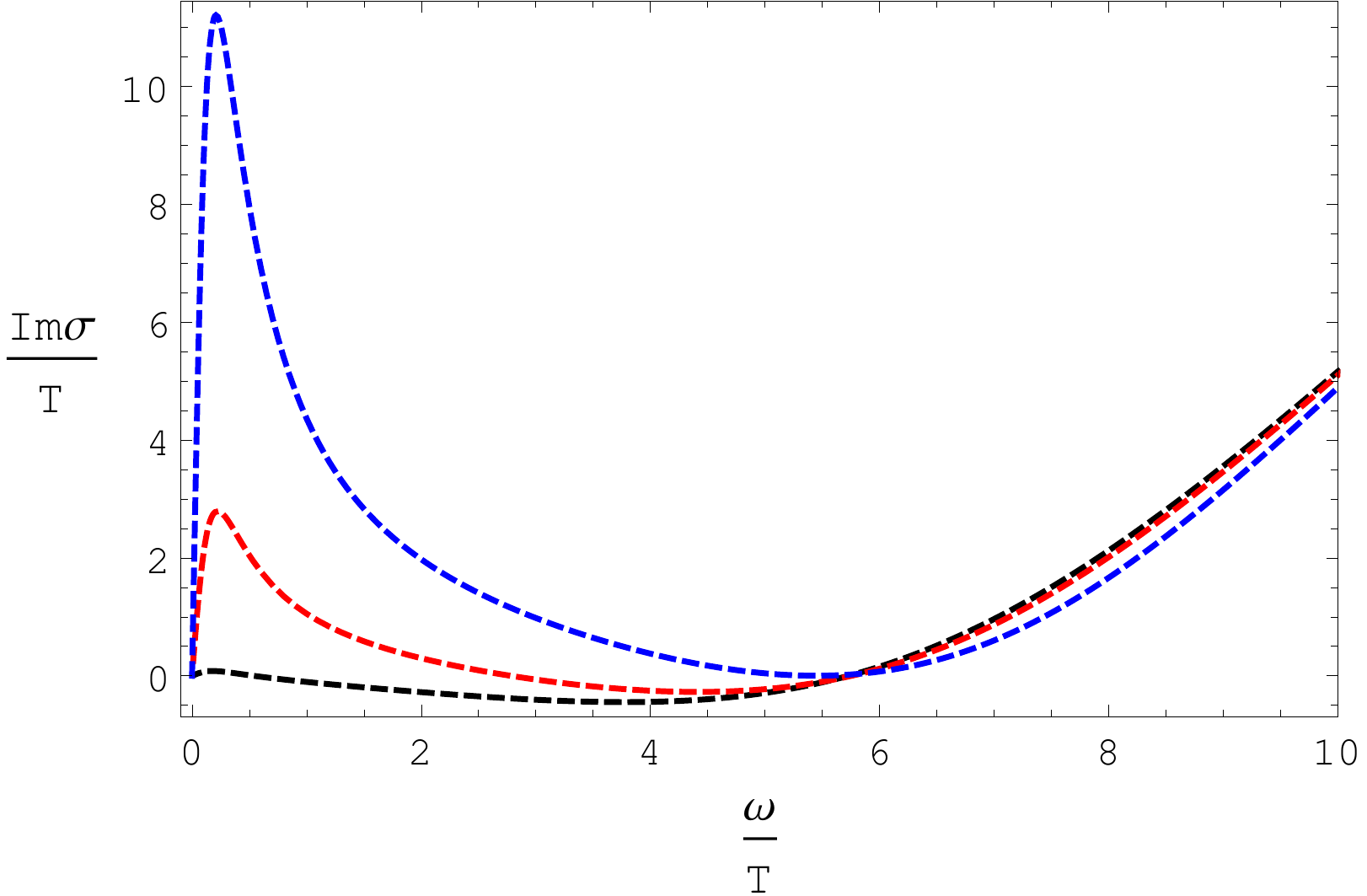}
\parbox{15cm}{\caption{\small The behaviour of both real and imaginary parts of AC longitudinal magnetoconductivity as a function of $\omega/T$ for different values of the source $qM/T$ and the magnetic field $8B\alpha/\pi^2T^2.$  Top plots:  $8B^2\alpha^2/(\pi^2T^2)=0.1$ with $qM/T=2$ (black), $0.5$ (red), $0.1$ (blue). Bottom plots: $qM/T=1$, with $8B^2\alpha^2/(\pi^2T^2)=1$ (blue), $0.5$ (red), $0.1$ (black). We found that the Drude behaviour becomes less obvious (or equivalently $\tau_5$ decreases) when we increase strength of $U(1)_A$ breaking $M$ or decrease the magnetic field $B.$ Moreover we have checked numerically that the sum rule $\frac{d}{dB} \int_0^\infty \text{Re}[ \sigma(\omega,B)] d\omega =0 $ holds.}
\label{fig:conduc}}
\end{center}
\end{figure}

We will identify the coefficient in front of $1/r^2$ of the fall-offs of $v_z$ as the dual response, i.e. we are using the covariant current with  $\langle J_z J_z\rangle_R=\frac{2 v_z^{(2)}}{v_z^{(0)}}.$\footnote{Note that we chose the renormalisation scheme (\ref{eq:ct}) such that the response is totally determined by the coefficient of the subleading term. } We plot the AC longitudinal magnetoconductivity in Fig. \ref{fig:conduc}. We find that the Drude peak behaviour becomes less obvious if we increase strength of $U(1)_A$ breaking $M$ or decrease the magnetic field $B.$ The appearance of the Drude peak is caused by the quasinormal mode with purely imaginary frequency $\omega = -i/\tau_5$ approaching the real axis. Since we expect  no new quasinormal
modes to appear or vanish upon varying the magnetic field a sum rule of the form
\be\label{eq:sumrule}
 \frac{d}{dB} \int_0^\infty \text{Re}[\sigma(\omega,B)] d\omega =0\,,
\ee
is suggested to hold. By considering differences between AC conductivies at different values of the magnetic field we have checked numerically that this sum rule indeed holds.

\subsubsection{DC conductivity: negative magnetoresistivity}
\label{subsec:dccon-explicitU1}

From the hydrodynamic calculations in  \cite{Landsteiner:2014vua}   the formula for the {\em longitudinal} magnetoconductivity in the $\frac{\mu}{\epsilon+p}=\frac{\mu_5}{\epsilon+p}=0$ limit is\footnote{In order that we can treat $J^\mu_A$ as an approximately ``conserved" quantity, we should have $\tau_5T\gg 1$. We also use the fact that the anomaly constant $c=8\alpha$ for the covariant current. }
 \be\label{2.21}
\sigma=\sigma_E+\frac{i}{\omega+\frac{i}{\tau_5}}\frac{(8B\alpha)^2}{\chi_{5}}\,,
\ee
where $\chi_{5}=\frac{\partial \rho_5}{\partial \mu_5}$ denotes the static axial susceptibility and $\tau_5$ is the axial relaxation time. In the probe limit, as long as we have axial charge dissipation the resulting DC conductivity will be finite. The most straightforward way to see the effect of axial charge dissipation in the longitudinal magnetoconductivity is to go to the DC limit directly and in this limit we have
\be\label{sigmadcfor}
\sigma=\sigma_E+\frac{(8B\alpha)^2 \tau_5}{\chi_{5}}\,.
\ee Note that this hydrodynamic formula only applies in the hydrodynamic limit $\tau_5 T \gg 1$, however, holographic calculations below will apply even beyond this hydrodynamic limit.

Instead of obtaining the DC conductivity by studying the AC conductivity and then considering its $\omega\to 0$ limit, there is an analytic way to directly obtain the DC conductivity following  \cite{Donos:2014uba}\footnote{For a recent application and development of this method see e.g.   \cite{Blake:2014yla,{Blake:2015ina}}. As will be shown in the subsection \ref{subsec:matching}, this DC method in our case is equivalent to directly taking the $\omega\to 0$ limit for small $\omega$ AC conductivity obtained from near-far matching calculations.}  using radially conserved quantities \cite{Iqbal:2008by}. The idea is to perform the holographic calculations for conductivity in the exact DC limit, i.e. $\omega =0$. Then we can consider the following perturbations around (\ref{eq:bgansatz}):\footnote{Note that one can also work in the radial gauge $a_r=0$ and the same results should be obtained. Here we include $a_r$ in order to totally ignore equations related to $\phi_2$.}
\be\label{perturbationdc}
\delta V_\mu=(v_t(r), 0,0,-Et+v_z(r),0)\,,~~~
\delta A_\mu=(a_t(r), 0,0,a_z(r),a_r(r))\,,~~~
\delta\phi=\phi_1(r)+i \phi_2(r)\,.
\ee
Substituting these perturbations into the equations (\ref{eq:eomtwou1-1} - \ref{eq:eomtwou1-3}) we obtain their equations which can be found in the appendix \ref{equationDC}. Note that these equations do not depend on the background fields $V_t, V_z$.  For zero axial density solution $\mu_5=0$, i.e. $A_t=V_z=0$, we have
\bea
\label{eq:dc1} a_t''+\frac{3}{r}a_t'-\frac{2 q^2\phi^2}{r^2 f}a_t+\frac{8B\alpha }{r^3}v_z'&=&0\,,\\
\label{eq:dc2} v_z''+\bigg(\frac{3}{r}+\frac{f'}{f}\bigg)v_z'+\frac{8B\alpha}{r^3f}a_t'
&=&0\,.
\eea
Near conformal boundary $r\to\infty,$ we have 
\be a_t=a_t^{(0)}+\frac{a_t^{(1)}}{r^2}\ln r +\frac{a_t^{(2)}}{r^2}+\dots\,,~~~
v_z=v_z^{(0)}+\frac{v_z^{(2)}}{r^2}+\dots\ee
with $a_t^{(1)}=-a_t^{(0)}q^2M^2.$ From the equation (\ref{eq:dc2}) we have the conserved quantity along the radial direction $J=-r^3 f v_z'-8B\alpha a_t$ with $\partial_r J=0$, which means $J|_{r\to \infty}=J|_{r\to r_0}.$ 
When $r\to \infty,$ we impose the sourceless boundary condition for $a_t$, i.e. $a_t^{(0)}=0$. Thus the electric current which is the response
of the external electric field can now be calculated at the horizon to be $j=J|_{r\to \infty}=J|_{r\to r_0}.$ 

The next thing that we need to do is to get $J|_{r\to r_0}$ from the near horizon boundary conditions. At the horizon we have 
\bea\label{dcnearhorizonbc}
v_z&=&-\frac{E}{4\pi T}\ln(r-r_0)+\cdots\,,\\
\label{dcnearhorizonbc2}
a_t&=&-\frac{2E(8B\alpha)}{(4\pi T)q^2 r_0^2{\phi_0}^2}+\cdots \,.
\eea
Note that near horizon the coefficient of the subleading term in $a_t$ is a free parameter which is precisely the shooting parameter that can be used to determine the sourceless condition for $a_t$ near boundary.  Also note that the near boundary condition (\ref{dcnearhorizonbc2}) seems to be inconsistent with (\ref{eq:bndat}), however, this is due to the different gauge: if we work in the gauge $\phi_2=0$, we should keep $a_r$ and the leading order in (\ref{eq:bndat}) should be the same as here in the limit $\omega\to 0.$ 
It follows immediately  
\be\label{eq:dccond-analytical0}
\sigma_\text{DC}=\frac{j}{E}=\pi T+\frac{32 B^2\alpha^2}{\pi^3 T^3 q^2\phi_0^2}\,
\ee
with $\phi_0$ the horizon value of the background scalar field.  We should emphasise that the above expression does not depend on the mass of the scalar filed in the bulk and it also applies for the case $\mu\neq 0$ as long as $\mu_5=0$. We emphasise that the $B^2$ dependance is exact.
and the recent experiments \cite{Li:2014bha,Huang,Zhang,Xiong} also found this $B^2$ dependence in the longitudinal magnetoconductivity.

For the simple case of $\mu=0$, i.e. $V_t=V_z=A_t=A_z=0$, $\phi$ has an analytic solution (\ref{specialphi})  for $m_s^2=-3$, and we have 
(\ref{eq:hbrelation}), thus
\be\label{eq:dccond-analytical}
\frac{\sigma_\text{DC}}{T}=\pi +\frac{\Gamma[1/4]^2}{4\pi^2\Gamma[3/4]^2}\frac{(8B\alpha/T^2)^2}{(qM/T)^2}\simeq \pi +21.6\frac{\big(8B\alpha/(\pi^2T^2)\big)^2}{(qM/T)^2}\,.
\ee
This is one main results of our paper. In fact this DC magnetoconductivity can also be obtained analytically using a near far matching calculation for small $\omega$, in which situation the near horizon boundary conditions (\ref{dcnearhorizonbc}) come from solutions in the near region with infalling boundary conditions while the equations in the DC limit are exactly the far region equations. In the subsection \ref{subsec:matching} we show the near far matching calculation which produces the same DC result for both this explicit breaking case and the massive gauge field case. From the near far matching analysis it is natural to identify $\pi T$ as the quantum critical conductivity $\sigma_E$ in (\ref{2.21}) and (\ref{sigmadcfor}) in this $\frac{\omega}{r_0}\ll m^2$ limit, and the value of $\sigma_E$ strongly depends on the value of $\frac{\omega}{r_0 m^2}$. We will explain at the end of the next subsection and also in subsection \ref{subsec:matching} that the discrepancy between this $\sigma_E=\pi T$ and the $\sigma_E$ in the case without axial charge dissipations (\ref{sigmaE-prvious}) in \cite{Landsteiner:2014vua} comes from the noncommutative nature of the two limits $\omega\to 0$ and $\tau_5\to\infty$.

\begin{figure}[ht]
\begin{center}
\includegraphics[width=0.49\textwidth]{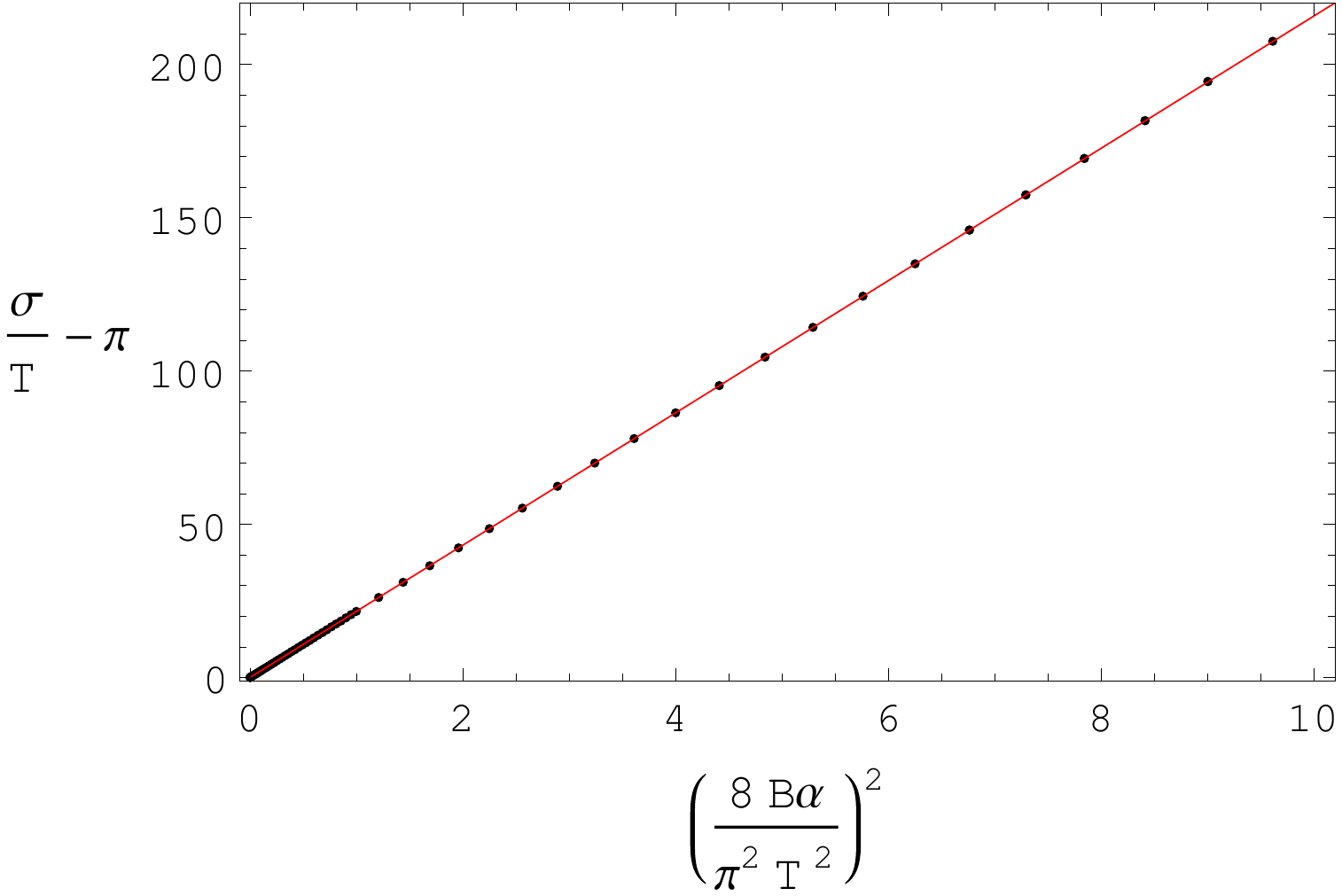}
\includegraphics[width=0.49\textwidth]{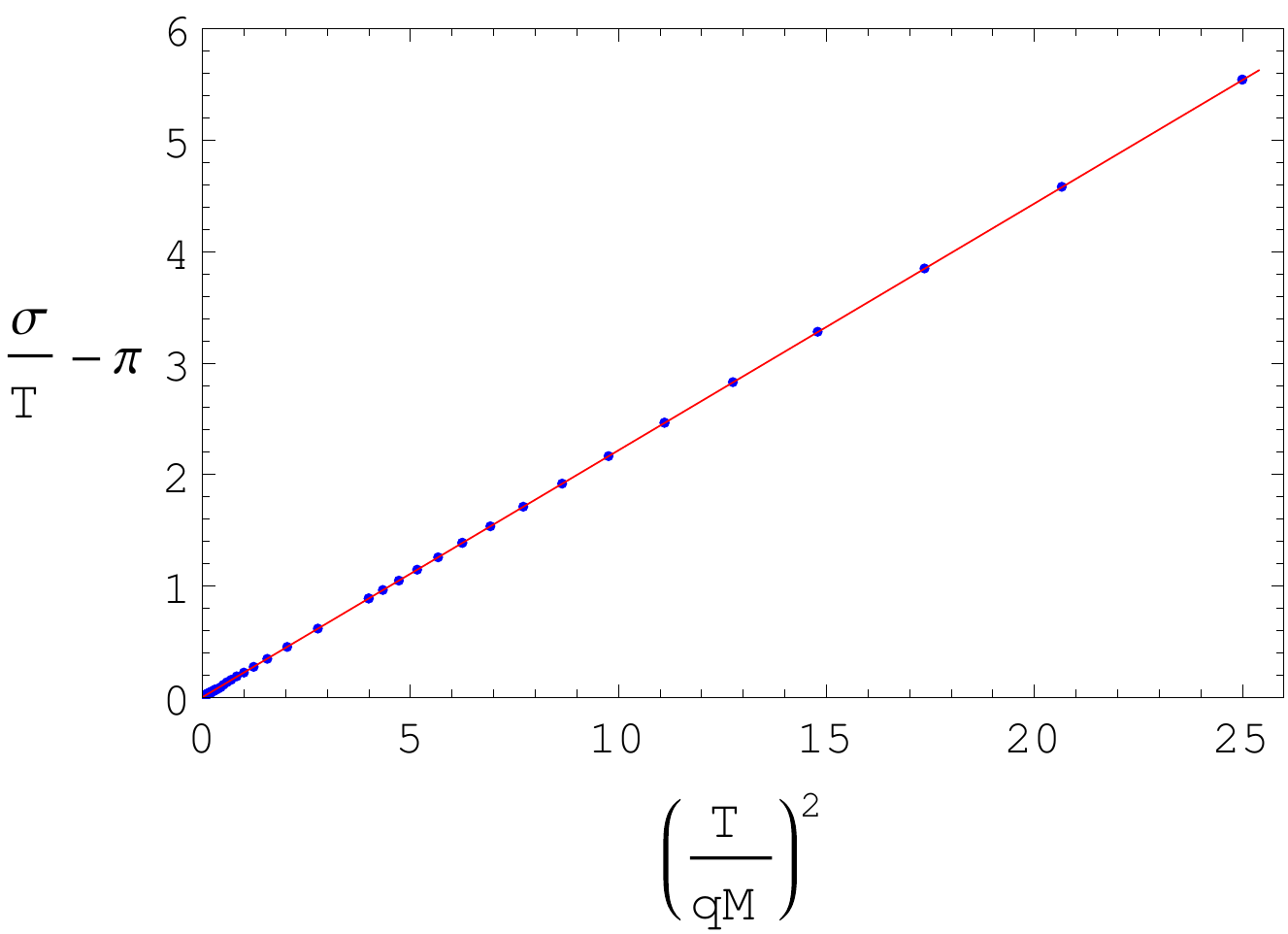}
\parbox{15cm}{\caption{\small The dotted lines are the numerical results for the DC conductivity obtained numerically (by setting $\omega/T=10^{-5}$ in the AC conductivity calculation). Note that if $\tau_5T$ is large, we should choose extremely small $\omega/T$ to obtain the DC conductivity numerically.  The solid lines are the best fitting for numerical data. Left plot is for $qM/T=1$ and the best fit is  $\sigma/T-\pi=0.00005+21.600  \big(8B\alpha/(\pi^2T^2)\big)^2/(qM/T)^2$. Right plot is for  $8B\alpha/(\pi^2T^2)=1/\pi^2$ and the best fit is  $\sigma/T-\pi=0.0007+21.574 \big(8B\alpha/(\pi^2T^2)\big)^2/(qM/T)^2$. The fitting formulae exactly reproduces (\ref{eq:dccond-analytical}) from analytical calculations. }
\label{fig:dccon}}
\end{center}
\end{figure}

From the formula (\ref{eq:dccond-analytical}) for the DC longitudinal magnetoconductivity, we can see that the negative magnetoresistivity has a universal $B^2$ behaviour even in the large $B$ limit (quantum regime). In the regime where the axial symmetry breaking is small this can be understood from the $B$ dependent behavior of the charge relaxation time $\tau_5$ such that $\tau_5/\chi_5$ in (\ref{sigmadcfor}) does not depend on $B$ which results in the universal $B^2$ behavior in (\ref{eq:dccond-analytical0}). Our numerical results obtained from the AC conductivity with $\omega\to 0$ also match quite well with the formulae (\ref{eq:dccond-analytical}) as can be seen in Fig. \ref{fig:dccon}.


\subsubsection{Calculating $\chi_5$ and $\tau_5$ independently}
\label{sec:eU1tauc}

With the analytic result (\ref{eq:dccond-analytical}) for the DC magnetoconductivity, we cannot identify $\tau_5$ and $\chi_5$ because there is not enough information in this formula.  Thus we will calculate the static susceptibility $\chi_5$ and the charge relaxation time $\tau_5$ independently and substitute them into (\ref{sigmadcfor}) to be compared with the analytic result  (\ref{eq:dccond-analytical}).

\begin{figure}[h]
\begin{center}
\includegraphics[width=0.49\textwidth]{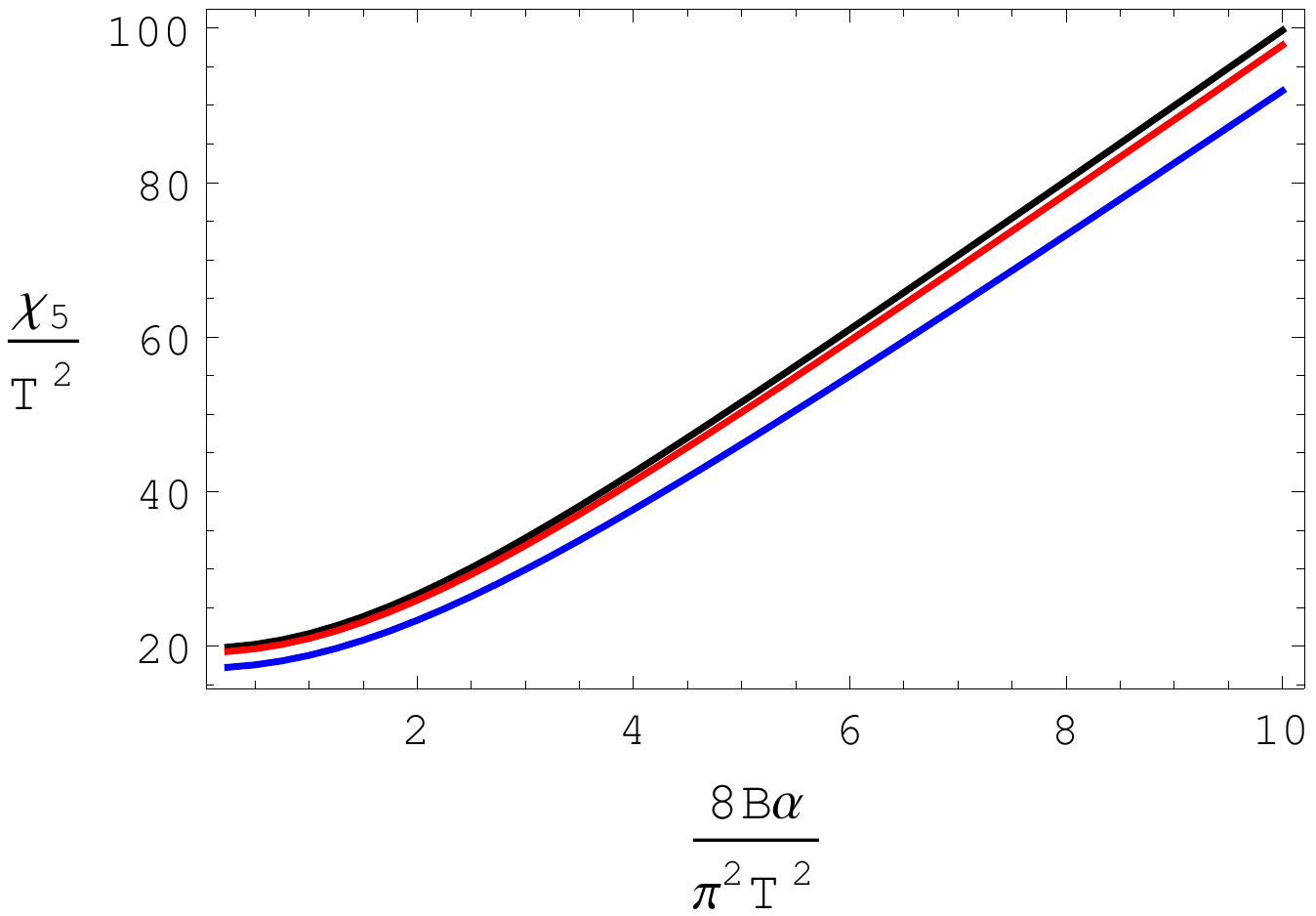}
\includegraphics[width=0.49\textwidth]{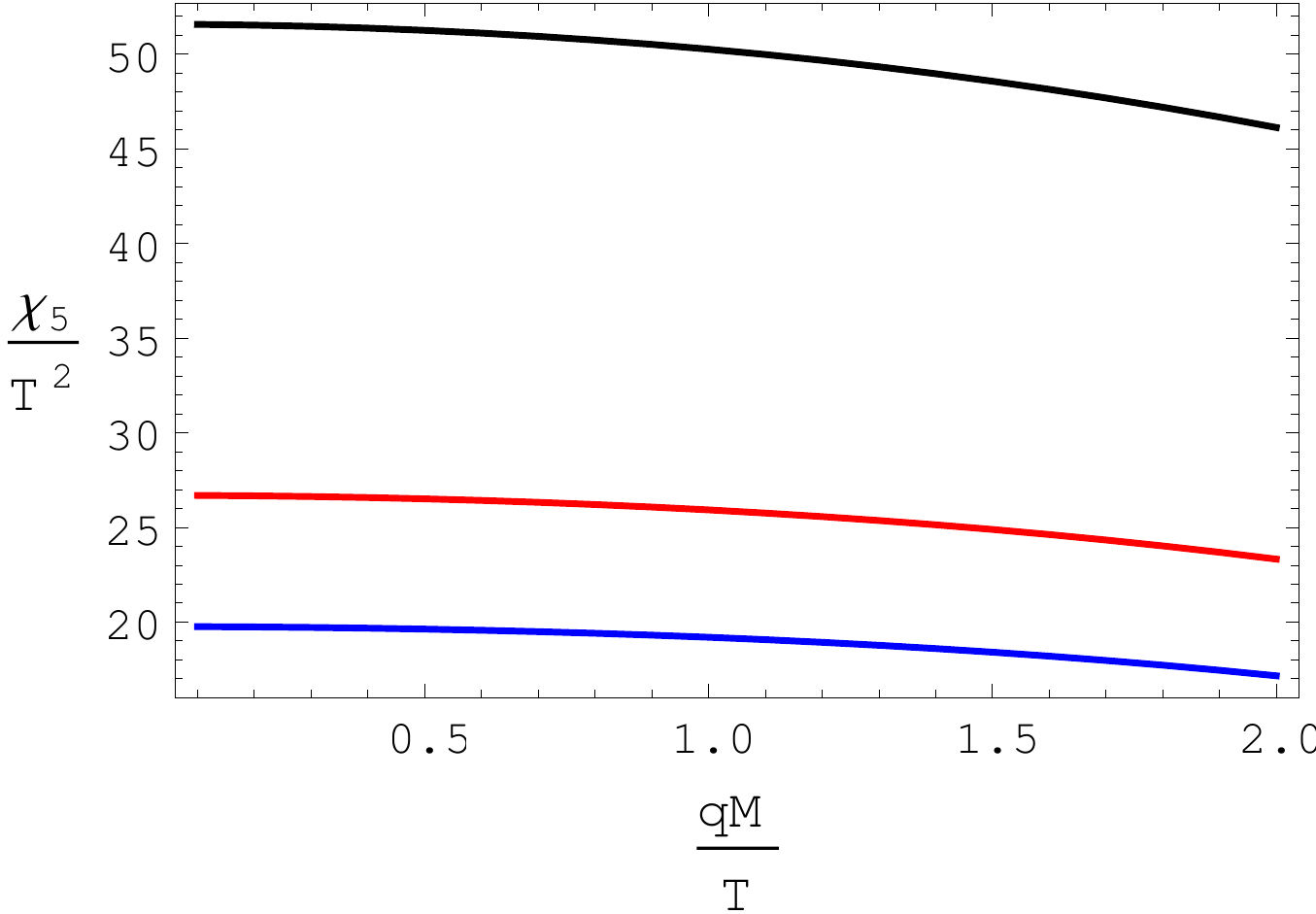}
\parbox{15cm}{\caption{\small Left: $\chi_5$ as a function of $B$ for different sources with $qM/T=0.1$ (black), $1$ (red), $2$ (blue). Right: $\chi_5$ as a function of $M$ for $8B\alpha/(\pi^2T^2)=5$ (black), $2$ (red), $0.1$ (blue). }
\label{fig:staticsus}}
\end{center}
\end{figure}

The static susceptibility $\chi_5=\lim_{\omega\to 0}\langle J_A^t J_A^t\rangle$ can be calculated from perturbations (\ref{perturbationdc}) while setting $E=0$, and the boundary conditions are now $a_t, v_z$ being regular at the horizon and $v_z$ being sourceless at the boundary, the latter of which can be fixed from gauge transformations. We need to solve the following equation
\be
a_t''+\frac{3}{r}a_t'-\frac{1}{r^2 f}\Big(2 q^2\phi^2+\frac{(8B\alpha)^2}{r^4}\Big)a_t=0\, ,
\ee
and this equation can only be solved numerically. The result is shown in Fig. \ref{fig:staticsus}. Note that different from the DC conductivity which can be totally determined by the near horizon data, the static susceptibility cannot be fully traced to horizon data which is due to the fact that $U(1)_A$ is no longer an exact symmetry in this explicit breaking model.


The next step is to calculate $\tau_5$ independently. This can be done by analyzing the quasinormal modes at $k=0$ on top of the $\mu=\mu_5=0$ background with no source for the electric field. We will solve the $(v_z, a_t, \phi_2)$ in (\ref{eq:flu1} - \ref{eq:flu3}) in the complex frequency plane with infalling near horizon boundary conditions for $a_t, v_z,\phi_2$: (\ref{infallingboundary}) with $c_0$ restored in $\phi_2$. As stated above there will be two linearly independent solutions for $c_0=0$ with boundary values $( v_z^{(0)I}, a_t^{(0)I}, qM c^{I})$ and $(v_z^{(0)II}, a_t^{(0)II},  qM c^{II})$, which are $\omega$ dependent. The residual gauge symmetry will give the third solution of boundary value $(0,-i\omega\Lambda, qM\Lambda)$ with radially independent $\Lambda$, and this represents the degree of freedom due to $c_0$. We can define the matrix for  boundary values as
\be
\text{M}_\text{bnd}=\begin{pmatrix}
v_z^{(0)I}~ & a_t^{(0)I}&~ qM c^{I} \\
v_z^{(0)II}~ & a_t^{(0)II}&~ qM c^{II} \\
0~&-i\omega\Lambda&~ qM\Lambda
\end{pmatrix}\,.
\ee
Following \cite{Amado:2009ts}, the QNM frequency is given by the zeros of the determinant of the fieled values at the boundary, i.e. 
$|\text{M}_\text{bnd}(-i\omega_I)|=0$. We define the axial charge dissipation time $\tau_5$ as $\tau_5=1/\omega_I.$ This is consistent with the hydrodynamic modes following the hydrodynamic equation $\partial_\mu J^\mu_V=0$ and $\partial_\mu J^\mu_A=-\frac{1}{\tau_5}J_A^0$ \cite{Jimenez-Alba:2014iia} although for small $\tau_5 T$ these equations may not apply anymore given that other non-hydrodynamic QNMs will dominate.

\begin{figure}[h]
\begin{center}
\includegraphics[width=0.48\textwidth]{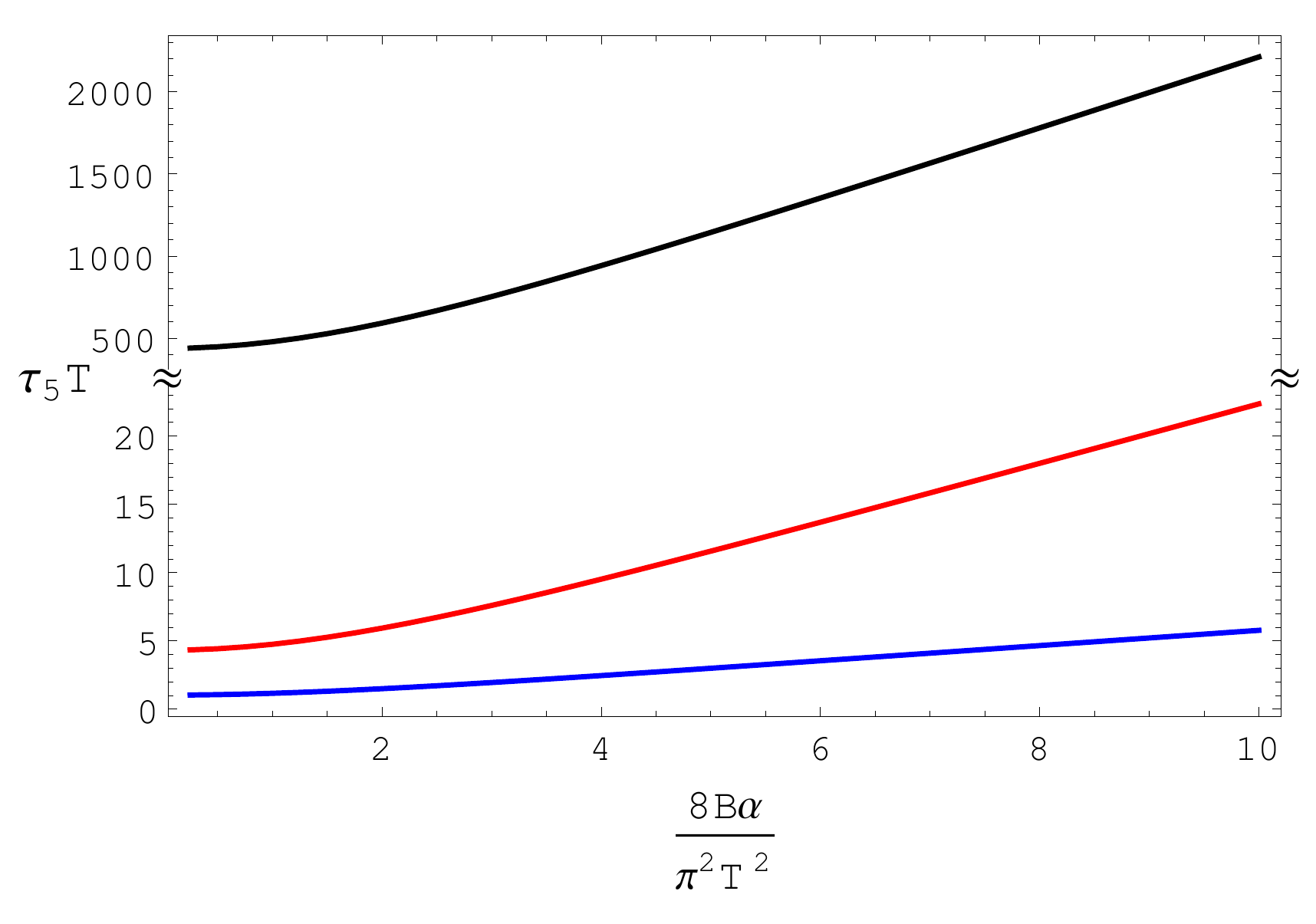}
\includegraphics[width=0.49\textwidth]{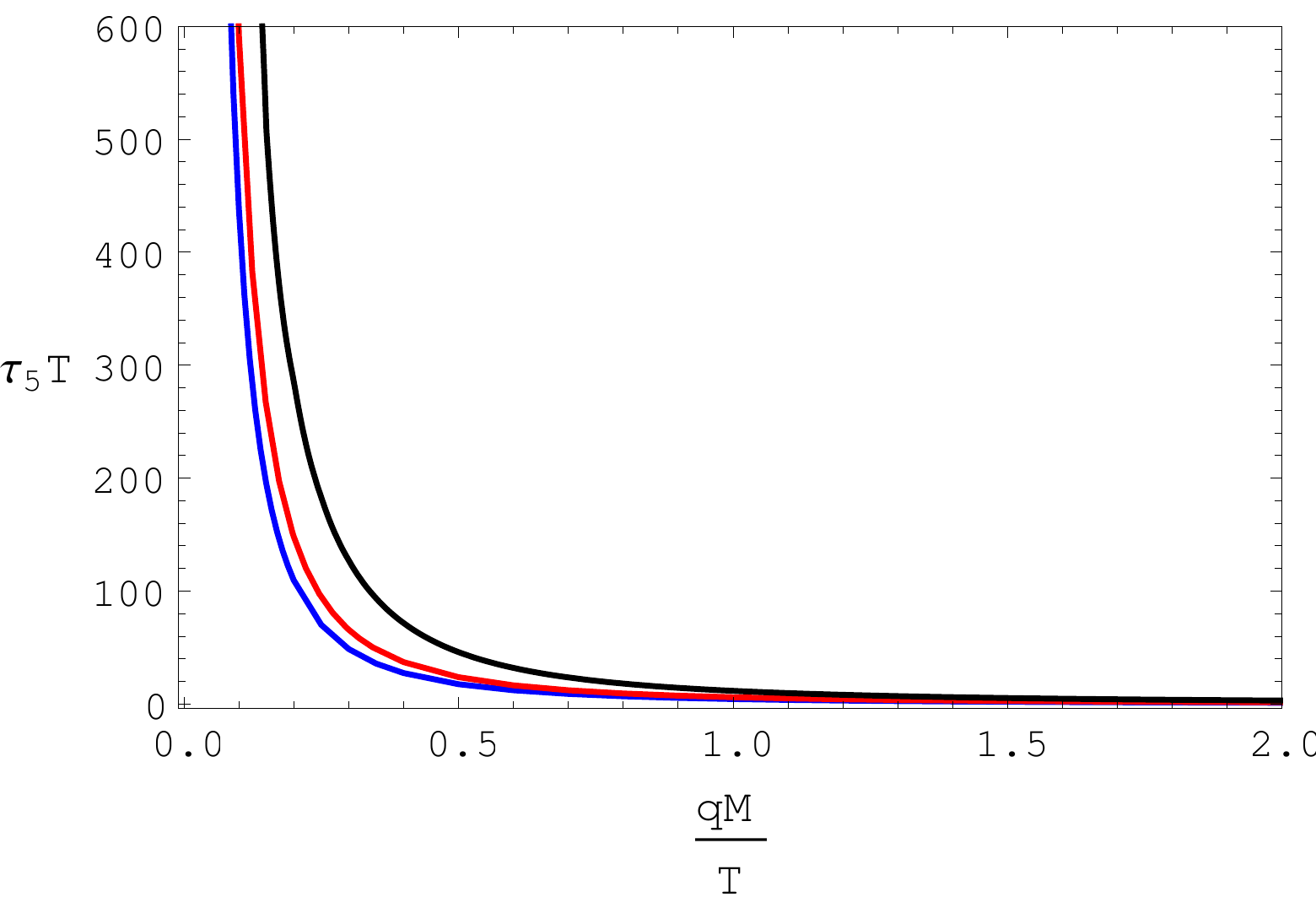}
\parbox{15cm}{\caption{\small Left: $\tau_5 T$ as a function of $B$ for different sources with $qM/T=0.1$ (black), $1$ (red), $2$ (blue). Right: $\tau_5 T$ as a function of $M$ for $8B\alpha/(\pi^2T^2)=5$ (black), $2$ (red), $0.1$ (blue). 
}\label{fig:tauc}
}

\end{center}
\end{figure}

The final numerical plot on $\tau_5$ is shown in
Fig. \ref{fig:tauc}.  One can see that $\tau_5T$ increases when we increase the magnetic field or decrease the value of the source that explicitly breaks $U(1)_A.$ When $\tau_5T\gg 1$ hydrodynamics applies\footnote{In this case in principle $\tau_5$ could be determined by memory matrix formalism as the three dimensional case \cite{Lucas:2015pxa}. We leave this interesting question for future investigation. }  and we are expecting a Drude peak behaviour  \cite{Landsteiner:2014vua} (i.e. coherent metal behaviour) and this is exactly what we found in the small frequency regime of the AC conductivity. When $\tau_5 T < 1$, the hydrodynamic description breaks down (incoherent metal behaviour) and the contribution from this hydrodynamic QNM will not be dominant \cite{Davison:2014lua}. Consequently there is no Drude peak in this regime in the AC conductivity.
Moreover the standard Boltzmann theory which is based on the the quasiparticle picture does not apply for the small $\tau_5$ case. Here we want to emphasize that even when the axial symmetry is strongly broken we still have negative magnetoresistivity as we have shown in the previous subsection.
We note that recently in a weak coupling context in \cite{sarma} it was pointed out that the ionic scattering can produce universal positive magnetoconductivity (or negative magnetoresistivity) with $B^2$ behaviour for a generic 3D metal in presence of parallel electric and magnetic fields.

\subsubsection{Scaling behaviours of $\chi_5$ and $\tau_5$ on $B$ and $M$}

We will analyse the scaling behaviour of $\chi_5$ and $\tau_5$ with $B$ and $M$ in this subsection and we will show that in the hydrodynamic limit the hydrodynamic formula reproduces the dependence of the DC conductivity on $B$. First, note that the formula \be\label{eq:dcfromhydro}
\sigma_\text{DC}=\sigma_E+\frac{\tau_5}{\chi_5}(8B\alpha)^2\,
\ee for the DC conductivity is only applicable in the hydrodynamic limit $B/T^2 \ll 1$ and $\tau_5 T\ll 1$ which means $M/T \ll 1$. However, in the following, we will show the scaling behaviour of $\chi_5$ and $\tau_5$ on $B$ for a large range of $B$, in which this formula still coincides with the holographic DC result. This also happen in the case without any
axial charge dissipation \cite{Landsteiner:2014vua}.

\begin{figure}[h]
\begin{center}
\includegraphics[width=0.49\textwidth]{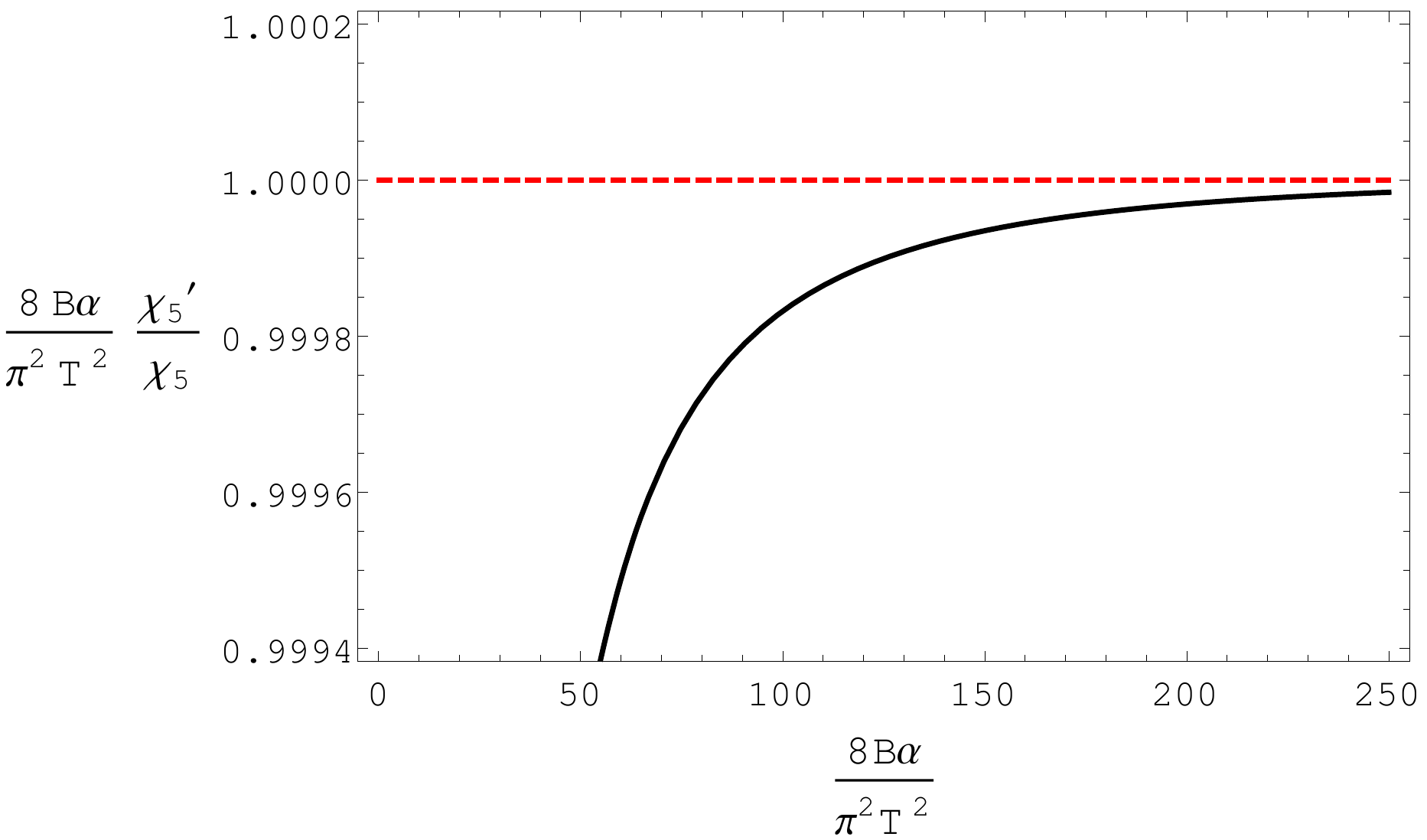}
\includegraphics[width=0.48\textwidth]{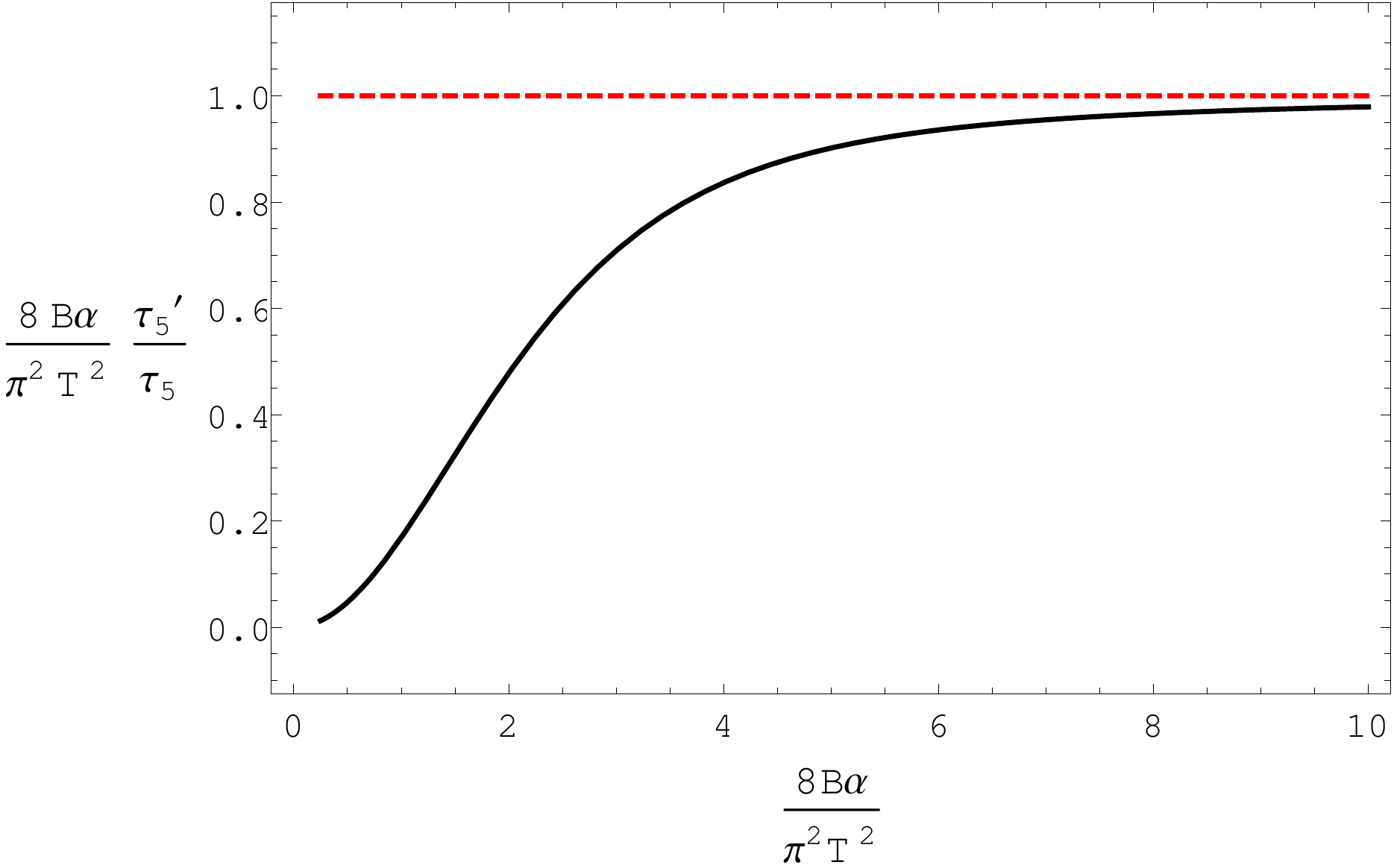}
\includegraphics[width=0.49\textwidth]{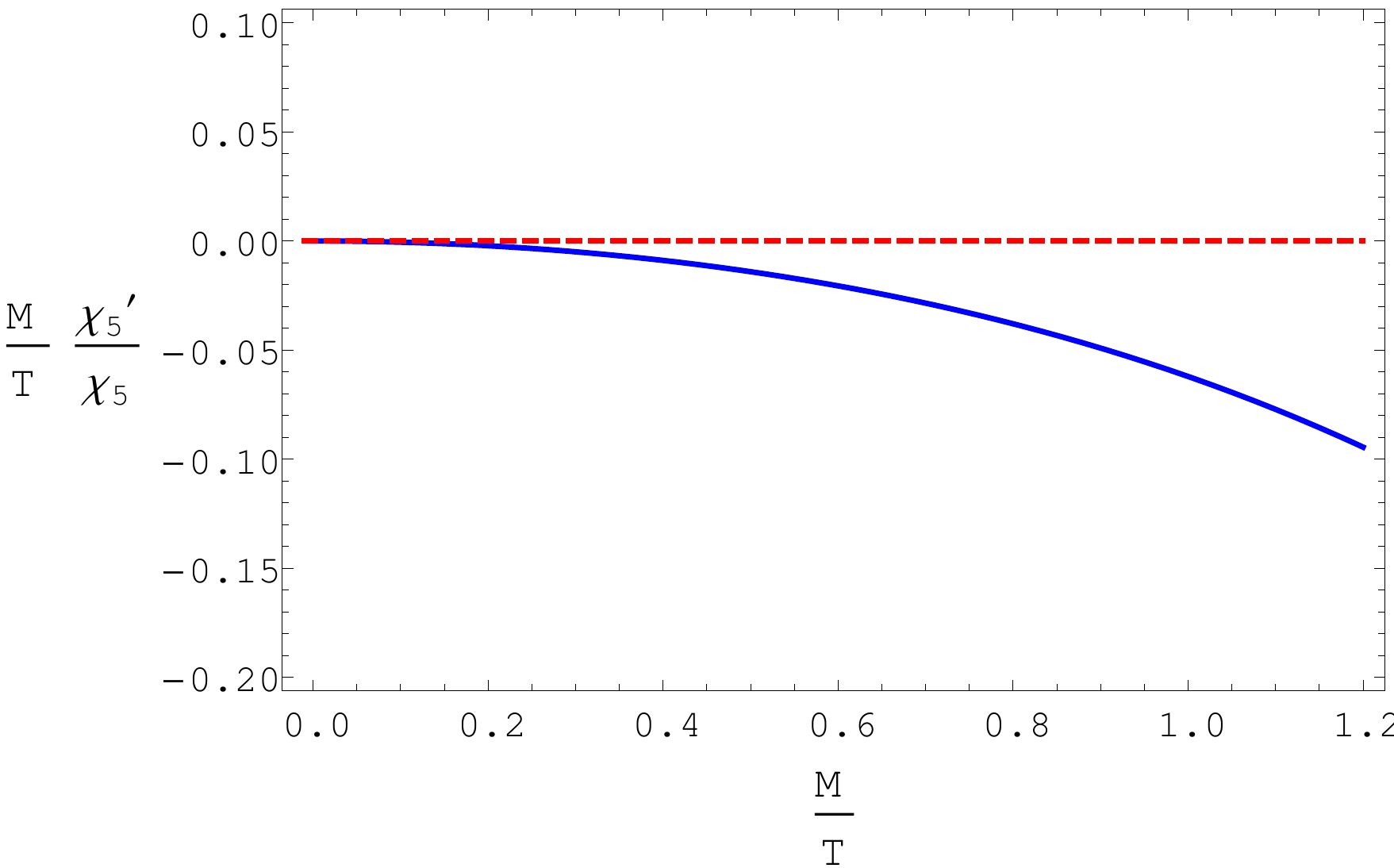}
\includegraphics[width=0.49\textwidth]{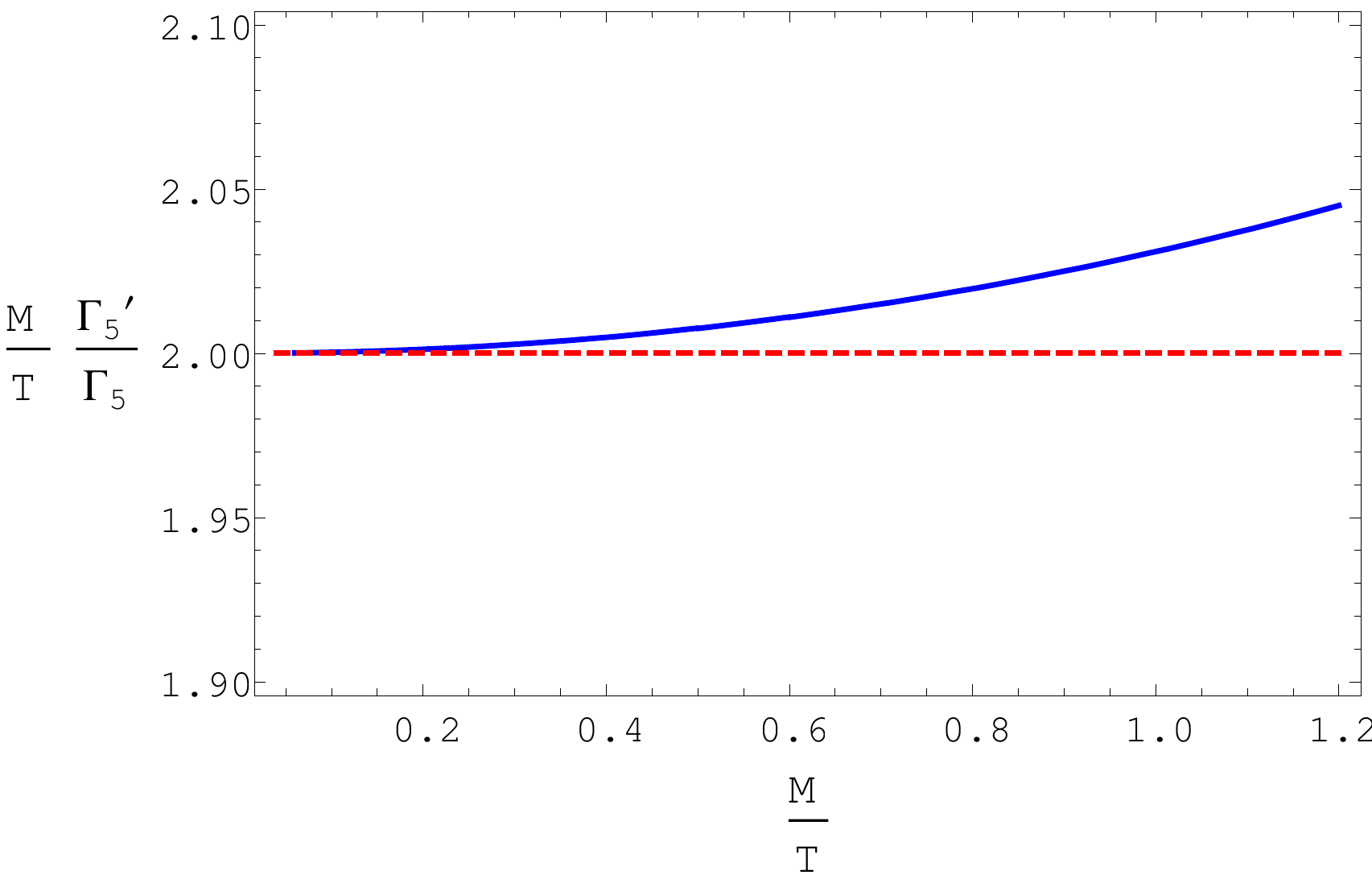}
\parbox{15cm}{\caption{\small Scaling behaviour of static susceptibility $\chi_5$ and relaxation time $\tau_5$. Here the prime means the derivative of the parameter in the horizontal axis. Black solid line is for $M/T=0.1$ while blue solid line is for $\frac{8B\alpha}{\pi^2 T^2}=0.1$.}
\label{fig:scaling}}
\end{center}
\end{figure}

\begin{figure}[h]
\begin{center}
\includegraphics[width=0.49\textwidth]{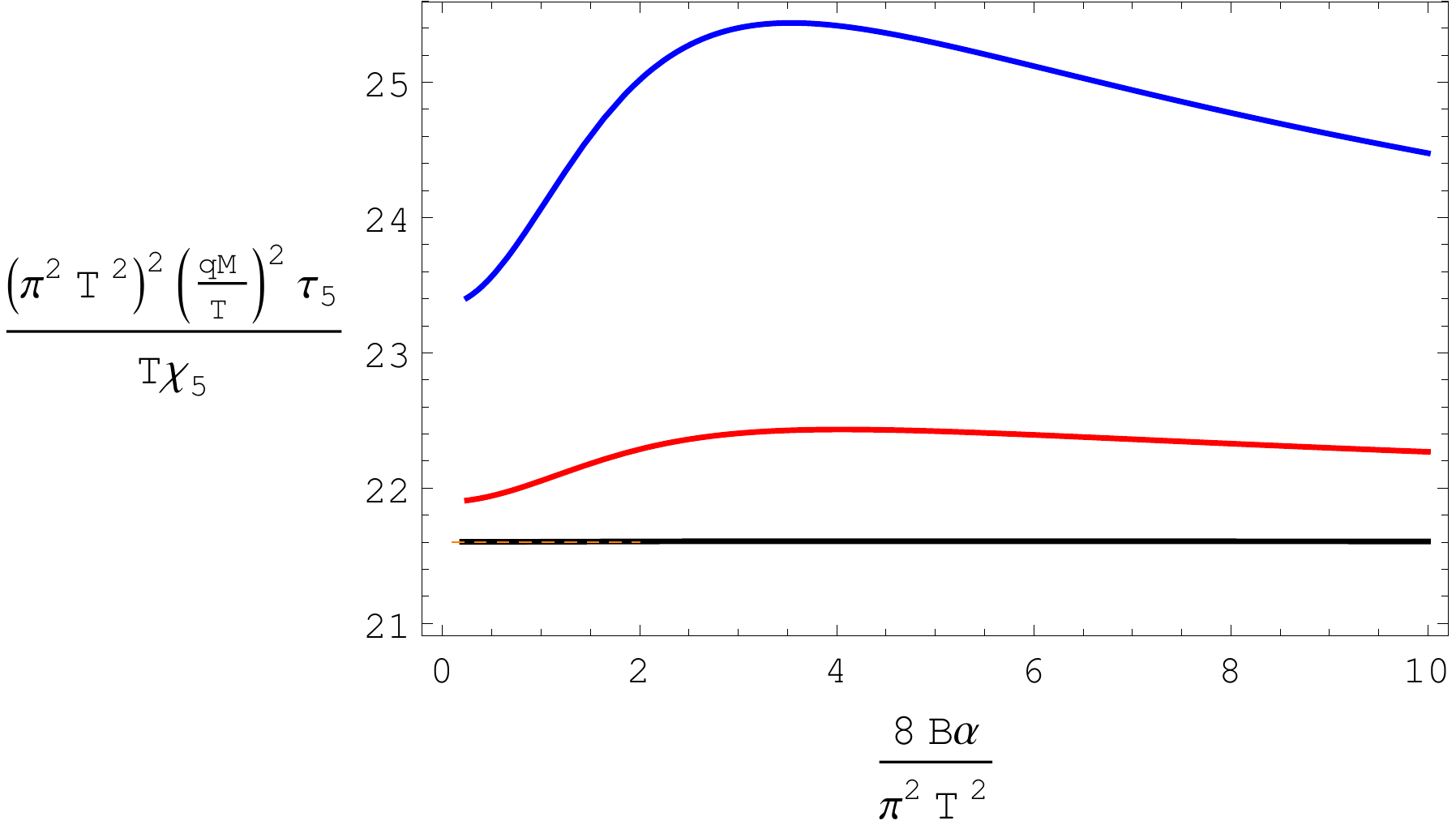}
\includegraphics[width=0.49\textwidth]{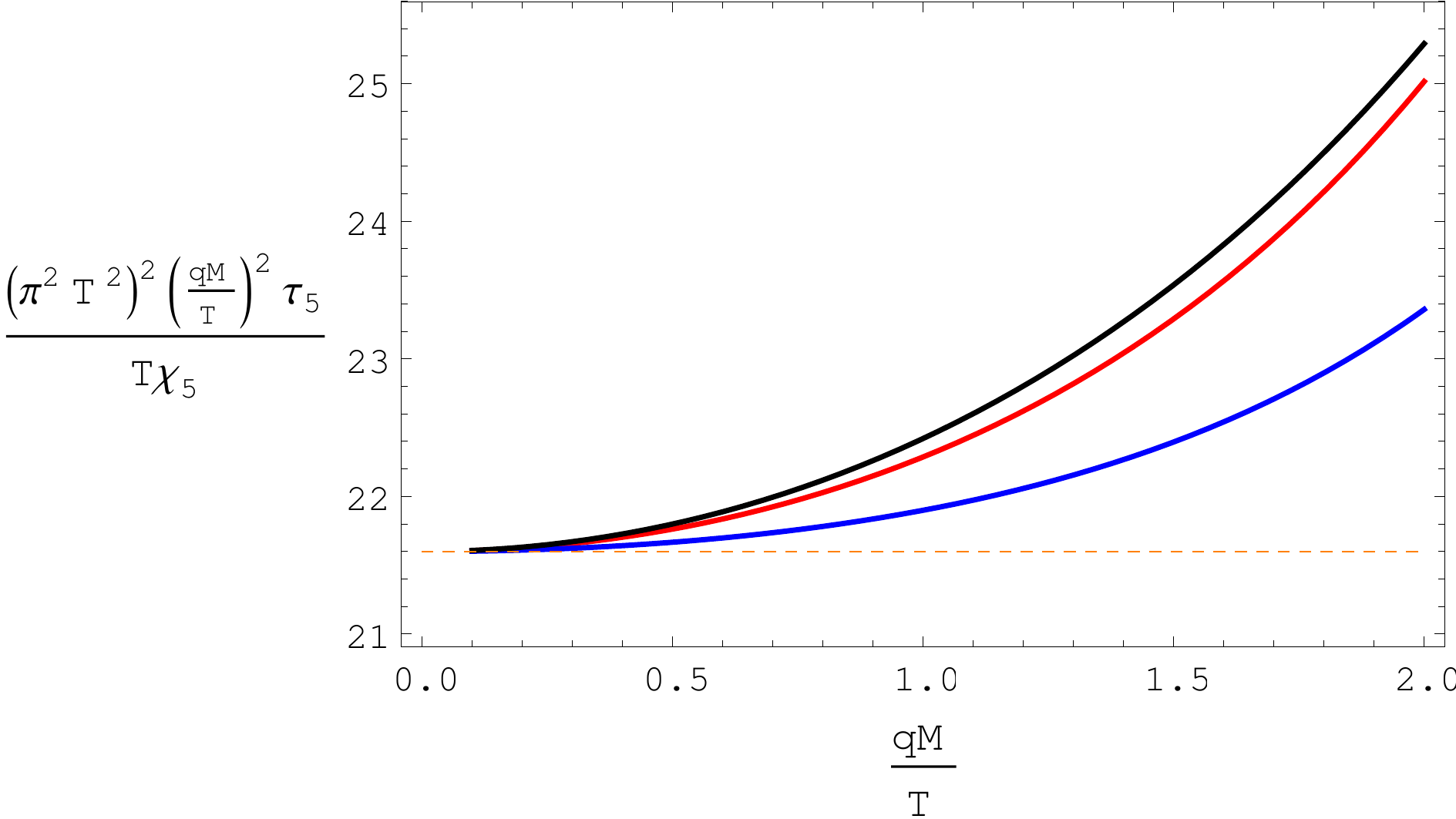}
\parbox{15cm}{\caption{\small  Left: $ \tau_5/\chi_5$ as a function of $B$ for different sources with $qM/T=0.1$ (black), $1$ (red), $2$ (blue). Right:  $\tau_5/\chi_5$ as a function of $M$ for $8B\alpha/(\pi^2T^2)=5$ (black), $2$ (red), $0.1$ (blue).  Orange dashed line is the constant coefficient $21.6$ in Eq. (\ref{eq:dccond-analytical}). This is consistent with the fact that our formula Eq. (\ref{eq:dcfromhydro}) is only valid when $\tau_5T\gg 1$. }
\label{fig:tauochi}}
\end{center}
\end{figure}

Fig. \ref{fig:scaling} shows that when $B$ is large (fixing $M$, $T$), both $\chi_5$ and $\tau_5$ are linear in $B$, while when $M/T$ is small (fixing $T$, $B$), $\chi_5$ is a constant and $\tau_5$ (or $\Gamma_5=1/\tau_5$) is proportional to $M^{-2}$ (or $M^2$). Fig. \ref{fig:tauochi} shows that when $M/T$ is small, $\tau_5/\chi_5$ does not dependent on $B$ in the large $B$ regime. From these figures we can see that the formula (\ref{eq:dcfromhydro}) is valid even for large magnetic field in the regime of small for small $M/T$.
When $\tau_5 T\gg 1$, we have $\sigma_E=\pi T$ and $\tau_5/\chi_5=21.6/(\pi^4 T^3 q^2M^2)$. At small $M$, $\tau_5/\chi_5$ is proportional to $M^{-2}$.


Now we explain why $\sigma_E$ that we observe here is different from the one in the case without any axial charge dissipation in \cite{Landsteiner:2014vua}. Note that if we take the limit $\tau_5\to \infty$ (by $M/T\to 0$) in the DC conductivity (\ref{eq:dccond-analytical0}), we can see that the quantum critical conductivity $\sigma_E=\pi T$ in this limit and we cannot see the effect from the anomaly and the magnetic field as in the case without axial charge dissipation \cite{Landsteiner:2014vua}, where $\sigma_E$ drops with $B$ at small $B$ and approaches zero at large $B$ (see (\ref{sigmaE-prvious})). This seeming discrepancy comes from the non-commutative nature of the two limits of $\omega\to 0$ and $\tau_5 \to \infty$. In the case without axial charge dissipation, the limit of $\tau_5 \to \infty$ was taken first while in our formula for the DC conductivity with axial charge dissipations, the limit of $\omega\to 0$ was taken first and some terms in $\sigma_E$ disappear in this limit. This means that $\sigma_E$ in the formula in \cite{Landsteiner:2014vua} can only be seen in the limit $1/\tau_5\ll \omega\to 0$.\footnote{A similar phenomenon was
observed in the momentum dissipation case. For four dimensional Reissner Nordstrom black hole solution without momentum dissipation, $\sigma_E$ behaves nontrivial at low temperature which can be found in \cite{Faulkner:2013bna}.  However, after introducing momentum dissipation through different mechanisms \cite{Blake:2013bqa,{Andrade:2013gsa},{Blake:2014yla}} $\sigma_E$ is constant.  The  reason should be the same as our case: it is exactly due to the non commutativity between two different limits $\tau_m\to\infty$ and $\omega\to 0$.}

\subsection{The anomalous transport coefficients}
\label{ssec:anocoe}

Another interesting transport property for the explicit axial symmetry breaking case is that we can still observe a similar anomalous transport as the chiral magnetic effect as in the normal chiral anomalous systems. To show the anomalous transport in this case, we should consider perturbations of the system at $\omega=0$ and small $k$ and use the Kubo formula as in normal chiral anomalous systems. We will consider the chiral magnetic conductivity ${\bf J} = \sigma_B {\bf B}$, the chiral separation conductivity ${\bf J}_5 = \sigma_\text{CSE} {\bf B}$ and the axial magnetic conductivity ${\bf J}_5 = \sigma_{55} {\bf B}_5$, where ${\bf B}_5$ is an axial magnetic field.

We consider the transverse fluctuations $a_x,a_y, v_x, v_y$ with $k$ along the $z$ direction and their equations can be found in the appendix \ref{equationCME}. It is convenient to introduce $a_\pm=a_x\pm i a_y, v_\pm=v_x\pm i v_y$. For the $\mu=\mu_5=0$  case $V_t=V_z=A_t=A_z=0,$ it is easy to find that  $\sigma_B=\sigma_{55}=\sigma_\text{CSE}=0$.

For the case with nontrivial axial charge density  (i.e. $\mu=0, \mu_5\neq 0$) we have
\bea
v_\pm''+\bigg(\frac{3}{r}+\frac{f'}{f}\bigg)v_\pm'-k^2\frac{v_\pm}{r^4f}\pm\frac{8k\alpha}{r^3f}A_t'v_\pm&=&0\,,\nonumber\\
a_\pm''+\bigg(\frac{3}{r}+\frac{f'}{f}\bigg)a_\pm'+\bigg(-\frac{k^2}{r^2}-2q^2\phi^2\bigg)\frac{a_\pm}{r^2f}
\pm\frac{8k\alpha}{r^3f}A_t'a_\pm &=&0\,.
\eea
Since the vector sector and axial sector decouple from each other, $\sigma_\text{CSE}=0.$  

Up to order $k$, the asymptotic series at conformal infinity is $v_\pm=v_0+\frac{v_1}{r^2}+\dots$ and $a_\pm=a_0+\frac{a_1}{r^2}\ln r+\frac{a_2}{r^2}+\dots$ with $a_1=-a_0q^2\lambda^2.$ 
The solution for $v_\pm$ is
\be
v_\pm=c_0\pm 8\alpha c_0\big(\int_r^\infty dx \frac{A_t}{x^3 f}\big)k+\mathcal{O}(k^2)\,.
\ee
Note that $\mp iG_{xy}+G_{xx}=G_{\pm}$ and $\sigma_B=\lim_{k\to 0}\frac{G_{xy}(\omega=0)}{ik}=\lim_{k\to 0}\frac{G_+-G_-}{2k}.$ Thus $\sigma_B=8\alpha \mu_5$.
We note that the exactly conserved ({\em consistent}) current that is defined via the functional variation of the on-shell action with respect to the gauge potential has
an additional contribution from the Chern-Simons part of the action $J^\mu_\text{CS} = 4\alpha\epsilon^{\mu\nu\rho\lambda} A_\nu \mathcal{F}_{\rho\lambda}$. This part is easy to calculate since it is completely determined by the boundary values of the fields and adds $-8\alpha\mu_5$ to the chiral magnetic conductivity. Therefore we find $\sigma_B=8\alpha \mu_5$ for
the current without the contribution of the Chern-Simons term ({\em covariant} current) and $\sigma_{B\text{(con)}}=0$ for the exactly {\em conserved} current. This is in line with the expectations from
recent arguments \cite{Yamamoto:2015fxa} that exactly conserved currents cannot have a non-vanishing expectation value in equilibrium.\footnote{For a detailed discussion of covariant vs. consistent
definition of currents in relation to anomalous transport see \cite{Landsteiner:2012kd}.}
 
Similarly $\sigma_{55}=\lim_{k\to 0}\frac{G^{a}_+-G^a_-}{2k}$ with $G^{a}_\pm$ the two point correlator for axial gauge field $a_\pm.$ We can expand $a_\pm=a_\pm^{(0)}+k a_\pm^{(1)}+k^2 a_\pm^{(2)}+\dots$ with
\bea
{a_\pm^{(0)}}''+\bigg(\frac{3}{r}+\frac{f'}{f}\bigg){a_\pm^{(0)}}'-2q^2\phi^2\frac{a_\pm^{(0)}}{r^2f}&=&0\,,\nonumber\\
{a_\pm^{(1)}}''+\bigg(\frac{3}{r}+\frac{f'}{f}\bigg){a_\pm^{(1)}}'-2q^2\phi^2\frac{a_\pm^{(1)}}{r^2f}&=&
\mp\frac{8\alpha}{r^3f}A_t'a_\pm^{(0)}\,.
\eea

\begin{figure}[h]
\begin{center}
\includegraphics[width=0.46\textwidth]{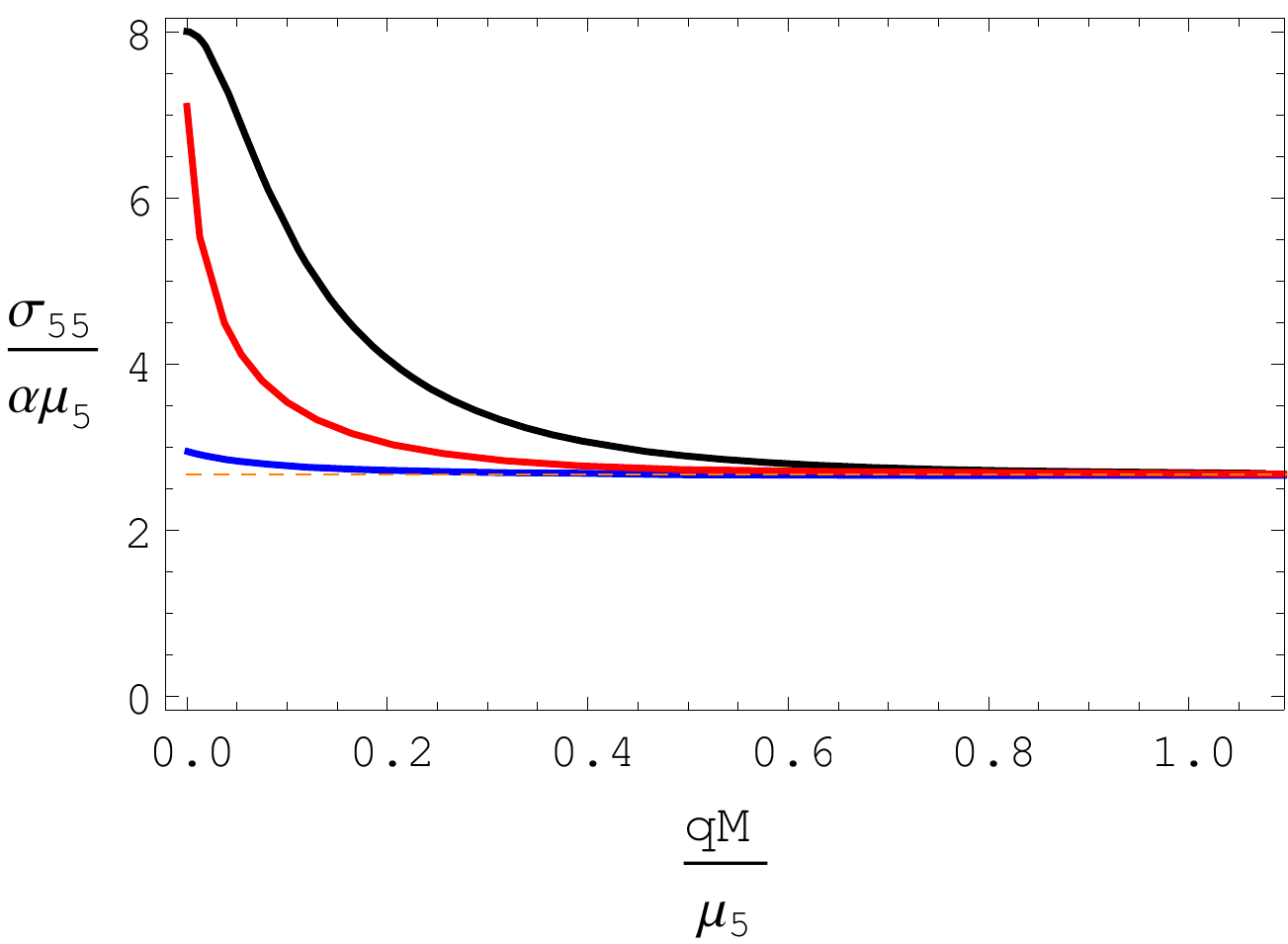}
\includegraphics[width=0.48\textwidth]{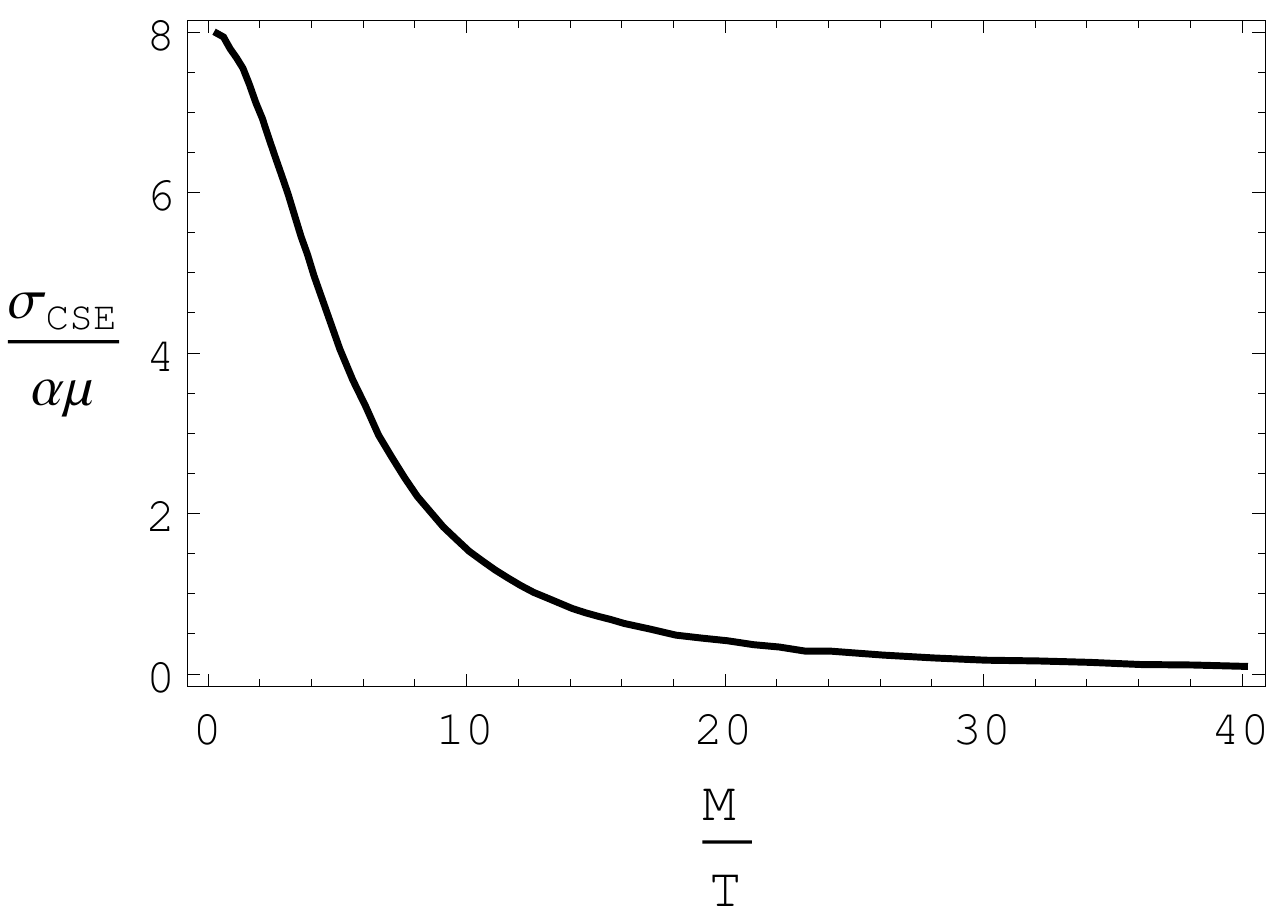}
\parbox{15cm}{\caption{\small Left plot: $\sigma_{55}$ as a function of the source $M$ for $8B\alpha/\pi^2T^2=0.1, \alpha=1$, $T/\mu_5=0.05$ (blue), $0.075$ (red), $0.1$ (black). Note that $T_c/\mu_5\simeq0.765$ for this case.  Dashed line $\sigma_{55}/(\alpha\mu_5)=8/3.$ Right plot: $\sigma_\text{CSE}$ as a function of $M/T$ for $B=0$ and 
$\mu_5=0$ while $\mu\neq 0$. When $M\to 0$, we have $\sigma_\text{CSE}=8\alpha\mu$. For large $M$, $\sigma_\text{CSE}\to 0$.}\label{fig:anomatran}}
\end{center}
\end{figure}

Since one can rescale $a_\pm^{(1)}\to \alpha\mu_5\tilde{a}_\pm^{(1)}$, we conclude that $\sigma_{55}$ is proportional to $\alpha\mu_5.$ The exact result can only be obtained numerically, which is shown in the left plot of Fig. \ref{fig:anomatran}.
We note that in the limit of large mass $M/\mu_5$, $\sigma_{55}$ approaches $\frac{8}{3}\alpha\mu_5$. Also note that this behaviour should not depend on $B$ for fixing $8B\alpha/\pi^2T^2$.  This is the same universal value that was observed in anomalous holographic superconductors
in \cite{Amado:2014mla, Jimenez-Alba:2014pea, {Melgar:2014fwa}} in the $T\rightarrow 0$ limit. We also note that for low temperatures we end up in the $M\rightarrow 0$ limit in the superconducting phase. Therefore the blue line in figure \ref{fig:anomatran} varies very little.
In order to interpret this result we again add the Chern-Simons contribution to the current in order to obtain the (consistent) current that couples axial gauge field $A_\mu$.
This time the Chern-Simons current is $J_{5,\text{CS}}^\mu = \frac{4\alpha}{3}\epsilon^{\mu\nu\rho\lambda} A_\nu F_{\rho\lambda}$. The factor $1/3$ can be understood by noting that this comes from the triangle
anomaly with three equal current on the vertices $\langle J_5 J_5 J_5 \rangle$ which implies a symmetry factor of $1/3$ compared to the triangle with one axial and two vector like currents.
The Chern-Simons term contributes now $-\frac{8\alpha}{3}\mu_5$. This means that in the large mass limit $M\rightarrow \infty$ in which the axial symmetry is maximally broken the
total axial current vanishes! This seems a very intuitive result. 
 
Finally we consider the case $\mu_5=0, \mu\neq 0$. For simplicity we will discuss the chiral separation conductivity in the linear response approximation, i.e. with
vanishing background magnetic field.
When $B=0$, we have a simple solution with non vanishing fields $V_t=\mu(1-\frac{r_0^2}{r^2})$ and $\phi$ as the same as (\ref{specialphi}). In this case $\sigma_B=\sigma_{55}=0$ while nonzero $\sigma_\text{CSE}$ which can be found in the right plot of Fig. \ref{fig:anomatran}. 
Let us make a comment on the behaviour of the chiral separation conductivity.
In this case there is no contribution due to the Chern-Simons term to the current since we only switch
on a chemical potential for the conserved vector like symmetry $\mu\neq 0$. We find that also in this case the axial current induced by a magnetic field vanishes in the limit of
maximal axial symmetry breaking. Since in this case there is no Chern-Simons current also the covariant current vanishes.

The important conclusion of this analysis is that in the limit of maximal axial symmetry breaking via the mass parameter $M$ the expectation value of the axial current ${\bf J}_5$ vanishes for both the chiral separation effect and the axial magnetic effect, but only if one uses the consistent definition of the currents. Since a vanishing axial current for
maximal axial symmetry breaking seems a physically plausible result we take this as an argument in support of using the consistent definition of currents.

\section{Holographic massive $U(1)_\text{A} \times U(1)_\text{V} $ model}
\label{sec3}

In this section we will concentrate on the massive $U(1)_A$ model and consider the presence of two $U(1)$ gauge fields in the bulk. Both fields are coupled via a Chern-Simons term as in the previous model. In order to implement (axial) charge dissipation we consider a constant mass term for one of the $U(1)$, which we refer to as the axial $U(1)_A$. This is achieved in a gauge invariant manner via the St\"uckelberg mechanism \cite{Gursoy:2014ela,{Jimenez-Alba:2014iia}}. The action reads 
\begin{align}
\label{actionu1u1}
\mathcal{S}=\int d^5x\sqrt{-g}\bigg(\frac{1}{2\kappa^2}\Big(R+\frac{12}{L^2}\Big)
-\frac{1}{4}F^2-\frac{1}{4}\mathcal{F}^2-\frac{m^2}{2}(A_\mu-\partial_\mu
\theta)(A^\mu-\partial^\mu
\theta)\nonumber\\
+\frac{\alpha}{3}\epsilon^{\mu\alpha\beta\gamma\delta}(A_\mu-\partial_\mu
\theta)\left(F_{\alpha\beta}F_{\gamma\delta}+3 \mathcal{F}_{\alpha\beta}\mathcal{F}_{\gamma\delta} 
\right)  \bigg)\,,
\end{align}
where $F=dA$ and $ \mathcal{F}=dV$. The St\"uckelberg field and the gauge field $A_\mu$ transform as $\theta\rightarrow \theta+\lambda,A_\mu\to A_\mu+\partial_\mu\lambda$, leaving the mass term invariant. The St\"uckelberg field it is not charged under the $U(1)_V$.

As in the previous section we will work in the probe limit with Schwartzschild AdS background. 
The equations of motion are
\begin{align}
\nabla_\mu F^{\mu \nu} - m^2 (A^\nu-\partial^\nu\theta ) +
\alpha\epsilon^{\nu \alpha \beta \gamma \rho} (F_{\alpha \beta}
F_{\gamma \rho}+\mathcal{F}_{\alpha \beta} \mathcal{F}_{\gamma \rho}) =0\,, \\
\nabla_\nu \mathcal{F}^{\nu \mu}+ 2 \alpha \epsilon^{\mu \alpha \beta \gamma \rho}
F_{\alpha \beta} \mathcal{F}_{\gamma \rho}  =0\,,\\
\nabla_\mu \left( A^\mu -\partial^\mu \theta   \right)=0\,.
\end{align}
The mass term affects the asymptotic behaviour of the axial gauge field $A_\mu$. The asymptotic expansion of the different fields reads
\begin{align}
A_{\mu}&\simeq A_{\mu(0)}r^\Delta+...+\tilde{A}_{\mu(0)}r^{-2-\Delta}+...\,,\\
V_\mu&\simeq V_{\mu(0)}+...+\tilde{V}_{\mu(0)}r^{-2}+...\,,\\
\theta&\simeq \theta_{(0)}+...+\tilde{\theta}_{(0)} r^{-4}+...\, ,
\end{align}
with $\Delta$ the conformal dimension of the source for dual axial current operator $\Delta=-1+\sqrt{m^2+1}$. 
$A_{\mu(0)}$, $V_{\mu(0)}$, $\theta_{(0)}$ are the coefficients of the non-normalisable modes and $\tilde{A}_{\mu(0)}$, $\tilde{V}_{\mu(0)}$, $\tilde{\theta}_{(0)}$ are the coefficients of the normalisable modes. 
As one can see the presence of the mass changes the dimension of the operator dual to the axial $U(1)_A$:  $[J^\mu_5]= 3+\Delta$. This implies that the dual axial charge is not conserved since in a conformal theory a conserved current must saturate the unitarity bound. We refer to $\Delta$ as the anomalous dimension of the axial current. Moreover the anomalous dimension has a bound, given by the condition $\Delta<1$. This is obtained by requiring the dual axial operator to be irrelevant in the UV. 

It is necessary to renormalise the theory in order to obtain finite observables \cite{Papadimitriou:2004ap}. The amount of divergent counterterms needed depends on the precise value of the mass and diverges as one reaches the marginal case $\Delta=1$. For this reason we will only consider masses such that $\Delta<1/3$, which keep this amount minimal. For these values of the mass the boundary term containing the counter terms reads
\begin{equation}
S_\text{ct}=\int_\partial d^4x\,\, \sqrt[]{-\gamma}\left( \frac{\Delta}{2}B_{\mu}B^{\mu}
- \frac{1}{4(\Delta+2)}(\partial_\mu B^\mu)^2 +
\frac{1}{8\Delta}F^2 +\frac{1}{8} \log r^2 \mathcal{F}^2   \right)\,,
\end{equation} 
with $B_\mu= A_\mu -\partial_\mu \theta$.\footnote{In addition to the divergent terms we have implicitly chosen a scheme that respects the axial symmetry by including the finite term $-\frac{\alpha}{3}\partial_\mu\theta\epsilon^{\mu \alpha \beta \gamma \delta} 3\mathcal{F}_{\alpha \beta}\mathcal{F}_{\gamma\delta}$ in (\ref{actionu1u1}).}
A comment is in order regarding the Ward identities of the currents. The vector current is conserved and therefore the divergence vanishes
\begin{equation}
\partial_\mu J^\mu = 0\,.
\end{equation}
With such a choice one could expect the divergence of the axial current to be explicitly proportional to $\alpha\,(\mathcal{F}\wedge\mathcal{F}+F\wedge F)$. However this intuition fails in the massive case. The axial symmetry is broken by the dynamical internal $SU(N)$ degrees of freedom. This implies that the ``current'' operator is to be considered as a non-conserved current which, therefore, lacks any constraint given by the symmetry. This is nicely seen in this model by computing the expectation for the divergence of this current. With the renormalised action and after using the asymptotic expansion one gets
\begin{align}
\partial_\mu J^\mu = 0\,,~~~~~~
\partial_\mu J_5^\mu = (2+2\Delta)\partial_\mu \tilde{A}^\mu_{(0)}\, ,
\end{align} 
with no explicit constraint for the axial current.

\subsection{Magnetoconductivity and relaxation time}

In this section we compute the electric DC conductivity, the static susceptibility and the axial charge diffusion time in presence of a background $B$ field. As background we switch on a spatial component of the vector field $V_x=By$ which trivially fulfils the background equations of motion and generates a constant magnetic field aligned to the $z$ direction. To compute the appropriated quantities we switch on perturbations on top of this background with finite frequency and momentum aligned to the $B$ field $\delta \theta =\eta(r)e^{-i\omega t + ikz}$,
$\delta A_{\mu} = a_{\mu}(r)e^{-i\omega t + ikz}$ and $\delta V_{\mu} =
v_{\mu}(r)e^{-i\omega t + ikz}$.  For our purposes we can just focus in the sector that contains $a_z\,, a_t\,, v_z\,, v_t\,,\eta $ which is decoupled from the other components in our background. The explicit form of the equations can be found in appendix \ref{sec:eqmassiveU1}. Here we list the $k=0$ sector,
\begin{align}\label{eq:massiveU1flu1}
a_t'' + \frac{3}{r}a_t' -\frac{m^2}{r^2f}a_t
+\frac{8\alpha B}{r^3}v_z' +\frac{i \omega
m^2}{r^2f}\eta=0\,,\\
\label{eq:massiveU1flu2} v_z''+\left( \frac{f'}{f}+\frac{3}{r}\right)v_z' +
\frac{\omega^2}{r^4f^2}v_z+\frac{8\alpha B}{f r^3}a_t'=0\,,\\
\label{eq:massiveU1flu3} -im^2 r^2 f\eta' + \omega \big(a_t'+\frac{8\alpha B}{r^3} v_z\big) =0\,.
\end{align}

In order to obtain two-point functions and quasinormal modes numerically\footnote{We refer the reader to \cite{Jimenez-Alba:2014iia} for a thorough explanation on how to numerically compute different quantities within this model.} we build the numerical, matrix valued, bulk to boundary propagator $F$ with the appropriated normalisation such that

\begin{equation}\label{eq:bbp}
\begin{pmatrix}
r^{-\Delta}a_i(r)\\
v_i(r)\\
\eta(r)
\end{pmatrix}= \mathbb{F}(r) 
\begin{pmatrix}
a_{j(0)}\\
v_{j(0)}\\
\eta_{(0)}\end{pmatrix}\,,
\hspace{1cm}
\mathbb{F}(\Lambda)=\mathbb{I}\,,
\end{equation}
where $\Lambda$ is the cutoff radius. This can be obtained imposing infalling boundary conditions and a set of orthonormal values of the amplitudes at the horizon.  With the usual holographic prescription different correlators can be written as a linear combination of derivatives of $\mathbb{F}(r)$ and quasinormal modes can be obtained from the zeros of the determinant of $\mathbb{F}^{-1}(r)$. Concretely for the electric DC conductivity and the axial static susceptibility we find 
\begin{align}
\chi_5&= \lim_{\Lambda\rightarrow \infty}  \Lambda^{3+\Delta} \mathbb{F}'_{a_t,a_t}(\Lambda)\bigg|_{\omega=k=0}\,,\\
\sigma_\text{DC}&= \lim_{\Lambda\rightarrow \infty}\lim_{\omega\rightarrow 0}\Lambda^3 \frac{1}{i\omega} \left( \mathbb{F}'_{v_z,v_z}(\Lambda)+\omega^2 \mathbb{F}_{v_z,v_z}(\Lambda) \log(\Lambda)\right)\bigg|_{k=0}\,,
\end{align}
where subscripts $a_t$, $v_z$ refer to the appropriated entry of the matrix. 

\subsubsection{Electric DC conductivity}
We compute the negative magnetoresistivity in this model and compare it to the results in the previous model. Note that the anomaly term proportional to the external electric and magnetic fields cannot be implemented in the hydrodynamic expansion in this model. This can easily seen from the conservation law since
$
\partial_\mu J_5^\mu = \frac{1}{\tau_5} J_5^0 + c{\bf E}\cdot {\bf B}\, ,
$
does no longer hold due to the anomalous dimension of the axial current. This ultimately implies that one cannot derive an equation analogous to (\ref{2.21}) in this situation. 

As we will show now the magnetic field dependence of the DC conductivity in this model still implies positive magnetoconductivity. We first obtain analytic formula for DC conductivity by means of the near horizon analysis \cite{Donos:2014uba}. The procedure is the same as the one in previous subsection \ref{subsec:dccon-explicitU1}.  Consider fluctuations 
\be
\delta V_\mu=(v_t(r), 0,0,-Et+v_z(r),0)\,,~~~
\delta A_\mu=(a_t(r), 0,0,a_z(r),a_r(r))\,,~~~\delta \theta=\eta\,,~~~
\ee
we need to consider 
\bea\label{eq:dc1-masivegau}
a_t'' + \frac{3}{r}a_t'-\frac{m^2}{r^2f}a_t
+\frac{8\alpha B}{r^3}v_z' &=&0\,,\\
\label{eq:dc1-masivegau2}
v_z''+\left( \frac{f'}{f}+\frac{3}{r}\right)v_z'+\frac{8\alpha B}{ r^3f}a_t'&=&0\,.
\eea

Near conformal boundary $r\to\infty,$ we have $a_t=a_t^{(0)}r^\Delta+\tilde{a}_t^{(0)}r^{-2-\Delta}+\dots, v_z=v_z^{(0)}+\tilde{v}_z^{(0)}r^{-2}+\dots$. From the equation (\ref{eq:dc1-masivegau}) we have conserved quantity $J=-r^3 f v_z'-8B\alpha a_t$ with $\partial_r J=0$. 
Thus $J|_{r\to \infty}=J|_{r\to r_0}.$ 
When $r\to \infty,$ we impose the sourceless boundary condition for $a_t$, i.e. $a_t^{(0)}=0$. Thus the electric current which is the response
of external electric field $j=J|_{r\to \infty}.$ 
\begin{figure}[t]
\begin{center}
\includegraphics[width=0.49\textwidth]{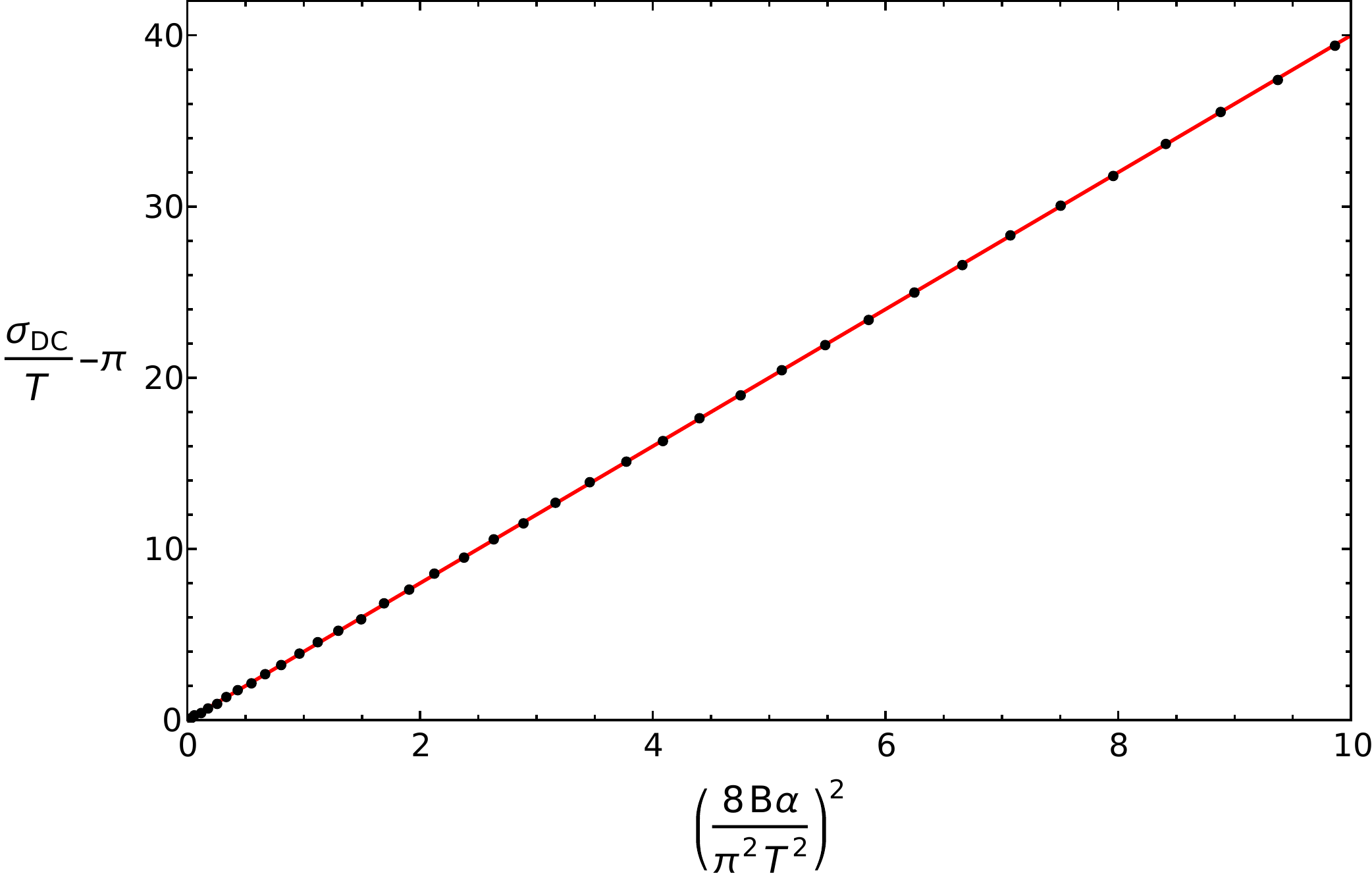}
\includegraphics[width=0.49\textwidth]{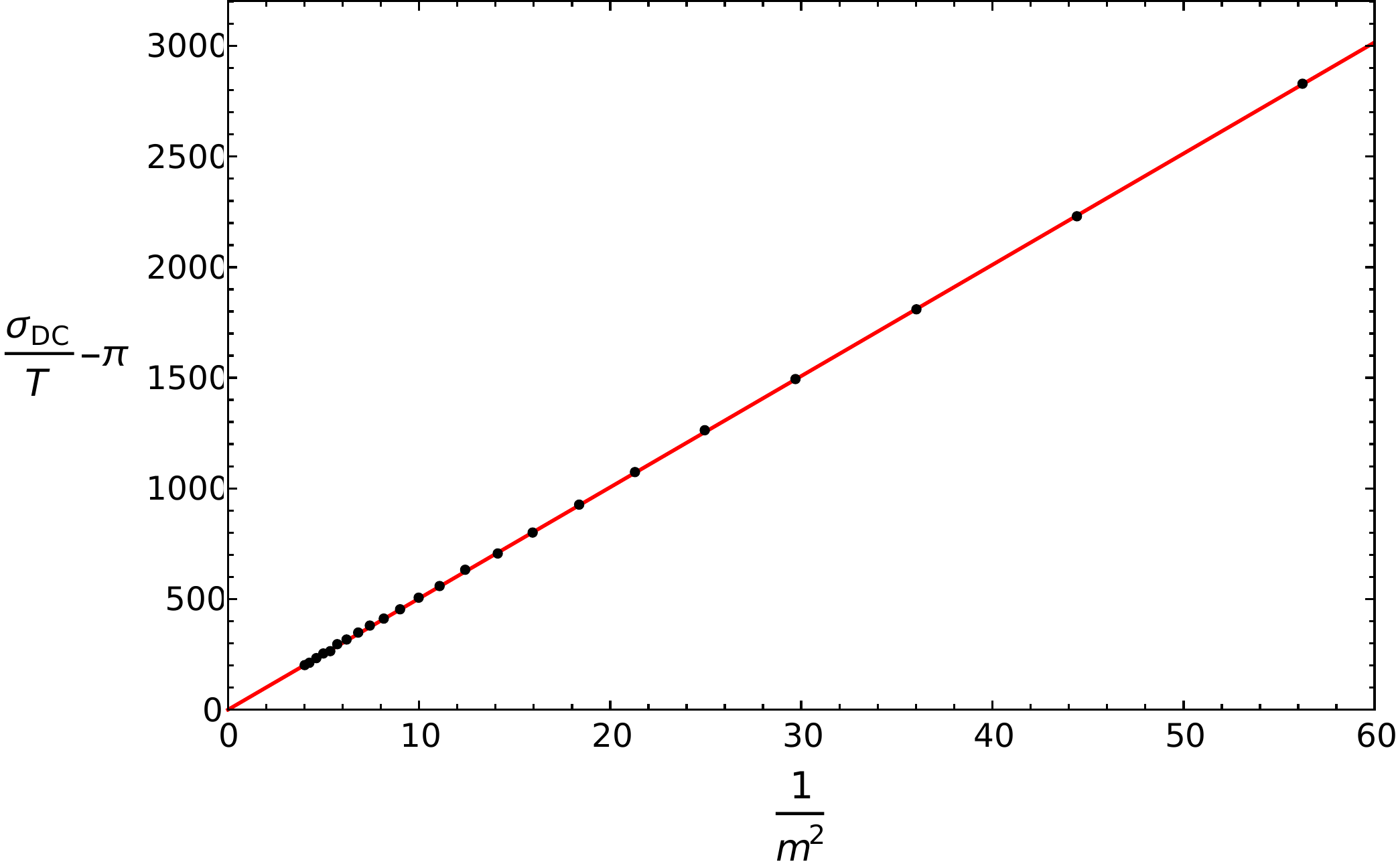}
\parbox{15cm}
{\caption{ \label{fig:dcconstd}\small Dots show numerical result for the DC conductivity obtained by setting $\omega/T=10^{-5}$ for the AC conductivity. Solid lines show the best fit for the data. Left: Fixed $m=0.5$ and the best fit is  $\sigma/T-\pi=0.0001+0.9994 \pi (8B\alpha/\pi^2T^2)^2/m^2$. Right: Fixed $8B\alpha/(\pi^2T^2)=4$ and the best fit is  $\sigma/T-\pi=0.0001+0.9994 \pi (8B\alpha/\pi^2T^2)^2/m^2$. The fitting formulae exactly reproduces (\ref{eq:dc-massiveU1}) from analytical calculations.}}
\end{center}
\end{figure}
We are looking for the solution which is regular near horizon.  Thus we have 
\bea
v_z&=&-\frac{E}{4\pi T}\ln(r-r_0)+\mathcal{O}(r-r_0)\,,\\
a_t&=&-\frac{4E(8B\alpha)}{(4\pi T)m^2 r_0^2}+\mathcal{O}(r-r_0)\,.
\eea
Note that near horizon the subleading term in $a_t$ is a free parameter which is precisely the shooting parameter that can be used to determine the sourceless condition for $a_t$ near boundary. 
It follows immediately  
\be\label{eq:dc-massiveU1}
\sigma_\text{DC}=\frac{j}{E}=\pi T+\frac{\pi T}{m^2}\bigg(\frac{8 B\alpha}{\pi^2 T^2}\bigg)^2\,.
\ee
This formula shows again the key point of this work: the DC longitudinal magnetoconductivity depends quadratically on the magnetic field independently of its strength when charge relaxation is built in the models. It is remarkable that this behaviour holds in this setup too, despite to the lack of a clear hydrodynamic prediction in terms of $\tau_5$ and $\chi_5$.
We also checked the formula numerically. Our results show perfect agreement with the analytic formula, as shown in Fig. \ref{fig:dcconstd}.

The AC conductivity in this model has been studied numerically already in \cite{Jimenez-Alba:2014iia}. Here we note that also in that model a sum rule of the form
(\ref{eq:sumrule}) was found to hold.


\subsubsection{Axial charge dissipation time}

We compute the dissipation time for the axial charge from the gap in the imaginary part of the lowest QNM associated to this symmetry. As shown in \cite{Jimenez-Alba:2014iia} and in \cite{Stephanov:2014dma} in chiral kinetic theory a diffusive mode for the vector charge and a gapped dissipative mode for the axial charge. It is this gap that we identify with the inverse of the axial relaxation time.

We compute the QNMs at zero momentum numerically using the numerical techniques explained in \cite{Amado:2009ts}. The QNMs are obtained from the zeros in the determinant of the inverse bulk to boundary propagator defined in the previous subsection (\ref{eq:bbp}). The explicit equations are shown in appendix \ref{sec:eqmassiveU1}.
In figure \ref{fig:fit_tau_stueck} we show our results for the relaxation time. As shown in \cite{Jimenez-Alba:2014iia} the relaxation time is found to be inversely proportional to the bulk photon mass. Moreover the behaviour of this relaxation with increasing magnetic field shows a transition to a linear regime for big enough magnetic fields, see Fig. \ref{fig:fit_tau_stueck} for fit analysis. This is analogous to what was found in the explicit $U(1)_A$ breaking model.\\
\begin{figure}[t!] 
\centering
\includegraphics[width=210pt]{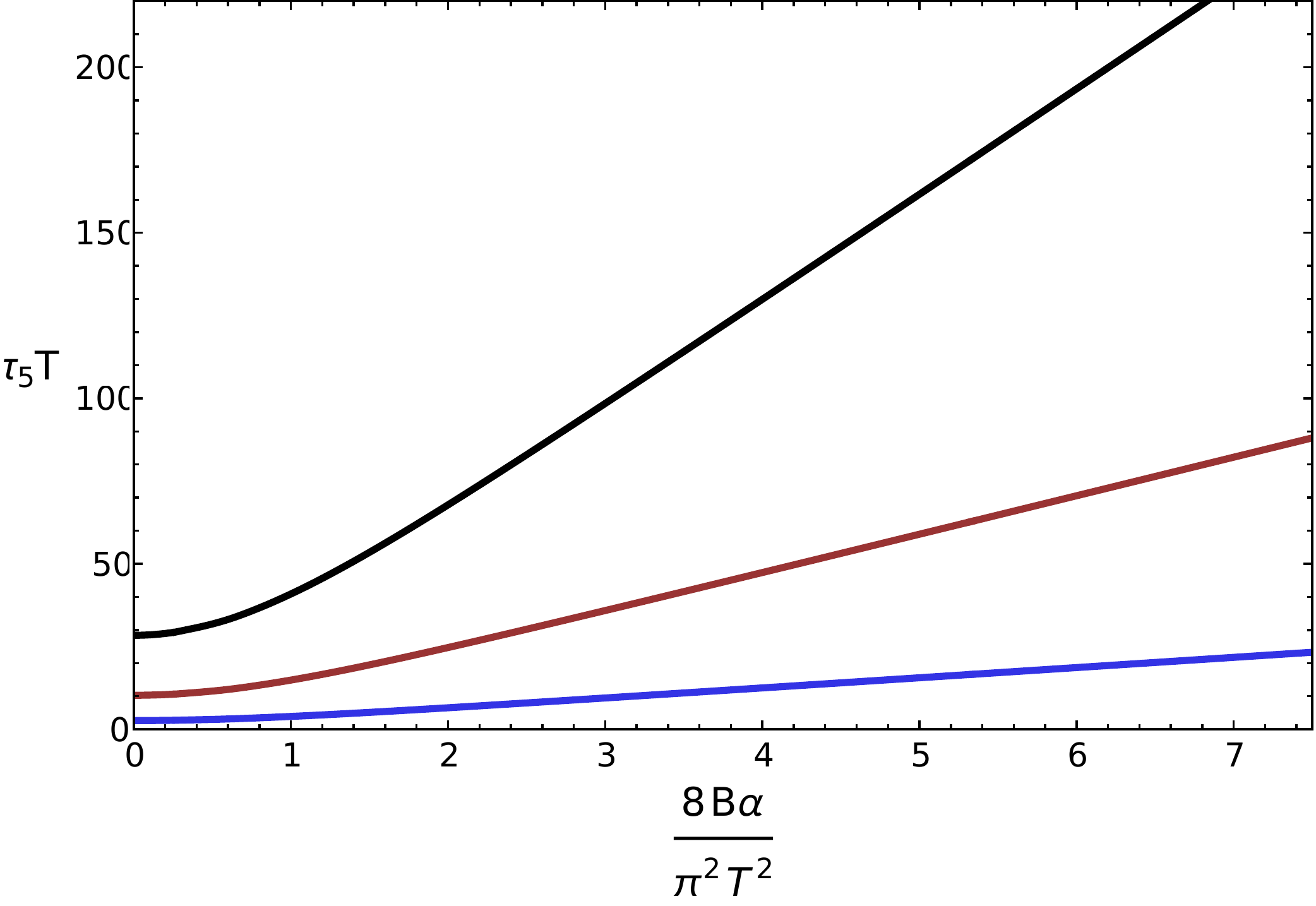}\hspace{0.5cm}
\includegraphics[width=210pt]{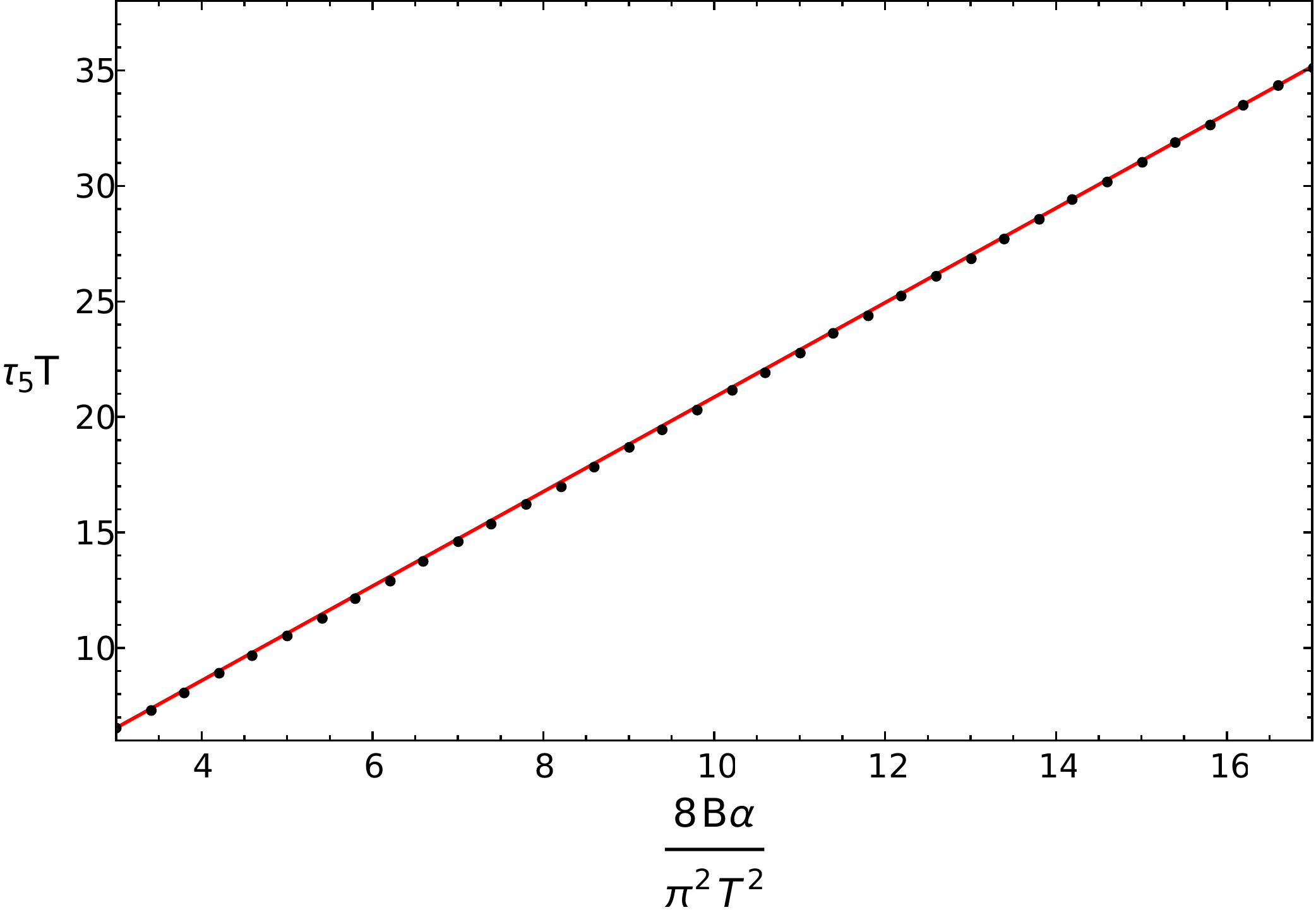}
\parbox{15cm}{\caption{\label{fig:fit_tau_stueck}\small Left panel: Axial charge relaxation time as a function of magnetic field for different values of the bulk mass $m=0.15$ (Black), $m=0.25$ (Red), $m=0.5$ (Blue). Right panel: Numerical data (dots) and linear fit (red) for $\tau_5$ at $m= 0.5$ and large magnetic field. The best fit in this region is $\tau_5 T= 0.0979+0.5030\frac{8B\alpha}{\pi^2 T^2m^2}$.}
}
\end{figure}

\subsubsection{Static axial susceptibility}

Although (\ref{2.21}) does not straightforwardly apply to this model too it is still interesting to compute the axial susceptibility and its dependence with the magnetic field. \\
To obtain the analytic result we set $E=0$, then we have 
\be a_t'' + \frac{3}{r}a_t'-\frac{m^2}{r^2f}a_t
+\frac{(8\alpha B)^2}{r^6 f}a_t =0\,.\ee
Near horizon we impose $a_t(r_0)=0.$ The solution is  
\bea
a_t&=&i\bigg(\frac{r_0}{r}\bigg)^{1+\sqrt{1-\beta^2}}\frac{\Gamma\big[1+\sqrt{1-\beta^2}/2\big]}{\Gamma\big[1-\sqrt{1+m^2}/4+\sqrt{1-\beta^2}/4\big]\Gamma\big[1+\sqrt{1+m^2}/4+\sqrt{1-\beta^2}/4\big]} \times\nonumber\\&&~~\,_2F_1\big[(-\sqrt{1+m^2}+\sqrt{1-\beta^2})/4, (\sqrt{1+m^2}-\sqrt{1-\beta^2})/4,1-\sqrt{1-\beta^2}/2, r^4/r_0^4\big]\nonumber\\
&&-i\bigg(\frac{r_0}{r}\bigg)^{1-\sqrt{1-\beta^2}}\frac{\Gamma[1-\sqrt{1-\beta^2}/2]}{\Gamma\big[1+\sqrt{1+m^2}/4-\sqrt{1-\beta^2}/4\big]\Gamma\big[1-\sqrt{1+m^2}/4-\sqrt{1-\beta^2}/4\big]} \times\nonumber\\&&~~ \,_2F_1\big[(-\sqrt{1+m^2}+\sqrt{1-\beta^2})/4,(\sqrt{1+m^2}+\sqrt{1-\beta^2})/4,1+\sqrt{1-\beta^2}/2, r^4/r_0^4\big]\,,\nonumber
\eea
where $\beta=\frac{8B\alpha}{\pi^2 T^2}.$ 
Near conformal boundary the above solution behaves as $a_t=a_t^{(+)}r^{-1+\sqrt{1+m^2}}+a_t^{(-)}r^{-1-\sqrt{1+m^2}},$ thus 
\bea
\chi_5&=&-2\sqrt{1+m^2}\frac{a_t^{(-)}}{a_t^{(+)}}\nonumber\\
&=&4 r_0^{2\sqrt{1+m^2}}\frac{\Gamma\big[1-\frac{\sqrt{1+m^2}}{2}\big]\Gamma\big[\frac{1}{4}\big(\sqrt{1+m^2}-\sqrt{1-\beta^2}\big)\big]\Gamma\big[\frac{1}{4}\big(\sqrt{1+m^2}+\sqrt{1-\beta^2}\big)\big]}{\Gamma\big[\frac{\sqrt{1+m^2}}{2}\big]\Gamma\big[-\frac{1}{4}\big(\sqrt{1+m^2}+\sqrt{1-\beta^2}\big)\big]\Gamma\big[\frac{1}{4}\big(-\sqrt{1+m^2}+\sqrt{1-\beta^2}\big)\big]}\,.\nonumber
\eea

Now let us study the behaviour of static axial charge susceptibility in large $B$ limit. When $\beta=\frac{8B\alpha}{\pi^2 T^2}\to \infty$, we have 
\be\label{eq:oracle}
\chi_5\to 4 r_0^{2\sqrt{1+m^2}}\frac{\Gamma[1-\frac{\sqrt{1+m^2}}{2}]}{\Gamma[\frac{\sqrt{1+m^2}}{2}]}\bigg(\frac{\beta}{4}\bigg)^{\sqrt{1+m^2}}\,.\ee

We check this numerically. We compute the axial static susceptibility $\chi_5$ and its dependence with the background magnetic field numerically by means of the the Kubo formula
\begin{equation}
\chi_5=\langle J^5_t J^5_t\rangle\bigg|_{\omega=k=0}\,.
\end{equation}

\begin{figure}[t!] 
\centering
\includegraphics[width=210pt]{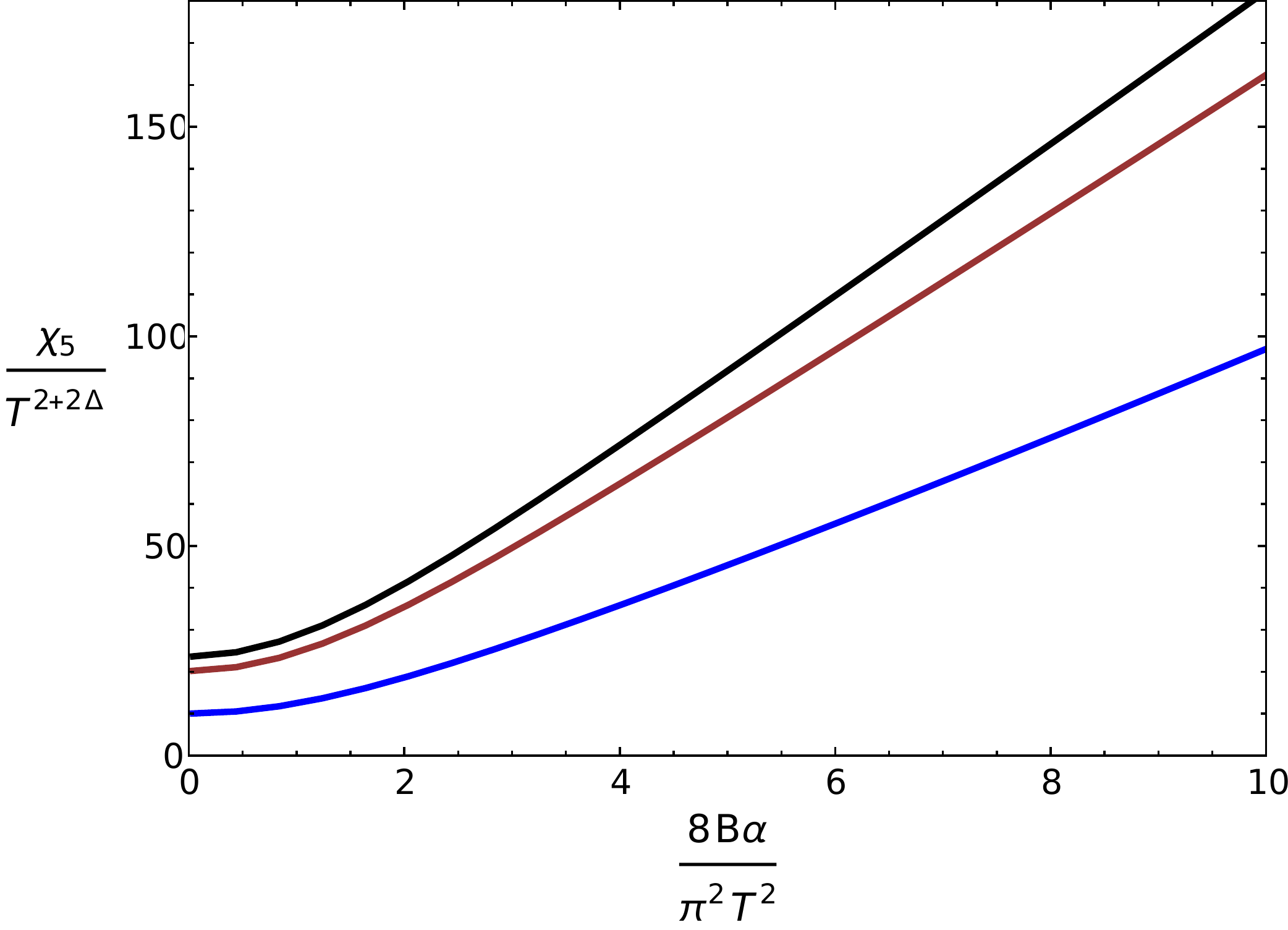}\hspace{0.5cm}
\includegraphics[width=210pt]{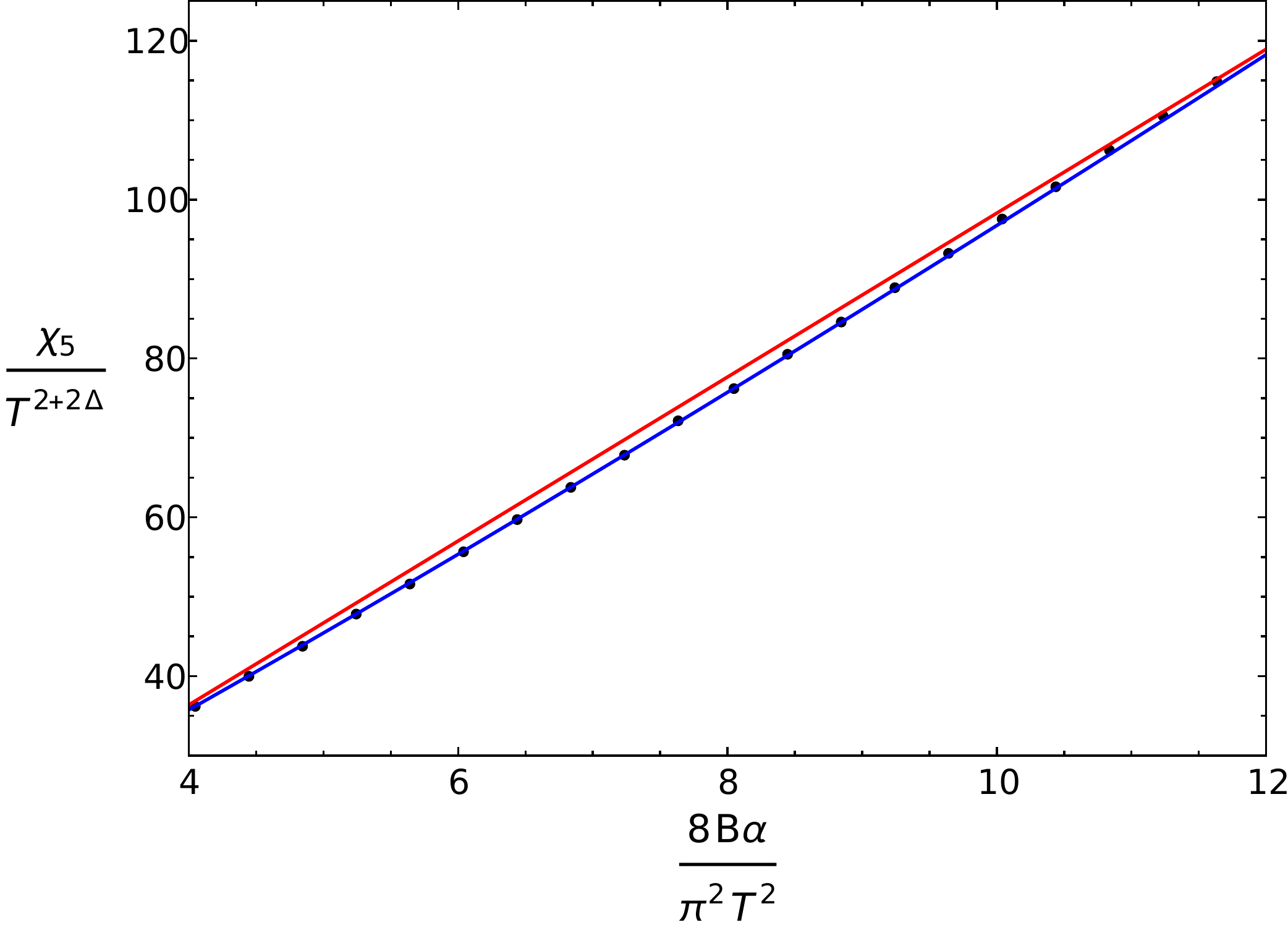}
\parbox{15cm}
{\caption{\label{fig:suscep_stueck}\small Left: Static susceptibility as a function of magnetic field for different values of the bulk mass $m=0.15$ (Black),
$m=0.25$ (Red),$m=0.5$ (Blue). Right: $\chi_5$ for
$m=0.5$. Red line corresponds to linear fit and blue line to  $(8\alpha B/T^2\pi^2)^{\sqrt{1+m^2}}$ as expected from (\ref{eq:oracle}).
 }}
\end{figure}

In figure \ref{fig:suscep_stueck} we show the behaviour of the static susceptibility against magnetic field. As expected there is a transition to a fixed scaling for large magnetic field. However, as predicted in (\ref{eq:oracle}) the exponent is now $\sqrt{1+m^2}$, see Fig. \ref{fig:suscep_stueck} for fit analysis. Our results show indeed that the ratio $\tau_5/\chi_5$ is not $B$ independent for large values of $B$, contrary to what was found in the previous model. 

\section{Remarks: relations of two models and DC results from small $\omega$ matching}
\label{sec4}
In this section we will make some important remarks on two separate aspects: the relations between our two holographic axial charge dissipation models and rederivation of the DC conductivity based on radially conserved quantities from the near-far matching calculation for AC conductivity at small frequencies.

\subsection{Comparison between the two models}

These two axial charge dissipation models look very different: one breaks the charge conservation symmetry by giving a mass to the gauge field and the other breaks this symmetry explicitly by a scalar operator source. However, if we look closer to these two models we will find that these two are in fact closely related. We can compare the actions or the equations of motion of these two models. It is easy to find that in the equations of motion for perturbations (\ref{eq:massiveU1flu1} - \ref{eq:massiveU1flu3}), if we replace $m^2$ in the massive gauge field model by $2 q^2 \phi^2$ and $\eta\to -\frac{\phi_2}{q\phi}$, we will reproduce exactly the same equations of perturbations for the explicit breaking model (\ref{eq:flu1} -\ref{eq:flu3}). It explains that our DC conductivities (\ref{eq:dccond-analytical0}) and (\ref{eq:dc-massiveU1}) in these two models have a universal formulae. This shows that the scalar operator in the explicit breaking model gives an effective mass to the gauge field.\footnote{Similar physics happens for the momentum relaxation case: as concluded in \cite{Blake:2013owa}  the explicit momentum breaking by a scalar field gives graviton a mass. } By solving the equations of the two models, we can even see that the massive gauge field model is in fact a special case for the explicit
breaking case in which the bulk mass for the scalar is set to zero and $\mu=\mu_5=0$. 
The massless scalar is dual to a marginal operator. A source for it does however change the scaling dimension of the axial current.  An advantage of the explicit breaking model is that we can choose the mass of the scalar field freely without changing the the scaling dimension of the axial current.

Note that in both models we can choose two kinds of gauges (or any combination of the two): one is to choose $\delta A_r=0$ and the other is to choose $\delta \eta=0$ in the massive gauge field case or $\delta \phi_2=0$ in the explicit breaking case. For both these two gauges, we will find that the equations for perturbations are the same with the mass of the massive gauge field replaced by an effective mass generated by the scalar source.

Finally, though at the level of actions or equations of perturbations the two are closely related or equivalent in some sense, there is a subtlety here that in the massive gauge model, there is no Higgs mode, i.e. $\eta$ is a real scalar while $\phi$ is a complex scalar field. This Higgs mode does not have any effect in transport coefficients, but it may have other effects.
In particular when $\mu\neq 0$ or $\mu_5 \neq 0$ the Higgs mode will not decouple.

\subsection{ Near far matching calculation for the AC conductivity at  low frequency}
\label{subsec:matching}

In this subsection we reproduce the DC magnetoconductivity result which was obtained using {\em radially conserved quantities} from the near far matching calculation for both the explicit $U(1)_A$ breaking and the massive gauge field case. 
As we emphasied in the previous subsection, the equations (\ref{eq:flu1} - \ref{eq:flu3}) and (\ref{eq:massiveU1flu1} - \ref{eq:massiveU1flu3}) are the same if we replace $m^2$ in the massive gauge field model by $2 q^2 \phi^2$ and $\eta\to -\frac{\phi_2}{q\phi}$. In the following we consider Eqs. (\ref{eq:massiveU1flu1} - \ref{eq:massiveU1flu3}). It is straightforward to apply to the replacement to (\ref{eq:flu1} -\ref{eq:flu3}). 

We work in the gauge $\delta \eta=0$ (or $\delta\phi_2=0$ gauge for the explicit breaking case) and in the coordinate $u=\frac{r_0^2}{r^2}$ for convenience.
The equations for the perturbations $\delta A_t=a_t(u)e^{- i \omega t}$, $\delta V_z=v_z(u)e^{- i \omega t}$, and $\delta A_r=a_r(u)e^{- i \omega t}$ are
\bea \label{eq:eqnearfar1}
4 u^2(1-u^2)(4\alpha B v_z'+r_0^2 a_t''+i r_0^2 \omega a_r')-m^2 r_0^2 a_t=0\,,\\ \label{eq:nearfar2}
(-m^2 r_0^4(1-u^2)+r_0^2u\omega^2)a_r-i u\omega(4\alpha B v_z+r_0^2 a_t')=0\,,\\ \label{eq:nearfar3}
4u(1-u^2)\Big(r_0^2\big((1-u^2)v_z'\big)'+4\alpha B a_t'\Big)+\omega^2 v_z+16 i \alpha B u\omega(1-u^2)a_r=0\,
\eea
with $'$ the derivative in $u.$

The near region is defined as $1-u\ll 1$ and the far region is defined as $1-u\gg \frac{\omega}{r_0}$. We will first solve the near region and the near horizon boundary conditions for the far region are provided by the near region solutions expanded at the matching region.

In the near region, we have the infalling boundary conditions and the solutions at leading order are
\bea
v_z &\simeq& (1-u)^{-i\omega/4 r_0}\big(1+\dots\big)\,,\\
a_t &\simeq&(1-u)^{-i\omega/4r_0}\big(\omega s_0 + s_1 (1-u)+\dots\big)\,,\\
a_r &\simeq& (1-u)^{-i\omega/4r_0-1}\big(\omega s_2 +s_3 (1-u)+\dots\big)\,,
\eea 
where $s_0$, $s_1$, $s_2$, $s_3$ are constants which may depend on $\omega$ and $m$ and using the equations there will be only one free parameter which we denote as $s_2$. The `$\cdots$' above denotes subleading terms which contribute to the same higher orders in the equations.

The following matching calculations are equivalent to defining new functions as $v_z=(1-u)^{-i\omega/4 r_0}v_z^f(u)$, $a_t=(1-u)^{-i\omega/4 r_0} a_t^f(u)$ and $a_r=(1-u)^{-i\omega/4 r_0} a_r^f(u)$ and solving these new functions with boundary conditions $a_t=0$ at $u=0$. Here we take the terminology of matching in order to present the results more clearly to be understood. With the near region leading order solutions, we then expand them at leading order in $\omega$ in the matching region $\frac{\omega}{r_0} \ll 1-u\ll 1$, which can be used as near horizon boundary conditions for the far region:
\bea
v_z &\simeq& 1-\frac{i\omega}{4r_0}\ln(1-u)\,,\\
a_t &\simeq& -4 r_0  s_2 \omega+(1-u)\Big[\big(\frac{4\alpha B}{r_0^2}-2 i m^2 r_0^2 s_2\big)+\big(\frac{i \alpha B}{r_0^3}+\frac{1}{2}m^2 r_0^2 s_2\big) \omega\Big]\, \nonumber\\
&+&(1-u)\ln (1-u) \bigg(\frac{4\alpha B}{r_0^2}-2 i m^2 r_0^2 s_2\bigg)\bigg(\frac{-i\omega}{4r_0}\bigg),
\eea 
where $s_2$ is the tuning parameter to make sure that the boundary condition $a_t=0$ at the boundary $u=0$ is satisfied. $a_r$ is decoupled in the far region and is not important here.

To derive the far region equations, we can drop all the terms in the equations at order $o(\omega)$ while order $\mathcal{O}(\omega)$ terms should be kept. In the equations above, we can solve $a_r$ from the second equation and substitute it to the first and the third equations and find that those $\mathcal{O}(\omega)$ order terms all become $\mathcal{O}(\omega^2)$ order terms and can be ignored in the far region. Note that in this procedure we have secretly assumed that $m^2 r_0^2/\omega \gg \omega u/(1-u)$ in the far region as can be seen from the coefficient of $a_r$ in the second equation. With a further constraint that the $a_r$ terms in (\ref{eq:eqnearfar1}) and (\ref{eq:nearfar3}) are at order $o(\omega)$, we know that the following far region equations are only valid for $m^2\gg \omega/r_0$ (this condition should be substituted by $M^2\gg \omega/r_0$ in
explicit breaking case), i.e. $\omega \tau_5 \ll 1$:
\bea
\big((1-u^2)v_z'\big)'+\frac{4\alpha B}{r_0^2} a_t'=0\,,\\
4u^2(1-u^2)\bigg(\frac{4\alpha B}{r_0^2} v_z'+a_t''\bigg)-m^2a_t=0\,,
\eea 
which are the same as we are studying in the paper (\ref{eq:dc1} - \ref{eq:dc2}) and (\ref{eq:dc1-masivegau} - 
\ref{eq:dc1-masivegau2}) for DC conductivity 
and the first equation can be integrated to give 
\be (1-u^2) v_z'+\frac{4\alpha B}{r_0^2} a_t=C_0\,, \ee 
where $C_0$ is an integration constant which can be decided by the near horizon analysis to be
\be C_0= \frac{i\omega}{2r_0} -\frac{16 \alpha B}{r_0} s_2 \omega\,,\ee at order $\mathcal{O}(\omega)$.

The next step is to see which $s_2$ can set the leading order coefficient of $a_t$ at the conformal boundary to be $0$. If $s_2$ has $1/\omega$ dependence at the leading order, then there is only one term ($1-u$) in $a_t$ at the order $1/\omega$ due to the $m^2$ term in the equation for $a_t$, which cannot be canceled by other terms at the boundary. Thus $s_2\sim s_{20}+\omega s_{21}\cdots$. 
Then for small $\omega$ we can see that there is one term $(1-u)$ in the near horizon expansion of $a_t$ at order $\mathcal{O}(1)$ while others are at $\mathcal{O}(\omega)$. $\mathcal{O}(\omega)$ and $\mathcal{O}(1)$ terms belong to linearly independent solutions and at the boundary they will lead to boundary values at order $\mathcal{O}(\omega)$ and $\mathcal{O}(1)$ separately because the far region equations do not depend on $\omega$. To make sure that the leading order in $a_t$ is $0$ 
at the conformal boundary, we have to impose that the $\mathcal{O}(1)$ order coefficient in front of $(1-u)$ vanish, which gives $s_{20}=-\frac{2 i \alpha B}{m^2 r_0^4}$. $s_{21}$ is now the tuning parameter to shoot the boundary value of $a_t$ to $0$. 

From the far region solution we know that $v_z'(u=0)=C_0$ and $v_z=1+ib_1 \omega/m^2$ where $b_1$ should be a constant at order $\mathcal{O}(1)$ in both $\omega$ and $m^2$. Note that in $v_z'(0)$ there should be $\omega^2$ order corrections which can only be obtained by considering the subleading order equations in the far region. Thus the final result for the magnetoconductivity should be 
\bea
\sigma&=&\frac{2 r_0^2}{i\omega}\frac{ C_0+ b_2 \omega^2}{1+ i \frac{b_1\omega}{m^2} }\\ \label{eq:condnearfar}
&=& \frac{r_0+\frac{64\alpha^2B^2}{m^2}r_0^3-i  b_2 \omega}{1+ i b_1 \frac{\omega}{m^2} },
\eea where $b_2$ comes from the $\omega^2$ corrections to $v_z'(0)$. With a simple estimate $b_2\sim 1/m^4$ for small $m$. This is consistent with the fact that we are working in the small $\omega/m^2T$ limit, in which the subleading order corrections compared to the leading order coefficients in the numerator and the denominator  are usually at the same order and $b_1$ can not be identified as related to the relaxation time $\tau_5$.  Note that the terms in the numerator at order $\mathcal{O}(\omega^0)$ is accurate and non-perturbative in $m^2$.  First we can see that when $\omega\to 0$ we reproduce the DC results:  Eq. (\ref{eq:dccond-analytical0}) in Sec. \ref{sec2} for the $U(1)_A$ explicit breaking model and Eq. (\ref{eq:dc-massiveU1}) in Sec. \ref{sec3} for the massive gauge field case. 
Second, it is easy to see that when we take first $\omega\to0$ and then $m^2\to 0$ limit, $r_0$ will give the value of $\sigma_E$. In fact there is no good definition for $\sigma_E$ from the holographic result because it may depend on the value of $\omega/m^2$ and as we will explain at the end of this subsection there will also be contributions to the DC conductivity at order $\sigma_E$ from other quasinormal modes than the one that we focus here. 


There are several comments on this result which we list as follows.
\begin{itemize}

\item
The calculation above is valid for $m^2\gg \omega/r_0$, i.e. $\omega\tau_5\ll 1$. If we want to go to the opposite limit $m^2\ll \omega/r_0\ll 1$ in this holographic framework we can still perform this calculation except that the far region equations now are totally different. This can be seen from the second equation (\ref{eq:nearfar2}): now the $\omega^2$ coefficient in front of $a_r$ is more important. The result in this limit would recover our old result in \cite{Landsteiner:2014vua}, including the different form of $\sigma_E$. Thus one important conclusion is that the DC calculation in the main text based on radially conserved quantities is only consistent with the AC result in the limit $\omega/r_0\ll m^2.$

\item

Note that here $b_1$ does not give us the value of $\tau_5$ because this is the $\omega\tau_5\ll 1$ limit while $\tau_5$ should be determined from the pole in the $\omega/r_0 \sim m^2$ limit. In this limit, we can also solve the equations by assuming $m^2= \lambda_1 \omega/r_0$ where $\lambda_1$ is an order $1$ number.  In this limit, there will be order $\mathcal{O}(\omega)$ terms in the far region equations and we should solve these equations order by order in $\omega$, i.e. $\mathcal{O}(\omega)$ order solutions come from two parts: $\omega$ order corrections to the $\mathcal{O}(1)$ order solutions and one linearly independent part. This is beyond the calculation in this paper and we will leave it for future investigation.

\item

The hydrodynamic formulas only capture the physics of the one quasinormal mode that we focus on, which has $\tau_5 T\to \infty$. However, the DC holographic result has contributions from all quasinormal modes. This means there will be an order $1/\tau_5 T$ difference in the holographic and hydrodynamic results for the DC conductivity.

\item Finally note that these considerations go through with little change for the explicit breaking case. The result is given by (\ref{eq:condnearfar}) upon substituting
 $ m^2 = 2 q^2 \phi_0^2$.

\end{itemize}

\section{Conclusion and discussion}
\label{sec5}

We have considered two holographic models to encode axial charge dissipation in the probe limit.
They are dual to four dimensional strongly coupled anomalous systems in presence of background magnetic field.
In our first holographic model,the axial charge dissipation is realised by a charged scalar non-normalisable mode.
At weak coupling this corresponds to introducing a fermionic mass term. 
In our second holographic model, the so called St\"uckelberg massive $U(1)_A$ model, the $U(1)_A$ is broken by giving the axial current an anomalous dimension. As we have argued both models are closely related. Indeed if one choses the scalar mass $m_s=0$ and $\mu=\mu_5=0$ the relevant equations coincide
in both models.

We found  that in both these two models, positive magnetoconductivity is exactly quadratic in in the magnetic field strength. Moreover this remains true even in the case of small relaxation times when the axial charge can not be considered to be approximately conserved. 

This is consistent with the recent experiments \cite{Li:2014bha,  Huang, Zhang, Xiong}.
A recent weakly coupled theoretical proposal for ionic scattering \cite{sarma} also found exact quadratic scaling of the magnetoconductivity with magnetic field. We note that our results, beyond being valid at strong coupling, are also different in that we consider high temperature and low chemical potentials. In particular our results suggest that at weak coupling but high temperature there should still be positive magnetoconductivity quadratic in $B$ even when the fermi energy does not intersect any Landau level but lies in the gap region. Our models should also be of relevance for application to non-central heavy ion collisions quark gluon plasma where strong magnetic fields are present.

We have introduced the axial charge relaxation by switching on constant sources. Anther interesting way
to induce it should be via a random source that averages to zero, i.e. introducing disorder. We leave this possibility for future investigation.

\subsection*{Acknowledgments}
We thank F. Pena-Benitez, A. Donos, L. Melgar, X -L. Qi, K. Schalm, and J. Zaanen for discussions. A.J. and Y.L. would like to thank the Galileo Galilei Institute for Theoretical Physics for the hospitality and the INFN for partial support during the completion of this work. 
This work was supported in part by the Spanish MINECO's ``Centro de Excelencia Severo Ochoa" 
Programme under grant SEV-2012-0249.

\appendix

\section{Appendix}

In this appendix we will list the details of the previous holographic result on the conductivity for the case without axial charge dissipation \cite{Landsteiner:2014vua} and the equations mentioned in the main text for reference. 

\subsection{The quantum critical conductivity $\sigma_E$ in the holographic model without axial charge dissipation}
\label{app:sigmaE}
\begin{figure}[t] 
\centering
\includegraphics[width=200pt]{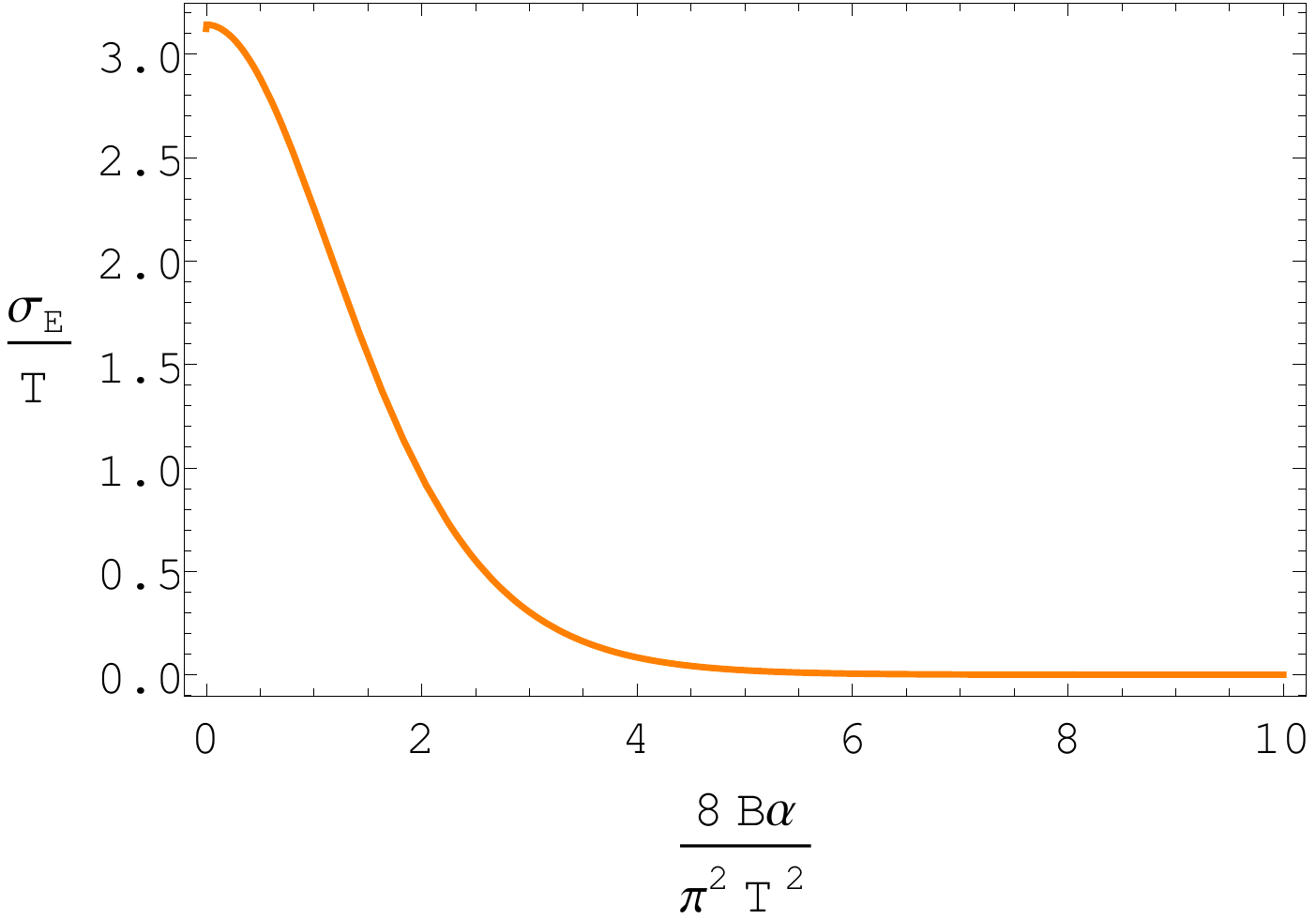}\hspace{0.3cm}
\includegraphics[width=220pt]{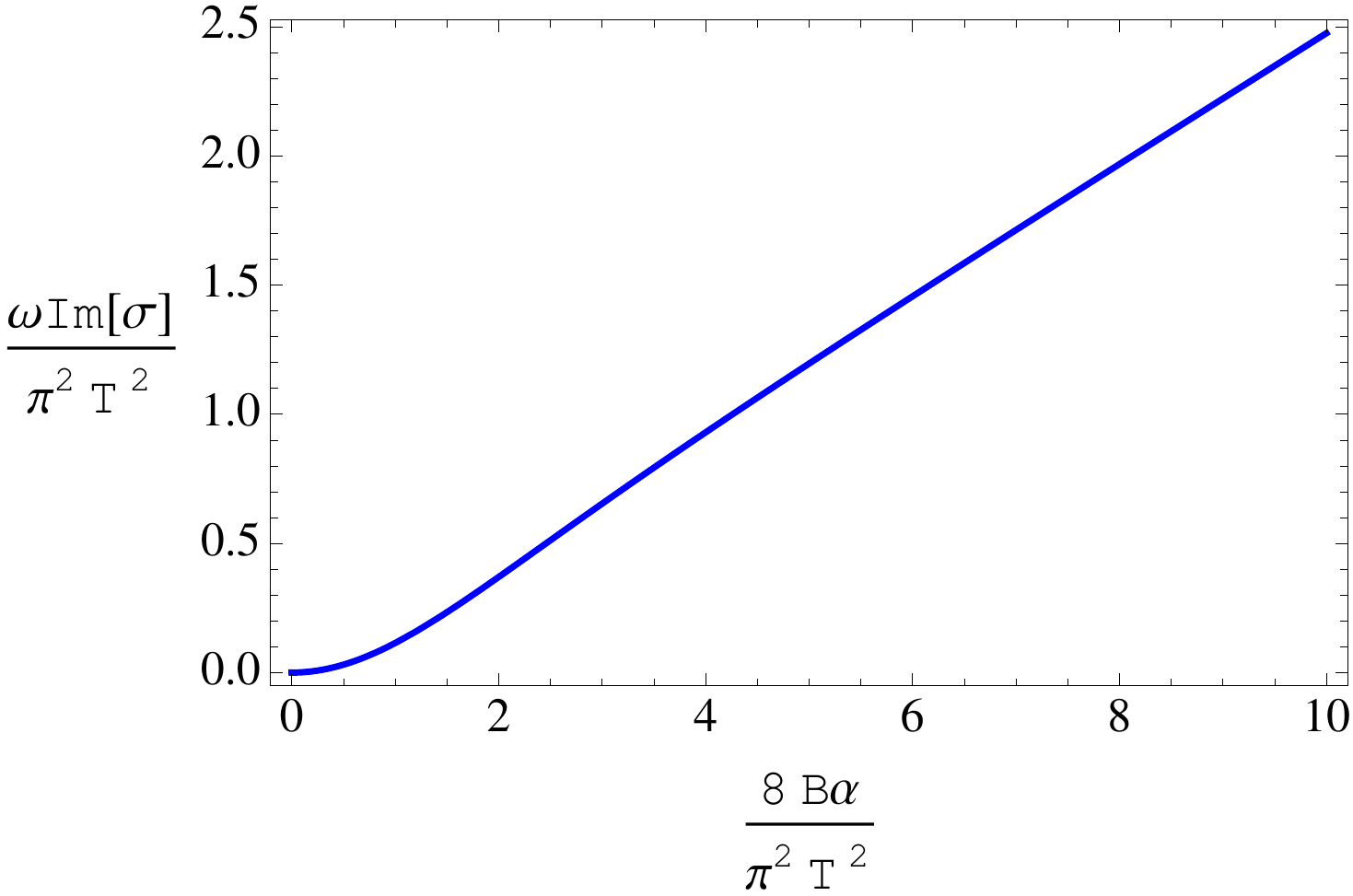}
\parbox{15cm}{\caption{\label{fig:sigmaEwc}\small The real part and the imaginary part of $\sigma$ in the holographic model without axial charge dissipation \cite{Landsteiner:2014vua}.}
}
\end{figure}

In this subsection, we recall the result of the quantum critical conductivity $\sigma_E$ and its plot for reference \cite{Landsteiner:2014vua}. 
From holography in the $U(1)_V\times U(1)_A$ model with a background $B$ field, we have 
\be
\sigma=\sigma_E+\frac{i}{\omega}\frac{(8B\alpha)^2}{\partial\rho_5/\partial\mu_5}
\ee
where 
\bea
\label{sigmaE-prvious}
\sigma_E&=&\frac{\pi^2 T}{8}\beta^2\sec\left(\frac{\pi}{2}\sqrt{1-\beta^2}\right)\frac{\Gamma\big[\frac{3-\sqrt{1-\beta^2}}{4}\big]\Gamma\big[\frac{3+\sqrt{1-\beta^2}}{4}\big]}{\Gamma\big[\frac{5-\sqrt{1-\beta^2}}{4}\big]\Gamma\big[\frac{5+\sqrt{1-\beta^2}}{4}\big]}\,,\\
\frac{\partial\rho_5}{\partial\mu_5}&=&\frac{\pi^2 T^2}{4}\beta^2\frac{\Gamma\big[\frac{3-\sqrt{1-\beta^2}}{4}\big]\Gamma\big[\frac{3+\sqrt{1-\beta^2}}{4}\big]}{\Gamma\big[\frac{5-\sqrt{1-\beta^2}}{4}\big]\Gamma\big[\frac{5+\sqrt{1-\beta^2}}{4}\big]}\,,
\eea
with $\beta=\frac{8B\alpha}{\pi^2 T^2}$. The plot for the real part ($\sigma_E$) and the imaginary part of $\sigma$ can be found in Fig. \ref{fig:sigmaEwc}. For large $B$, we have $\sigma_E\to 0.$

\subsection{Equations for the background in the explicit $U(1)_A$ breaking model}
\label{app:eqexplicit}

Substituting the ansatz (\ref{eq:bgansatz}) into (\ref{eq:eomtwou1-1}-\ref{eq:eomtwou1-3}) we find the the following background equations of motion
\bea
\label{eq:twou1forbg1} A_t''+\frac{3}{r}A_t'-\frac{2q^2 \phi^2 }{r^2f}A_t+\frac{8B\alpha}{r^3}V_z'&=&0\,,\\
\label{eq:twou1forbg2} A_z''+\Big(\frac{3}{r}+\frac{f'}{f}\Big)A_z'-\frac{2q^2\phi^2}{r^2 f}A_z+\frac{8B\alpha}{r^3 f} V_t'&=&0\,,\\
\label{eq:twou1forbg3} V_t''+\frac{3}{r}V_t'+\frac{8B\alpha}{r^3}A_z'&=&0\,,\\
\label{eq:twou1forbg4}V_z''+\Big(\frac{3}{r}+\frac{f'}{f}\Big)V_z'+\frac{8B\alpha}{r^3f}A_t'&=&0\,,
\\
\label{eq:twou1forbg5}
\phi''+\bigg(\frac{5}{r}+\frac{f'}{f}\bigg)\phi'+\bigg(\frac{q^2 A_t^2}{r^4 f^2}-\frac{q^2 A_z^2}{r^4 f}-\frac{m^2}{r^2f}\bigg)\phi&=&0\,.
\eea

We focus on the solution which is regular on the black hole event horizon. Thus $A_t(r=r_0)=0$. We set $V_t(r=r_0)=0.$ Equations (\ref{eq:twou1forbg3}) and (\ref{eq:twou1forbg4}) can be simplified as 
$(r^3 V_t'+8B\alpha A_z)'=0$ and $(r^3 f V_z'+8B\alpha A_t)'=0.$ Thus we can integrate the two equations to get $r^3 V_t'+8B\alpha A_z=c_0$ and 
$r^3 f V_z'+8B\alpha A_t=0.$ 
Eq. (\ref{eq:twou1forbg1}) can be written as 
\be\label{eq:twou1forbg1a}
A_t''+\frac{3}{r}A_t'-\frac{1 }{r^2f}\bigg(2q^2 \phi^2+\frac{(8B\alpha)^2}{r^4}\bigg)A_t=0\,.
\ee
Note that $V_z$ can be totally fixed by adding normalisable boundary condition.

\subsection{Equations for the fluctuations in the explicit $U(1)_A$ breaking model}

\subsubsection{Equations of motion for longitudinal fluctuations at zero momentum}

\label{secEqfluc1}
We have the following equations of motion for the {\em longitunial} fluctuations $a_t, a_z, v_t, v_z,\phi_1,\phi_2$ at zero momentum on the top of the background (\ref{eq:bgansatz})
\bea
 a_t''+\frac{3}{r}a_t'-\frac{2 q^2\phi^2}{r^2 f}a_t+\frac{8B\alpha }{r^3}v_z'-\frac{2q\phi}{r^2f}\Big(2qA_t\phi_1+i\omega\phi_2\Big)&=&0\,,\nonumber\\
 a_z''+\bigg(\frac{3}{r}+\frac{f'}{f}\bigg)a_z'+\Big(\omega^2-2 q^2r^2f\phi^2\Big)\frac{a_z}{r^4 f^2}+\frac{8B\alpha}{r^3f} v_t'-\frac{4q^2 A_z\phi}{r^2 f}\phi_1&=&0\,,\nonumber\\
\omega a_t'+\frac{8B\alpha \omega}{r^3}v_z+2i q r^2 f\big(-\phi_2\phi'+\phi\phi_2'\big)&=&0\,,\nonumber\\
v_z''+\bigg(\frac{3}{r}+\frac{f'}{f}\bigg)v_z'+\frac{\omega^2}{r^4 f^2}v_z+\frac{8B\alpha}{r^3f}a_t'
&=&0\,,\nonumber\\
v_t'+\frac{8B\alpha}{r^3}a_z
&=&0\,,\nonumber\\
  \phi_1''+\bigg(\frac{5}{r}+\frac{f'}{f}\bigg)\phi_1'+\Big(\omega^2+ q^2A_t^2-m^2r^2f-q^2fA_z^2\Big)\frac{\phi_1}{r^4 f^2}~~~&&\nonumber\\~~~~~~~~
  +\frac{1}{r^4f^2}
 \bigg(2q^2 \phi (A_t a_t-a_zfA_z)+2iq\omega \phi_2 A_t\bigg) &=&0\,,\nonumber\\
\phi_2''+\bigg(\frac{5}{r}+\frac{f'}{f}\bigg)\phi_2'+\Big(\omega^2
  + q^2A_t^2-m^2r^2f-q^2 f A_z^2\Big)\frac{\phi_2}{r^4 f^2}-\frac{iq\omega}{r^4f^2}
 \big(\phi a_t+2 A_t \phi_1\big)&=&0\,.\nonumber
\eea

\subsubsection{Equations of motion for transverse fluctuations at zero momentum}
\label{secEqfluc2}

The equations of motion for the {\em transverse} fluctuations $a_x, a_y, v_x, v_y$ at zero momentum  on the top of background (\ref{eq:bgansatz}) are the follows
\bea
a_x''+\bigg(\frac{3}{r}+\frac{f'}{f}\bigg)a_x'+\bigg(\frac{\omega^2}{r^2f}-2q^2\phi^2\bigg)\frac{a_x}{r^2f}+\frac{8i\omega\alpha}{r^3f}\bigg(V_z'v_y+A_z'a_y\bigg)&=&0\,,\\
a_y''+\bigg(\frac{3}{r}+\frac{f'}{f}\bigg)a_y'+\bigg(\frac{\omega^2}{r^2f}-2q^2\phi^2\bigg)\frac{a_y}{r^2f}-\frac{8i\omega\alpha}{r^3f}\bigg(V_z'v_x+A_z'a_x\bigg)&=&0\,\\
v_x''+\bigg(\frac{3}{r}+\frac{f'}{f}\bigg)v_x'+\frac{\omega^2}{r^4f^2}v_x+\frac{8i\omega\alpha}{r^3f}\bigg(V_z'a_y+A_z'v_y\bigg)&=&0\,,\\
v_y''+\bigg(\frac{3}{r}+\frac{f'}{f}\bigg)v_y'+\frac{\omega^2}{r^4f^2} v_y-\frac{8i\omega\alpha}{r^3f}\bigg(V_z'a_x+A_z'v_x\bigg)&=&0
\eea

After defining $a_\pm=a_x\pm i a_y, v_\pm=v_x\pm i v_y$, we have 

\bea
a_\pm''+\bigg(\frac{3}{r}+\frac{f'}{f}\bigg)a_\pm'+\bigg(\frac{\omega^2}{r^2f}-2q^2\phi^2\bigg)\frac{a_\pm}{r^2f}\pm\frac{8\omega\alpha}{r^3f}\bigg(V_z'v_\pm+A_z'a_\pm\bigg)&=&0\,,\\
v_\pm''+\bigg(\frac{3}{r}+\frac{f'}{f}\bigg)v_\pm'+\frac{\omega^2}{r^4f^2}v_\pm\pm\frac{8\omega\alpha}{r^3f}\bigg(V_z'a_\pm+A_z'v_\pm\bigg)&=&0\,,
\eea

For zero density case $V_t=A_t=V_z=A_z=0,$ by repeating the calculation in the appendix of \cite{Landsteiner:2014vua}, we have $\sigma_{xx}=\sigma_{yy}=\pi T$ and the Hall conductivity $\sigma_{xy}=0.$  
When $A_t=V_z=0$, i.e. with $\mu_5=0$ and $\mu\neq 0$, we have $\sigma_{xx}=\sigma_{yy}=\pi T$ and $\sigma_{xy}=\frac{\rho-\rho_h}{B}$ which is the same as the case without axial charge dissipation  \cite{Landsteiner:2014vua}. We do not have analytical solutions for other cases. 

\subsubsection{Equations for DC conductivity calculation}
\label{equationDC}

We have seven ODEs for the fields around the background (\ref{eq:bgansatz}) in which we assume the most general case with background $\mu$ and $\mu_5$. The equations are the following
\bea
 a_t''+\frac{3}{r}a_t'-\frac{2 q^2\phi^2}{r^2 f}a_t+\frac{8B\alpha }{r^3}v_z'-\frac{4q^2\phi A_t}{r^2f}\phi_1&=&0\,,\nonumber\\
 a_z''+\bigg(\frac{3}{r}+\frac{f'}{f}\bigg)a_z'-\frac{2 q^2\phi^2}{r^2 f}a_z+\frac{8B\alpha}{r^3f} v_t'-\frac{4q^2  A_z\phi}{r^2 f}\phi_1&=&0\,,\nonumber\\
-\frac{8B\alpha E}{r^5f}-2q^2\phi^2a_r+2q \big(-\phi_2\phi'+\phi\phi_2'\big)&=&0\,,\nonumber\\
v_t''+\frac{3}{r}v_t'+\frac{8B\alpha}{r^3}a_z'&=&0\,,\nonumber\\
v_z''+\bigg(\frac{3}{r}+\frac{f'}{f}\bigg)v_z'+\frac{8B\alpha}{r^3f}a_t'
&=&0\,,\nonumber\\
  \phi_1''+\bigg(\frac{5}{r}+\frac{f'}{f}\bigg)\phi_1'+\Big( q^2 A_t^2-m^2r^2f-q^2f A_z^2\Big)\frac{\phi_1}{r^4 f^2}
  +\frac{2q^2 \phi}{r^4f^2}
 \big(  A_t a_t-a_zf A_z\big) &=&0\,,\nonumber\\
\phi_2''+\bigg(\frac{5}{r}+\frac{f'}{f}\bigg)\phi_2'+\Big(q^2 A_t^2-m^2r^2f-q^2 f  A_z^2\Big)\frac{\phi_2}{r^4 f^2}-\bigg(\frac{5}{r}+\frac{f'}{f}\bigg)qa_r\phi-q\phi a_r'-2q a_r\phi'&=&0\,.\nonumber
\eea

\subsubsection{Equations for transverse fluctuations at finite $\omega$ and $k$}
\label{equationCME}

We consider transverse fluctuations $a_x, a_y, v_x, v_y$ with finite frequency and momentum on top of  (\ref{eq:bgansatz}) and the equations are
\bea
a_x''+\bigg(\frac{3}{r}+\frac{f'}{f}\bigg)a_x'+\bigg(\frac{\omega^2}{r^2f}-\frac{k^2}{r^2}-2q^2\phi^2\bigg)\frac{a_x}{r^2f}+\frac{8i\omega\alpha}{r^3f}\bigg(V_z'v_y+A_z'a_y\bigg)+\frac{8ik\alpha}{r^3f}\bigg(A_t'a_y+V_t'v_y\bigg)&=&0\,,\nonumber\\
a_y''+\bigg(\frac{3}{r}+\frac{f'}{f}\bigg)a_y'+\bigg(\frac{\omega^2}{r^2f}-\frac{k^2}{r^2}-2q^2\phi^2\bigg)\frac{a_y}{r^2f}-\frac{8i\omega\alpha}{r^3f}\bigg(V_z'v_x+A_z'a_x\bigg)-\frac{8ik\alpha}{r^3f}\bigg(A_t'a_x+V_t'v_x\bigg)&=&0\,,\nonumber\\
v_x''+\bigg(\frac{3}{r}+\frac{f'}{f}\bigg)v_x'+\bigg(\frac{\omega^2}{r^2f}-\frac{k^2}{r^2}\bigg)\frac{v_x}{r^2f}
+\frac{8i\omega\alpha}{r^3f}\bigg(V_z'a_y+A_z'v_y\bigg)+\frac{8ik\alpha}{r^3f}\bigg(A_t'v_y+V_t'a_y\bigg)&=&0\,,\nonumber\\
v_y''+\bigg(\frac{3}{r}+\frac{f'}{f}\bigg)v_y'+\bigg(\frac{\omega^2}{r^2f}-\frac{k^2}{r^2}\bigg)\frac{v_y}{r^2f}-\frac{8i\omega\alpha}{r^3f}\bigg(V_z'a_x+A_z'v_x\bigg)-\frac{8ik\alpha}{r^3f}\bigg(A_t'v_x+V_t'a_x\bigg)&=&0\,.\nonumber
\eea
After defining $a_\pm=a_x\pm i a_y, v_\pm=v_x\pm i v_y$, we have 
\bea
a_\pm''+\bigg(\frac{3}{r}+\frac{f'}{f}\bigg)a_\pm'+\bigg(\frac{\omega^2}{r^2f}-\frac{k^2}{r^2}-2q^2\phi^2\bigg)\frac{a_\pm}{r^2f}\pm\frac{8\omega\alpha}{r^3f}\bigg(V_z'v_\pm+A_z'a_\pm\bigg)
\pm\frac{8k\alpha}{r^3f}\bigg(A_t'a_\pm+V_t'v_\pm\bigg)&=&0\,,\nonumber\\
v_\pm''+\bigg(\frac{3}{r}+\frac{f'}{f}\bigg)v_\pm'+\bigg(\frac{\omega^2}{r^2f}-\frac{k^2}{r^2}\bigg)\frac{v_\pm}{r^2f}\pm\frac{8\omega\alpha}{r^3f}\bigg(V_z'a_\pm+A_z'v_\pm\bigg)
\pm\frac{8k\alpha}{r^3f}\bigg(A_t'v_\pm+V_t'a_\pm\bigg)&=&0\,,\nonumber
\eea

\subsection{Equations of fluctuations for massive $U(1)_A$}
\label{sec:eqmassiveU1}

The linearised equations of motion for the fluctuations at general $\omega$ and $k$ in the massive model read
\begin{align}
a_t'' + \frac{3}{r}a_t' - \left( \frac{k^2}{f r^4}+\frac{m^2}{r^2f}\right)a_t
-\frac{\omega k}{f r^4}a_z+\frac{8\alpha B}{r^3}v_z' +\frac{i \omega
m^2}{r^2f}\eta=0\,,\\
v_t'' + \frac{3}{r}v_t' -\frac{k^2}{f r^4}v_t -  \frac{\omega k}{f
r^4}v_z+\frac{8\alpha B}{r^3}a_z' =0\\
a_z''+\left( \frac{f'}{f}+\frac{3}{r}\right)a_z' + \left(
\frac{\omega^2}{r^4f^2}-\frac{m^2}{r^2f}\right)a_z+\frac{\omega k}{r^4f^2}a_t+\frac{8\alpha B}{f r^3}v_t'-\frac{i k m^2}{r^2f}\eta=0\,,\\
v_z''+\left( \frac{f'}{f}+\frac{3}{r}\right)v_z' +
\frac{\omega^2}{r^4f^2}v_z+\frac{\omega k}{r^4f^2}v_t+\frac{8\alpha B}{f r^3}a_t'=0\,,\\
\eta'' + \left(\frac{5}{r} +
\frac{f'}{f}\right)\eta'+\left(\frac{\omega^2}{r^4f^2}- \frac{k^2}{r^2f} \right)\eta
+\frac{i \omega}{r^4f^2}a_t + \frac{i k}{f r^3}a_z=0\,,
\end{align}

with two constraint equations
\begin{align}
\omega a_t' + k fa_z' +\frac{8 B\alpha}{r^3}\left( \omega v_z+k
v_t\right)-i m^2r^2 f \eta '=0\,,\\
\omega v_t' + k fv_z' +\frac{8 B\alpha}{r^3}\left( \omega a_z+k
a_t\right)=0\,.
\end{align}



\end{document}